\documentclass[lettersize,journal]{IEEEtran}

\usepackage{xurl}
\usepackage{cite}
\usepackage[dvipsnames]{xcolor}
\usepackage{titlesec}
\usepackage[utf8]{inputenc}
\usepackage[T1]{fontenc}
\usepackage{textcomp}
\usepackage{stfloats}

\usepackage{amsmath}
\usepackage{amsfonts}

\usepackage{newtxmath}

\usepackage{amsthm}
\usepackage{array}
\usepackage{multirow}
\usepackage{graphicx}
\usepackage{subcaption}
\usepackage{tablefootnote}
\usepackage[ruled,vlined,lined,commentsnumbered, algo2e]{algorithm2e}
\usepackage{algorithm}
\usepackage{algorithmicx}
\usepackage{esvect}
\usepackage{booktabs}
\usepackage{tabularx}
\usepackage{anyfontsize}
\usepackage[all]{nowidow}
\usepackage[normalem]{ulem}
\usepackage{xcolor}
\usepackage{colortbl}
\usepackage{enumitem}
\usepackage{makecell}
\usepackage[switch]{lineno}
\usepackage{framed}
\usepackage{wrapfig}
\usepackage{placeins} 
\usepackage{longtable}
\usepackage{diagbox}

\usepackage{gensymb}

\newcolumntype{Y}{>{\centering\arraybackslash}X}

\newcommand\timedomain{\ensuremath{\mathbb{T}}}

\usepackage{pifont}

\usepackage{listings}

\usepackage{framed}
\usepackage[most]{tcolorbox}
\usepackage{tikz}
\definecolor{light-gray}{gray}{0.9}
\usepackage[framemethod=TikZ]{mdframed}
\mdfsetup{skipabove=5pt,skipbelow=3pt}
\usepackage{lipsum}
\mdfdefinestyle{RQFrame}{%
	linecolor=black,
	outerlinewidth=0.15pt,
	roundcorner=3pt,
	innertopmargin=2pt,
	innerbottommargin=2pt,
	innerrightmargin=4pt,
	innerleftmargin=4pt,
	backgroundcolor=light-gray}

\definecolor{javagreen}{rgb}{0.25,0.5,0.35}

\newcommand\synt[1]{\textsf{#1}}

\newcommand{\app}[0]{GenTV}

\newtheorem{definition}{Definition}

\usepackage{hyperref}

\usepackage{balance}

\newcounter{commentnumber}

\begin{document}

\title{Automated Test Validators for Flaky Cyber-Physical System Simulators: Approach and Evaluation}

\author{Baharin~A. Jodat, Khouloud~Gaaloul, Mehrdad~Sabetzadeh, and Shiva~Nejati
\thanks{B.~A. Jodat, S. Nejati and M. Sabetzadeh are with University of Ottawa, Canada.\protect\\
E-mail: \{balia034, snejati, m.sabetzadeh\}@uottawa.ca}
\thanks{K.~Gaaloul is with University of Michigan at Dearborn, USA.\protect\\
E-mail: kgaaloul@umich.edu.}
}

\twocolumn

\maketitle
\begin{abstract}
Simulation-based testing of cyber-physical systems (CPS) is costly due to the time-consuming execution of CPS simulators. In addition, CPS simulators may be flaky, leading to inconsistent test outcomes and requiring repeated test re-execution for reliable test verdicts.  Many test inputs within the input space of CPS may not effectively exercise the behaviour of the system under test (SUT)  -- for instance, those that violate system preconditions, exceed operational design domain (ODD) limits, or represent inherently safe scenarios.
In this article, we propose to use test validators to filter out such test inputs before execution.  We describe two methods for generating test validators: one using genetic programming~(GP) that employs well-known spectrum-based fault localization (SBFL) ranking formulas, namely Ochiai, Tarantula, and Naish, as fitness functions; and the other using  decision trees (DT) and decision rules (DR). We evaluate our  test validators through case studies in the domains of aerospace, networking and autonomous driving. We show that test validators generated using GP with Ochiai are significantly more accurate than those generated using GP with Tarantula and Naish or using DT or DR. Moreover, this accuracy advantage remains even when accounting for the flakiness of the simulator. We further show that our test validators generated by GP with Ochiai are robust against flakiness with only 4\% average variation in their accuracy results across four different network and autonomous-driving systems with flaky behaviours. Finally, we show that, on average, 88.7\% of the assertions inferred by our approach align with or overlap with requirements precondition violations, ODD-limit violations, and   nominal safe conditions extracted from technical standards and empirical results in the literature.\\
Our full \textbf{replication package} is available online~\cite{github}.
\end{abstract}

\begin{IEEEkeywords}
    Cyber-Physical Systems, Test Validators, Flakiness, Environment Assumptions, Genetic Programming, Spectrum-Based Fault Localization
\end{IEEEkeywords}

\section{Introduction}
\label{sec:intro}
Testing cyber-physical systems (CPS)   involves exploring vast, multidimensional input spaces. However, many test inputs within these spaces fail to meaningfully exercise the system under test (SUT). Some violate the SUT's requirements preconditions, causing tests to pass or fail irrespective of the system's behaviour; others represent situations outside the system's operational design domain (ODD)~\cite{J3016_202104} or correspond to inherently low-risk, nominal scenarios. Such inputs can waste testing resources and distort confidence in the system's trustworthiness. To ensure that only meaningful tests are executed, we require a mechanism that filters out inputs violating preconditions, exceeding ODD limits, or producing vacuous results. We refer to such a mechanism as a \emph{test input validator}, or \emph{test validator} for short.

An input filtered out by the validator may be well-formed and within the system's input domain, yet not useful for exercising the SUT's behaviour. This notion of validity aligns with recent work in deep learning (DL) testing~\cite{WhenWhyTest2022,delaram,Dola_2021_DAIV}, where tests outside the DL model's training distribution are considered invalid.  Even though an out-of-distribution test may be meaningful on its own, any failure resulting from it would likely be due to the DL model's unfamiliarity, not a fault.

Validity for DL systems has primarily been studied for image inputs~\cite{WhenWhyTest2022,delaram,Dola_2021_DAIV,Stocco_2020_selforacle}.  We extend this notion to CPS, which are typically tested using simulators that make testing expensive and time-consuming.
Moreover, simulator non-determinism from environmental variability and stochastic processes can cause identical inputs to produce different outputs, leading to \emph{flaky} test outcomes~\cite{birchler2023machine, nguyen2021salvo, khatiri2024simulation, luo2014empirical, parry2021survey,strandberg2020intermittently, dutta2021flex, osikowicz2025empirically}. For flaky tests -- 
tests that pass or fail non-deterministically -- a single execution is often insufficient to determine the true outcome. The need for repeated executions and costly CPS simulations highlights the importance of a test validator that filters out non-essential test inputs for the SUT.

Tests that vacuously pass because they violate the SUT's preconditions, or trivially pass because they correspond to the SUT's nominal and low-risk conditions, can create a false sense of confidence in the system's trustworthiness. For example, vacuity may occur when testing an autopilot function in a scenario where the autopilot is never engaged, while a low-risk case may involve testing an autonomous driving system (ADS) collision requirement in a scenario where there are no nearby objects with which a collision is possible. Similarly, tests that fail because they violate the SUT's preconditions or exceed the ODD limits misrepresent trustworthiness by generating spurious errors. For example, to ensure the ascent requirement of an autopilot, a precondition is that sufficient initial throttle must be provided; otherwise, the test will fail. Likewise, an ADS travelling over 90 km/h at night in dense traffic exceeds its ODD limits and requires driver intervention. 

Existing standards and specifications often describe preconditions and ODDs qualitatively but rarely define them with the precision needed to construct automated test validators.  For example, a router specification may state that performance degrades under heavy streaming traffic without specifying which differentiated-services classes are affected or the threshold at which such degradation begins. To determine constraints and threshold bounds needed for constructing test validators, one must empirically infer input constraints that delineate nominal, boundary, and invalid regions of operation.

Prior data-driven approaches have inferred environmental assumptions and preconditions from simulation data using techniques such as genetic programming (GP)~\cite{metaheuristicsbook} and interpretable machine learning (ML) models (e.g., decision trees)~\cite{kapugama2022human, gopinath2020abstracting, kampmann2020does, tosem, gaaloul2021combining, enrich}.  However, these methods typically (1) treat inferred conditions as independent rules rather than as components of a test validator, and (2) overlook the impact of simulator flakiness on the reliability of the inferred rules. To our knowledge, no prior work has examined how label inconsistencies resulting from flaky test outcomes affect the robustness of learned input conditions.

\textbf{Contributions.}
We propose a data-driven method for constructing test validators. Each validator is a set of assertions -- arithmetic and logical predicates over system inputs -- that explains the pass/fail outcomes observed in simulation-based testing. 
Assertions with high predictive confidence of pass or fail verdicts are likely to delineate input regions corresponding to nominal or low-risk conditions, violations of ODD boundaries, or unmet preconditions.  We refer to our test validators as \emph{assertion-based test validators} and  infer them using (1)~GP, which evolves assertions through fitness optimization, and (2)~interpretable ML methods, namely decision trees (DT) and decision rules (DR), which extract human-readable conditions from training data.  As described below, we ensure three key properties of these validators: (1)~consistency of inferred verdicts, (2)~effectiveness in learning interpretable assertions from training data, and  (3)~applicability to signal-based CPS:

\textbf{(1)} Combining assertions into a set requires ensuring that the set remains consistent in the verdicts it issues. Assertions from DT are consistent by construction, whereas those from DR and GP may be conflicting, since these techniques generate assertions for passing and failing behaviours independently. \emph{We propose a pruning mechanism to prevent test validators from assigning conflicting verdicts (Section \ref{subsec:consistency}).}
    
\textbf{(2)} \emph{To improve the effectiveness of the test validators generated using GP,  we use spectrum-based fault-localization (SBFL) ranking formulas as the fitness functions of GP (Section \ref{subsec:failurerule}).} Specifically, we adopt three well-known SBFL ranking formulas from the literature~\cite{ochiai, tarantula, naish,LiuLNBB16}, namely \emph{Ochiai},  \emph{Tarantula}, and  \emph{Naish}. These formulas rank the suspiciousness of program statements based on their involvement in passing and failing executions. This ranking mechanism aligns with our goal of deriving assertions from training data that differentiate passing from failing behaviours.
    
\textbf{(3)} \emph{To show the applicability of our test validators to signal-based CPS, we present a formal characterization of the expressive power of these validators in capturing common CPS signal properties (Section \ref{sec:example}).} We formally show that, for CPS with piecewise-constant input signals, our assertions capture all logical operators as well as the ``globally'' temporal operator from Signal Temporal Logic (STL)~\cite{maler2004monitoring}. This level of expressiveness is sufficient to capture 85 of the 98 requirements in the Lockheed Martin benchmark~\cite{lockheedmartin}. The assumption of piecewise-constant input signals is widely adopted in CPS benchmarks and case studies in the literature~\cite{lockheedmartin, cruisecontroller, clutchlockup, guidancecontrol, dcmotor, khandait2024arch}.

\textbf{Findings.} Our evaluation focuses on two aspects: (1) the soundness and effectiveness of the proposed test validators in identifying violations of ODD limits, preconditions, and low-risk situations; and (2) their robustness to flakiness in training data obtained from simulators. To assess soundness and effectiveness, we measure both \emph{accuracy} -- how accurately the inferred test validators capture the input conditions that cause the SUT to pass or fail   -- and \emph{alignment} -- the degree to which the learned assertions align with the descriptions of precondition violations, ODD limits, and low-risk scenarios provided in  technical standards as well as in empirical and expert-validated studies~\cite{cakepaper, CAKE, ITU-T, J3016_202104, yoneda2019automated, chen2024modeling, lou2024autonomous, wang2025openlka, wang2025empirical, bhandari2020driving, boyapati2023automated, abdessalem2018testing, autopilotbenchmark, federal2009pilot}. We  measure \emph{robustness} as the variation in prediction accuracy across multiple test validators trained on datasets with identical test inputs but inconsistent pass or fail verdicts due to flakiness. Our robustness analysis aims to determine whether flaky tests -- due to potentially unreliable verdicts -- should be excluded from the initial training set used to develop test validators, or whether they can remain because their inclusion has minimal impact on the accuracy of the resulting test validator. 

Our evaluation compares GP-based and ML-based approaches for constructing test validators with respect to accuracy, robustness, and alignment, across five case-study systems spanning networking, autonomous driving, and aerospace domains (Section~\ref{sec:eval}). The main findings are summarized as follows:

\emph{Accuracy:} Using GP with Ochiai is most effective for generating assertion-based test validators. Test validators generated by GP with Ochiai are significantly more accurate than those generated by GP with Tarantula or Naish, as well as those generated by DT and DR. Further, GP with Ochiai misclassifies fewer failing tests as passing compared to other techniques, thus reducing the likelihood of masking failures as passes. 
Overall, our test validators with confidence above 90\% achieve an average accuracy of 89\% across case studies in aerospace, networking, and autonomous driving.

\emph{Robustness:} When using GP with Ochiai, the average variation in test validator accuracy is approximately 4\%. This indicates that the impact on the prediction accuracy of test validators, caused by pass or fail label inconsistencies, is on average 4\%.
    If such variation is acceptable, practitioners can forgo removing flaky tests -- which require multiple test executions -- from the initial training set, thus reducing costs.
 
\emph{Alignment:} Based on a systematic human-subject study, we show that, on average, 80\% of test-validator assertions with confidence above 90\% inferred by GP align with descriptions of preconditions, nominal conditions, and ODD-limit violations in technical standards and empirical or expert-validated studies~\cite{cakepaper, CAKE, ITU-T, J3016_202104, yoneda2019automated, chen2024modeling, lou2024autonomous, wang2025openlka, wang2025empirical, bhandari2020driving, boyapati2023automated, abdessalem2018testing, autopilotbenchmark, federal2009pilot}. Another 8.7\% partially match these sources, while the remaining 11.2\% describe conditions not covered in the documentation but deemed plausible low-risk situations or ODD-violation scenarios by the human evaluators in our study. Finally, none of the assertions with confidence higher than 90\% in our study contradicted the documentation.

Our full \textbf{replication package} is available online~\cite{github}.

\section{Assertion-based Test Validators}
\label{sec:assertionbasedtestoracles}
We define and illustrate assertion-based test validators, followed by an overview of their construction and evaluation.

\textbf{Defining Test Validators.} An \textit{assertion} for a system $S$ is an expression of the form $\mathit{condition} \Rightarrow  \mathit{verdict}$,  where  $\mathit{condition}$ is a logical predicate over the inputs of $S$ and $\mathit{verdict}$ is either pass or fail. For a test input $t$, the assertion $\mathit{condition} \Rightarrow  \mathit{fail}$ implies that if $t$ satisfies $\mathit{condition}$, then the verdict for $t$ is fail. Similarly, the assertion $\mathit{condition} \Rightarrow  \mathit{pass}$ implies that  if $t$ satisfies $\mathit{condition}$, then the verdict for $t$ is pass. We refer to an assertion with a fail verdict as a \emph{fail assertion} and an assertion with a pass verdict as a \emph{pass assertion}.

\begin{figure*}[t]
    \centering
    \hspace*{-1.5em}\includegraphics[width=0.85\linewidth]{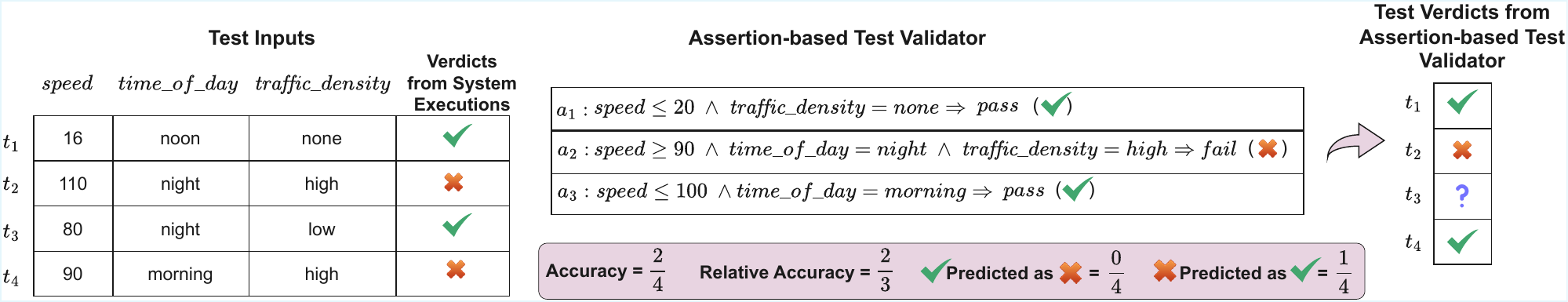}
    \caption{An assertion-based test validator for a simplified ADS with inputs $\mathit{speed}$, $\mathit{time\_of\_day}$ and $\mathit{traffic\_density}$, along with an example set of test inputs for this system. In this figure, \ding{51} indicates the pass verdict, \ding{55} indicates the fail verdict and \textbf{?} indicates that the test validator is inconclusive. }
    \label{fig:assertion}
    \vspace*{-.4cm}
\end{figure*}

Figure~\ref{fig:assertion} shows two passing assertions ($a_1$ and $a_3$) and one failing assertion ($a_2$) for a simplified ADS. These assertions are defined over three ADS inputs: ego-vehicle speed ($\mathit{speed}$), time of day ($\mathit{time\_of\_day}$), and traffic density ($\mathit{traffic\_density}$). For instance, the pass assertion $a_1$ specifies that when the ego vehicle is driving at $\leq 20~\mathrm{km/h}$ and there are no nearby vehicles, the scenario is low risk and the ADS is very unlikely to violate the collision requirement. Similarly, the fail assertion $a_2$ indicates that when the ego vehicle is driving faster than $90~\mathrm{km/h}$ at night under heavy traffic, the scenario is high risk, with a high likelihood of collision, because the situation is likely beyond the control of the underlying ADS.

\begin{definition}[\textbf{Test validator}]
\label{def:testoracle}
A \emph{test validator} $\mathcal{V}$ for a system $S$ is a set of assertions defined for $S$ whose members are pairwise \emph{consistent}.
Two assertions $\mathit{cnd} \Rightarrow \mathit{v}$ and $\mathit{cnd}' \Rightarrow \mathit{v'}$ are \emph{consistent} if $v \neq v'$ implies that the conjunction $\mathit{cnd} \wedge \mathit{cnd}'$ is unsatisfiable (UNSAT).
\end{definition}

Definition~\ref{def:testoracle} ensures that a test validator is consistent in the verdicts it produces. While a test input \( t \) may satisfy the conditions of multiple assertions, all such assertions must agree on the verdict -- either all pass or all fail. For example, the assertions $a_1$, $a_2$ and $a_3$ in Figure~\ref{fig:assertion} constitute a test validator as  they are pairwise consistent. However, assertion \( a_4: \mathit{speed} > 15 \land \mathit{time\_of\_day} = \mathit{morning} \Rightarrow \mathit{fail} \), not shown in the figure,  is inconsistent with \( a_3 \), since the conjunction of the conditions of \( a_3 \) and \( a_4 \) is satisfiable, and they predict opposing verdicts (pass and fail, respectively).

Given a test validator $\mathcal{V}$ and a test input $t$, if $t$ satisfies the condition of \textit{any} pass assertion ($\mathit{cnd} \Rightarrow \mathit{pass}$) in $\mathcal{V}$, then the verdict for $t$ is \textit{pass}. Conversely, if $t$ satisfies the condition of \textit{any} fail assertion ($\mathit{cnd} \Rightarrow \mathit{fail}$) in $\mathcal{V}$, then the verdict for $t$ is \textit{fail}. If $t$ does not satisfy the condition of any assertion in $\mathcal{V}$, the verdict is considered \emph{inconclusive}. 
In other words, inconclusive tests are those that the validator cannot classify as passing or failing. These tests are suitable candidates for execution on the SUT, as they are likely to exercise its behaviour.
For example, Figure~\ref{fig:assertion}  shows four test inputs $t_1$ to $t_4$  for the ADS.  The test validator in the figure implies that $t_1$ and $t_4$ pass (satisfying $a_1$ and $a_3$), $t_2$ fails (satisfying $a_2$), and $t_3$ is inconclusive.

\textbf{Constructing Test Validators.}  We use a data-driven method to learn assertions from the training data of test inputs labelled  with pass/fail verdicts. Each assertion is assigned a confidence score equal to the precision of the assertion as measured on the training set. We add an assertion to the test validator only if its confidence score is above a user-defined \emph{verdict threshold} as defined below.

\begin{definition}[\textbf{Verdict threshold}]
\label{def:verdictthreshold}
Let $\mathcal{V}$ be an assertion-based test validator. The \textit{verdict threshold} $\theta$ is a minimum confidence level such that only assertions whose confidence levels meet or exceed $\theta$ are retained in $\mathcal{V}$.
\end{definition}

\textbf{Metrics for Test Validators.} 
As discussed in Section~\ref{sec:intro}, we evaluate  test validators in terms of their accuracy, robustness, and alignment. Alignment is assessed \emph{qualitatively} by comparing  test validators with  system specifications, technical standards and empirical results in the literature.  We assess accuracy and robustness \emph{quantitatively} against test sets obtained through system execution using the following metrics:

We measure prediction accuracy in two ways: (1) using the \emph{accuracy} metric, which measures the percentage of correctly predicted verdicts across all tests (capturing losses from both incorrect and inconclusive predictions), and (2) using the \emph{relative accuracy} metric, which measures the percentage of correctly predicted verdicts restricted to conclusive predictions (capturing only incorrect predictions). In addition, we report the \emph{misprediction rate} metric, defined as the percentage of incorrectly predicted verdicts across all tests, distinguishing between (1) passing tests wrongly predicted as failing and (2) failing tests wrongly predicted as passing.

For example, based on the verdicts from the SUT execution shown in Figure~\ref{fig:assertion}, the test validator correctly predicts the verdicts for $t_1$ and $t_2$, incorrectly predicts the verdict for $t_4$, and cannot conclusively predict the verdict for $t_3$.  Thus, the accuracy of the test validator in Figure~\ref{fig:assertion} is $\frac{2}{4}$ and its relative accuracy is $\frac{2}{3}$. In addition, the rate of the pass tests wrongly predicted as fail and the rate of fail tests wrongly predicted as pass  are $\frac{0}{4}$ and $\frac{1}{4}$, respectively.

For robustness, we measure the variation in prediction accuracy across multiple test validators trained on datasets with identical test inputs but inconsistent pass/fail verdicts due to flakiness. A robust test validator should maintain consistent accuracy, showing minimal variation despite such inconsistencies in the training data.

\section{Generating Assertion-based Test Validators}
\label{sec:approach}
\begin{figure}[t]
    \centering
    \hspace*{-1.2em}\includegraphics[width=1.1\columnwidth]{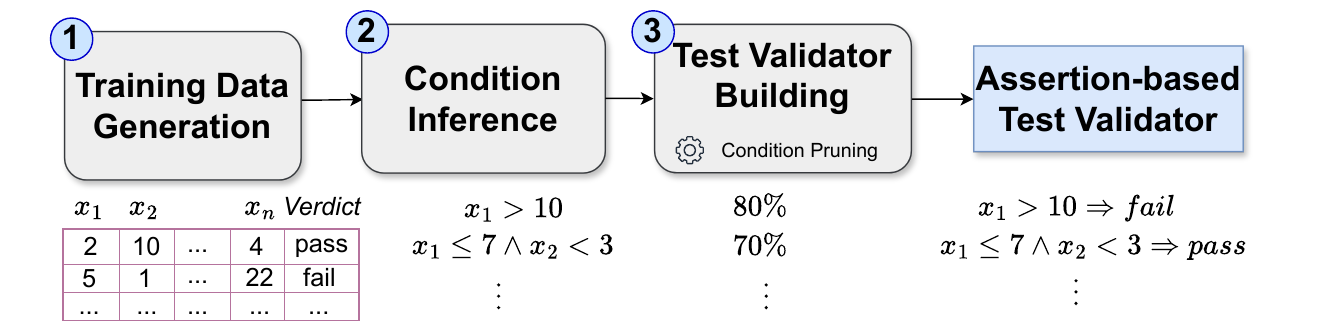}
    \caption{Our approach for  deriving  test validators, \app. }
    \label{fig:threeapproach}
    \vspace*{-.2cm}
\end{figure}

Figure~\ref{fig:threeapproach} shows an overview of \app, our framework for generating assertion-based test validators. The framework has three steps. The first step of \app\ uses adaptive random testing~\cite{metaheuristicsbook} to generate a training set of test inputs. Adaptive random testing randomly samples test inputs from the search space by maximizing the Euclidean distance between the sampled test inputs, hence increasing diversity among the generated tests.
To label these test inputs as either pass or fail, they are executed on the SUT.

The second step of \app\ uses the training data generated in the first step to learn conditions for assertions.  We consider  two alternatives for condition inference: (1)~using genetic programming (GP) presented in Sections~\ref{subsec:cndGP}, and (2)~using Decision Trees (DT) and Decision Rules (DR) presented in Section~\ref{subsec:cndDT}. The third and final step of \app, described in Section~\ref{subsec:consistency}, filters out assertions whose confidence levels fall below a specified verdict threshold $\theta$, and then applies a pruning algorithm to ensure that the remaining assertions in the test validator are mutually consistent.

\subsection{Condition Inference by Genetic Programming (GP)}
\label{subsec:cndGP}
The first alternative we consider for generating conditions defined over the SUT’s input variables is GP. GP requires a grammar to define its candidate solutions, which, in our work, are conditions over the SUT’s input variables. Below, we first present the grammar that \app\ uses for GP and then explain how GP learns assertion conditions based on the training set generated in the first step.

\subsubsection{Grammar Specification} 
\label{sec:grammar}
Since our work targets CPS, where inputs are typically numeric~\cite{tosem,gaaloul2021combining}, we adopt the grammar in Figure~\ref{fig:grammar} (denoted $\mathcal{G}$), previously used to specify environmental assumptions in CPS~\cite{gaaloul2021combining} and control logic in network systems~\cite{li2024using}. Grammar $\mathcal{G}$ generates conditions that are either conjunctions of relational expressions over arithmetic terms or disjunctions of such conjunctions. 

\begin{figure}[t]
\centering
\scalebox{0.77}{
\begin{tabular}{l@{\ }l@{\ } l@{\ } p{0.05mm} l@{\ } l@{\ } p{63mm}}
    \synt{or-term} &:: = & \synt{or-term} $\vee$ \synt{or-term} $|$ \synt{and-term} \\
    \synt{and-term} &:: = &
    \synt{and-term} $\wedge$ \synt{and-term} $|$ \synt{rel-term}
    \\
    \synt{rel-term} &:: = & 
    \synt{exp} $<$ $0$ $|$ \synt{exp} $\leq$ $0$ $|$ \synt{exp} $>$ $0$ $|$ \synt{exp} $\geq$ $0$ $|$ \synt{exp} $=$ $0$ $|$ \synt{exp} $\neq$ $0$
    \\
    \synt{exp} &:: = &  
    \synt{exp} $+$ \synt{exp} $|$ \synt{exp} $-$ \synt{exp} $|$ \synt{exp} $*$ \synt{exp} $|$ \synt{exp} $/$ \synt{exp} $|$ \synt{const} $|$ \synt{cp}
\end{tabular}
}
\caption{Syntactic rules of the grammar (denoted by $\mathcal{G}$) that define the assertions over system input variables. The symbol $|$ separates alternatives, \synt{const} is an ephemeral random constant generator and \synt{cp} represents 
an input variable of the SUT.}
\label{fig:grammar}
\vspace*{-.3cm}
\end{figure}

\subsubsection{Condition Learning}
\label{subsec:failurerule}
We use standard Genetic Programming (GP)~\cite{o2009riccardo} to infer conditions that explain the pass and fail verdicts in the training set $TS$, generated in the first step of \app\ (Figure~\ref{fig:threeapproach}). The core of GP is a fitness function $F$ that evaluates how well each candidate conditions explain the pass and fail verdicts in $TS$. Specifically, GP employs two complementary fitness functions: one aimed at generating conditions that explain the fail results, and another aimed at generating conditions that explain the pass results. We run GP separately with each fitness function to independently derive the pass and fail conditions.

Following the standard practice for expressing meta-heuristic search problems~\cite{Harman2012}, we define the individual representation, the genetic operators and fitness function of GP: 

\textit{Individual Representation. }A GP  individual represents a condition created by following the grammar $\mathcal{G}$ in Figure~\ref{fig:grammar}. The initial population is formed by randomly constructing parse trees employing the grow method~\cite{o2009riccardo}. 

\textit{Genetic Operators.} We use one-point crossover as well as one-point mutation for population breeding. These operators are adopted from the prior application of GP to similar applications, particularly in learning environmental assumptions for CPS~\cite{gaaloul2021combining}. To ensure that the generated candidate solutions comply with GP's grammar, we verify them during breeding and mutation and discard any invalid ones.

\textit{Fitness Functions.}  Figure~\ref{fig:rankingmetrics} presents the fitness functions we use for  GP-based condition inference. These fitness functions are adopted from the spectrum-based fault localization (SBFL) literature~\cite{landsberg2015evaluation,tarantula, ochiai, naish} and  are presented for the fail verdict, noting that the fitness functions for the pass verdict are the duals of those for the fail verdict. 

\begin{figure}[t]
    \centering
    \footnotesize
    \begin{align*}
        \mathit{Tarantula}(c) &= 
        \frac{\frac{c_f (c)}{|\mathit{TS}_f|}}
             {\frac{c_p (c)}{|\mathit{TS}_p|} + \frac{c_f (c)}{|\mathit{TS}_f|}}
        &
        \mathit{Ochiai}(c) &= 
        \frac{c_f (c)}
             {\sqrt{|\mathit{TS}_f| \cdot (c_p (c) + c_f (c))}}
        \\
        \mathit{Naish}(c) &= 
        \frac{c_f (c)}
             {|\mathit{TS}_f|} - 
        \frac{c_p (c)}
             {1 + |\mathit{TS}_p|}
    \end{align*}
        
    \caption{Fitness functions for \app, adopted from the spectrum-based fault localization (SBFL) literature.The $c_p (c)$ and $c_f (c)$ functions respectively represent the number of passing and failing tests in the training set $\mathit{TS}$  satisfying the condition~$c$. The $\mathit{TS}_f$ and $\mathit{TS}_p$ sets indicate the set  of failing and passing tests in $\mathit{TS}$, respectively. That is, 
    $ \mathit{TS} = \mathit{TS}_p \cup \mathit{TS}_f$.}
    \label{fig:rankingmetrics}
    \vspace*{-.3cm}
\end{figure}

SBFL aims to identify program statements most likely responsible for program failures. Given a program spectrum -- sequences of statements executed by test cases labelled as pass or fail -- an SBFL ranking function assigns a suspiciousness score to each statement. The most commonly used ranking functions in SBFL are Tarantula~\cite{tarantula}, Ochiai~\cite{ochiai}, and Naish~\cite{naish} shown in Figure~\ref{fig:rankingmetrics}. In these ranking functions, $c$ represents a program statement, and $c_p(c)$ and $c_f(c)$ represent the number of passing and failing tests that execute $c$, respectively. A statement with a high suspiciousness score is one executed by many failing tests but few passing ones.

\begin{figure}[t]
    \centering
    \includegraphics[width=\columnwidth]{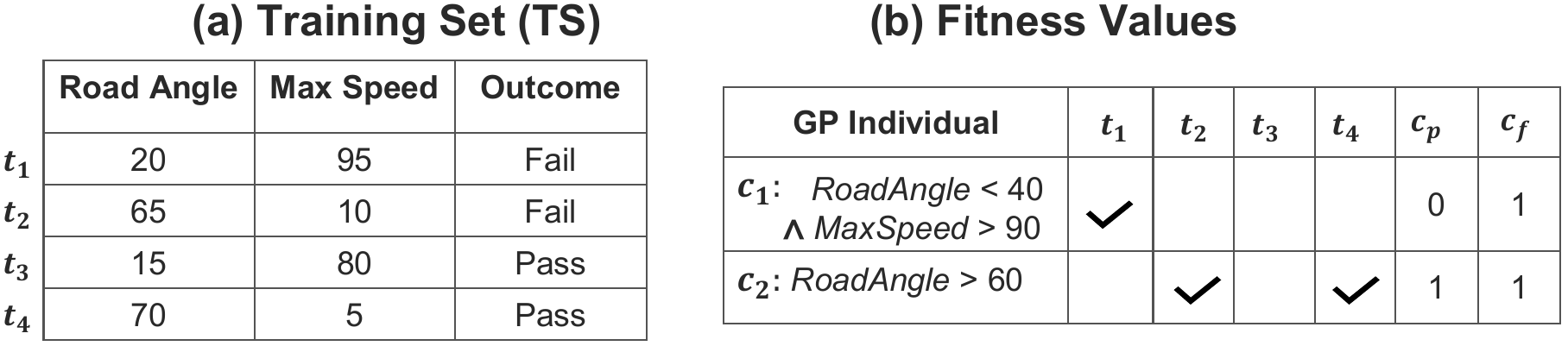}
    \caption{Computing $c_p(c)$ and $c_f(c)$ in SBFL fitness functions in Figure~\ref{fig:rankingmetrics} for our GP-based condition inference}
    \label{fig:susexample2}
    \vspace*{-.3cm}
\end{figure}

To use the functions in Figure~\ref{fig:rankingmetrics} for condition inference, we interpret $c$ as a candidate condition within the GP's population. The functions $c_p(c)$ and $c_f(c)$ then compute, respectively, the number of passing and failing test inputs in the training set $\mathit{TS}$ that satisfy $c$.
Figure~\ref{fig:susexample2} shows how  $c_p$ and $c_f$ are calculated for two GP individuals, namely $c_1$ and $c_2$.  
As shown in the figure, $t_1$ satisfies $c_1$, and  $t_2$ and $t_4$ satisfy $c_2$. Hence, we have $c_p(c_1)=0$, $c_f(c_1)=1$,  $c_p(c_2)=1$ and $c_f(c_2)=1$. Any of the SBFL ranking functions in Figure~\ref{fig:rankingmetrics} can then be used to compute a fitness value for conditions $c_1$ and $c_2$. 

SBFL functions assign high values to conditions met by many failing and few passing test inputs in the training set. By using SBFL ranking functions as fitness functions, our GP-based approach selects conditions that are more likely to explain and characterize failures effectively. For fitness functions that explain passing test cases, we swap $c_f$ with $c_p$ and swap $\mathit{TS}_p$ with $\mathit{TS}_f$ in the functions of Figure~\ref{fig:rankingmetrics}.

\subsection{Condition Inference by Interpretable ML}
\label{subsec:cndDT}
The second alternative we consider for generating conditions defined over the SUT’s input variables is interpretable ML. Specifically, we consider decision trees (DT) and decision rules (DR) as two forms of interpretable supervised ML models. Inferring conditions using either DT or DR involves two steps: (1)~\emph{Feature Engineering:} We define input features for learning, initially using the system's input variables as default features. However, if only these variables are used, DT and DR can learn only simple conditions that relate a single variable to a constant through a relational operator. For systems where the relationship between input variables and test outcomes is more complex, feature engineering becomes crucial. This process involves creating features that combine input variables using mathematical operators,
allowing DT and DR to learn conditions based on these arithmetic combinations, similar to those generated by the grammar $\mathcal{G}$ in Figure~\ref{fig:grammar}. (2)~\emph{Model Training and Condition Generation:} With the input features established, we train the DT or DR models using the training set $\mathit{TS}$ generated in the first step of \app. DT and DR generate conditions as disjunctions of conjunctions of expressions that relate input features to constants, similar to conditions generated by the grammar $\mathcal{G}$. 
These models can produce conditions for both pass and fail verdicts.

\subsection{Test Validator Building}
\label{subsec:consistency}
In the third and final step, we construct consistent assertion-based test validators from the conditions generated in the second step of \app. To do so, we first calculate each condition's confidence level based on its precision in classifying pass or fail tests in the training set. Specifically, if condition $\mathit{cnd}$ is associated with a fail (or pass) verdict, its confidence level is the percentage of actual fail (or pass) test inputs in the training set that satisfy $\mathit{cnd}$, relative to all test inputs that satisfy $\mathit{cnd}$. For instance, if the condition $c_2$  in Figure~\ref{fig:susexample2} is associated with a pass verdict, its confidence level is $50\%$ because among the two tests in the training set that satisfy $c_2$ (i.e., $t_2$ and $t_4$), only $t_4$ has a pass verdict.  Then, we retain only those conditions whose confidence levels meet or exceed the user-defined verdict threshold $\theta$. Recall from Definition~\ref{def:verdictthreshold} that $\theta$ specifies the minimum confidence level required for a condition to be included in the assertion-based test validator.

Next, we  apply a pruning strategy  to obtain a consistent set of assertions in the test validator. Recall from Definition~\ref{def:testoracle} that a consistent test validator ensures no conflicting assertions exist, where one assertion indicates a test input passes while another indicates it fails.
While DT-inferred assertions are consistent by construction, those inferred by GP or DR may require pruning to maintain consistency. The formalization and details of our pruning algorithm is provided in Appendix~\ref{apx:prune}. Below, we summarize the pruning process and exemplify it. 

To identify inconsistencies, we represent all assertions as a bipartite graph $\mathcal{B} = (V, E)$, where vertices represent pass ($V_p$) and fail ($V_f$) conditions, and edges connect pairs whose conjunction is satisfiable -- indicating a potential conflict. The graph thus encodes which pass and fail assertions can simultaneously be true.

Our pruning strategy iteratively removes assertions (i.e., vertices) to eliminate all conflicts (i.e., edges) while discarding as few conditions as possible. In each iteration, the algorithm prioritizes removal of the simplest and most conflicting assertions -- those with the shortest conditions and highest degrees in the graph. Preference is given to removing pass assertions when ties occur, as fail conditions are typically less prevalent and more informative. The process terminates when no conflicting pairs remain, yielding a consistent subset of assertions. Since, at each iteration, the algorithm removes exactly one vertex with at least one incident edge, it terminates after at most $|E|$ iterations, where $|E|$ is the number of edges in the bipartite graph. 

\begin{figure}[t]
    \centering
    \includegraphics[width=0.7\columnwidth]{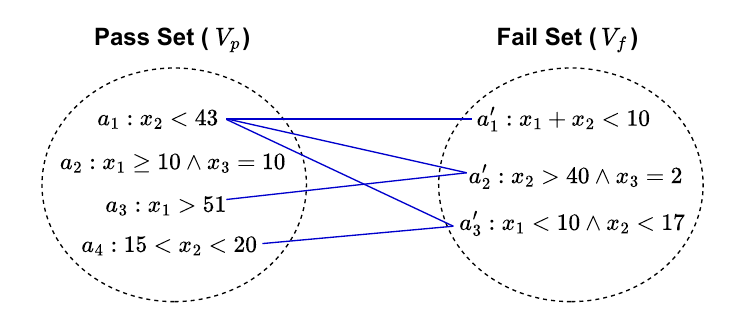}
    \caption{Pruning inconsistent assertions from test validators using a bipartite graph representation.}
    \label{fig:inconsistency}
    \vspace*{-.3cm}
\end{figure}

Figure~\ref{fig:inconsistency} illustrates a bipartite graph where pass and fail assertions are connected by edges representing inconsistent pairs of assertions. In this example, our pruning algorithm removes vertex $a_1$  first because it has the shortest condition and the highest degree. Next, $a_3$ is removed since it has the highest degree and a condition shorter than those of the vertices with the same degree, i.e., $a'_2$, $a_4$ and $a'_3$. Finally, $a_4$ is removed as it belongs to the pass class and has the same length and degree as $a'_3$. After removing $a_1$, $a_3$ and $a_4$, the remaining conditions are consistent.

\section{Test Validators for Signal-based CPS}
\label{sec:example}
In this section, we adapt Definition~\ref{def:testoracle} to signal-based CPS. While Definition~\ref{def:testoracle} defines assertion-based test validators for discrete-input CPS, signal-based systems require a formulation that accounts for inputs represented as  signals. In the following,  we introduce a running example to illustrate assertion-based test validators for such systems, explain how assertions are specified over signals, and discuss the expressiveness of these assertions in capturing signal properties.

\textbf{Motivating example.} To illustrate assertion-based test validators for CPS with signal-based inputs, we use a simplified autopilot controller, \textsc{Autopilot}, as our running example.  \textsc{Autopilot} has time-varying input signals: \emph{throttle} (engine power adjustment), \emph{pitchwheel} (nose tilt), and \emph{apeng} (autopilot engagement state). When \emph{apeng} indicates that the autopilot is engaged, \textsc{Autopilot} uses the throttle and pitchwheel signals to issue actuator commands that control the aircraft’s orientation and motion. \textsc{Autopilot} has a requirement stating that when autopilot is engaged, the aircraft should reach a specified altitude within 500 seconds.

\begin{figure*}[t]
    \centering
    \hspace*{-1.5em}\includegraphics[width=0.8\linewidth]{Figures/example2-signal-piecewise-500second-simplified.pdf}
    \caption{(a) Two test inputs, $t_1$ and $t_2$ for an autopilot system, with input signals \includegraphics[height=1.5ex]{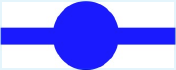} $\mathit{Throttle}$, \includegraphics[height=1.5ex]{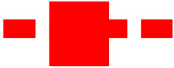} $\mathit{PitchWheel}$, and \includegraphics[height=1.5ex]{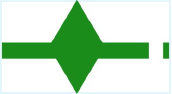} $\mathit{APEng}$; (b) An assertion-based test validator for this system. According to the test validator, $t_1$ fails and $t_2$ passes.}
    \label{fig:background}
    \vspace*{-.4cm}
\end{figure*}

Figure~\ref{fig:background} shows three assertions over \textsc{Autopilot}’s inputs: throttle ($\mathit{th}$), pitchwheel ($p$), and apeng ($e$). Note that these three signal variables ($\mathit{th}$, $p$, and $e$) are time dependent. Assertion $a_1$ characterizes a situation in which the autopilot is engaged but engine power is insufficient, violating the ascent requirement’s precondition of sufficient thrust. Assertion $a_2$ characterizes a situation in which the autopilot is engaged while the aircraft’s nose remains sharply pitched downward for 300 seconds, violating the precondition that the nose should point upward during ascent. Finally, assertion $a_3$ characterizes a safe situation in which the autopilot is disengaged, so the ascent requirement is vacuously satisfied.

Together, the assertions in Figure~\ref{fig:background}  form an assertion-based test validator for \textsc{Autopilot}. Figure~\ref{fig:background} further presents two test inputs ($t_1$ and $t_2$), each displaying signals for throttle, pitchwheel, and apeng. The assertions determine that $t_1$ fails while $t_2$ passes without executing the system.

\textbf{Assertions over signals.} 
 Let $M$ be a CPS with input signals. We denote each test input for $M$ as $\bar{u}=\{u_1, u_2, \ldots, u_m\}$ where each $u_i$ is a signal for an input of $M$ over a time domain $\timedomain=[0,b]$, i.e., $u_i: \timedomain \rightarrow \mathbb{R}$. For example, each test input for the \textsc{Autopilot} example in Figure~\ref{fig:background} represents the values for the three signals: $\mathit{th}$, $p$ and $e$,  over the time domain $\timedomain=[0,500s]$.  To generate signals,  we use \emph{control-point encoding}~\cite{tuncali2019requirements, arrieta2019pareto} --  a common approach in signal processing and control theory that captures signals in a compact and efficient representation. Specifically, we encode signals as sequences of equally spaced control points, where each index denotes a fixed time interval and each corresponding value approximates the signal's value during that time interval. Following this encoding, to generate signals, it suffices to generate the control points of the signal. Provided with the control points, the actual signals are constructed through interpolation.
An interpolation function (e.g., piecewise constant, linear, or cubic) connects the control points to form a signal~\cite{tuncali2019requirements, arrieta2019pareto,  miningassumption}.

In this article, we assume the interpolation function for system inputs is piecewise constant. Our evaluation of major CPS benchmarks and case studies from the literature indicates that most assume inputs to be piecewise constant. Notably, the Lockheed Martin industrial CPS benchmark and the MathWorks' publicly available CPS models for automotive driving systems~\cite{cruisecontroller,clutchlockup,guidancecontrol,dcmotor} 
 use piecewise-constant input signals. The ARCH benchmark~\cite{khandait2024arch}, which includes seven hybrid CPS models, compares arbitrary piecewise continuous and piecewise constant interpolations. Results from ARCH-COMP19~\cite{ernst2019arch} show minimal performance differences, supporting our assumption of piecewise-constant signals.

Let $u: \timedomain \rightarrow \mathbb{R}$ be an input signal. We encode $u$ using $n_u$ control points, i.e., $c_{u,0}$, $c_{u,1}$, \ldots, $c_{u,n_u-1}$, 
equally distributed over the time domain $\mathbb{T}=[0,b]$, i.e., positioned at a fixed time distance $I=\frac{b}{n_u-1}$.
Let $c_{x,y}$ be a control point, $x$ is the signal the control point refers to, and $y$ is the position of the control point.  The control points $c_{u,0}$, $c_{u,1}$, \ldots, $c_{u,n_u-1}$ respectively contain the values of the signal $u$ at time instants $0,I,2\cdot I, \ldots ,(n_u-1) \cdot I$.  For example, in the test inputs in Figure~\ref{fig:background},  pitchwheel ($p$) and apeng ($e$) are represented using six control points (denoted by $c_{p,0}, \ldots, c_{p,5}$ for the pitchwheel and $c_{e,0}, \ldots, c_{e,5}$ for the apeng), while throttle ($\mathit{th}$) is represented using three control points ($c_{th,0}, \ldots, c_{th,2}$).

Assertion conditions over signals are generated according to the syntactic rules of grammar $\mathcal{G}$ in Figure~\ref{fig:grammar} by using \synt{cp} to represent signal control points.  Furthermore, to ensure the well-formedness of the conditions, we constrain each arithmetic expression, \synt{exp},  to contain only signal control points at the same position.  For example, the condition \hbox{$(c_{e,0} - c_{p,0} \ge 0) \;\land\; (c_{e,1} + c_{th,1} < 1)$} can be generated by our grammar and satisfies the constraint that each arithmetic expression must involve control points associated with the same position; $c_{e, 0}$ and $c_{p, 0}$ in the first expression are control points at position 0, and $c_{e,1}$ and $c_{th,1}$ in the second expression are control points at position 1.

\textbf{Expressive power of assertions over signals.} 
We characterize the expressive power of the assertions generated using the grammar $\mathcal{G}$ (Figure~\ref{fig:grammar}) over signal control points through the following two steps:

\emph{First,} we translate assertion conditions defined over control points into constraints defined directly over signal variables. This translation is carried out using two sets of rewriting rules presented in Appendix~\ref{subapx:translation}. For example, using these rewriting rules, the following condition defined over control points of pitchwheel ($p$) and apeng ($e$):

$c_{p, 2} < -0.5 \land c_{p, 3} < -0.5 \land c_{p, 4} < -0.5 \land c_{e, 0} = 1.0 \land c_{e, 1} = 1.0 \land c_{e, 2} = 1.0 \land c_{e, 3} = 1.0 \land c_{e, 4} = 1.0 $ 

is converted into the following logical formula defined over the signal variables $p$ and $e$: $(\forall t \in [200, 500) : p(t) < -0.5) \land (\forall t \in [0, 500) : e(t) = 1.0)$

\emph{Second,} after applying the rewriting rules in Appendix~\ref{subapx:translation}, the generated formulas are expressible within the following logic fragment, denoted by $\mathcal{L}$:

\scalebox{.95}{\parbox{.5\linewidth}{%
\begin{align}
& \psi \Coloneqq &&  \psi_1 \vee \psi_2  \mid \phi & \nonumber \\
&\phi  \Coloneqq && \rho  \sim 0 \mid \phi_1 \wedge \phi_2  \mid \forall t \in \langle n_1, n_2 \rangle \colon \phi 
   & \nonumber \\
&\rho   \Coloneqq &&   u(t) \mid r \mid  \rho_1 + \rho_2 \mid \rho_1 - \rho_2 \mid   \rho_1 \times \rho_2 \mid  \rho_1 / \rho_2& \nonumber 
\end{align}}}

where $u$ is a signal over time domain \timedomain,  
$t$ is a  time variable, $n_1, n_2$ are non-negative real numbers including zero, $r \in \mathbb{R}$,  $\sim$~is a relational operator in $\{< , \leq, >, \geq, =, \neq \}$, and $\langle n_1, n_2 \rangle$ is a \emph{time interval} of \timedomain\ (i.e., $\langle n_1, n_2 \rangle \subseteq \timedomain$). 
The symbols $\langle$ and $\rangle$  are equal to $[$ or $($, respectively to $]$ or $)$, depending on whether  $n_1$, respectively  $n_2$, are included  or excluded from the interval. 

Any formula obtained by  grammar $\mathcal{G}$ and modified through our rewriting rules is a formula in $\mathcal{L}$. Conversely, any formula $\varphi \in\mathcal{L}$ that satisfies the following two conditions corresponds to a condition that can be generated by the grammar $\mathcal{G}$: (1)~$\varphi$ is closed, i.e., does not contain any free occurrence of the variable $t$, and (2)~$\varphi$  does not involve any nested use of the $\forall$ quantifier. The proof is provided in Appendix~\ref{subapx:equivalence}.

The logic $\mathcal{L}$ is able to express the temporal operator globally, i.e., $G$, from Signal Temporal Logic (STL)~\cite{maler2004monitoring}. In addition, $\mathcal{L}$ can express arithmetic operations within predicates, which are not part of the core STL formula syntax, thus extending STL with explicit arithmetic expression support. To demonstrate that $\mathcal{L}$ is capable of expressing common CPS properties, we assessed a dataset of 98 industrial CPS requirements previously used by Menghi et al.~\cite{menghi2019generating}. Menghi et al. formalized this dataset in restricted signal first-order logic, a logic fragment proposed in their study. Of the 98 formalized requirements, 85 can be expressed in our logic $\mathcal{L}$. The remaining 13 cannot, as they rely on existential quantifiers ($\exists$) or nested universal quantifiers ($\forall$), which lie outside the scope of $\mathcal{L}$. The fact that $86$\% of the industrial CPS requirements can be expressed in $\mathcal{L}$ suggests that the logic fragment $\mathcal{L}$  has the level of expressiveness necessary to capture a wide range of real-world CPS properties.

\section{Evaluation}
\label{sec:eval}
We address the following four research questions:

\textbf{RQ1 (Existence of Flakiness)} \textit{How flaky are our case-study systems?} We assess the level of flakiness in our network, ADS and aerospace case studies by calculating the percentage of inconsistent test verdicts from multiple re-executions of randomly selected test inputs.

\textbf{RQ2 (Accuracy)} 
\textit{How accurate are the assertion-based test validators inferred by our approach using different condition-inference methods?} In RQ2, we evaluate the accuracy  of test validators generated using the condition-inference techniques described in Sections~\ref{subsec:cndGP} and~\ref{subsec:cndDT}.

\textbf{RQ3 (Robustness to Flakiness)} \textit{How is the accuracy of test validator assertions impacted when using training sets from different executions of the SUT?}  We assess whether flakiness in training sets impacts the accuracy of test validators. Since a flaky SUT can yield different outcomes across runs, the generated validators should remain robust regardless of which run the training data comes from. We examine the robustness of our assertion-generation technique to ensure its accuracy is consistent across different SUT executions.

\textbf{RQ4 (Alignment)} \textit{To what extent do the assertion-based test validators align with violations of ODD limits, violations of preconditions, and low-risk scenarios in the real world?} We evaluate how closely the generated test-validator assertions for our case studies align with the descriptions of precondition violations, ODD limits, and low-risk scenarios provided in the reference documentation. This documentation includes technical standards as well as empirical and expert-validated results from the literature that define the operational limits and system requirements for our case studies. To determine the alignment of the test-validator assertions, we translate a representative set of assertions from our experiments into natural language and assess their alignment with the corresponding reference documentation through a systematic human-subject study.

\subsection{Case-Study Systems}
\label{sec:studysubjects}

We use five networking,  aerospace and ADS systems as our case-study systems. We present these systems below:

\textbf{Router system.} Our first case-study system is a router optimized for real-time streaming applications, such as video conferencing and online meetings. The router uses priority-based flow management, dividing the incoming traffic into different priority classes.  The router's real-world deployment involves both hardware and software. We evaluate this router using an open-source virtual testbed we have developed in our earlier work~\cite{enrich}, which accurately simulates the router's operational environment, allowing for large-scale, high-fidelity experimentation. The inputs to the router are bandwidths of data flows passing through the router's priority classes. To ensure realism in our experiments, we ensure that the total bandwidth of data flows does not exceed the system’s capacity. Further, we consider different traffic profiles -- for example, small, frequent UDP packets for VoIP traffic, and larger, bursty TCP flows for background traffic such as file transfers.
 The router testbed enables us to assess whether the user experience for streaming services is satisfactory (pass) or unsatisfactory (fail).  Each test execution takes approximately $4.5$ minutes and is compute-intensive. In addition, there is non-determinism in the test results due to fluctuations in network bandwidth, latency, jitter, asynchrony in network flows, and the CPU and memory load on the machine hosting the testbed. 

\textbf{Aircraft autopilot system.} We use an autopilot model of a De Havilland Beaver aircraft from a public-domain benchmark of Simulink specifications provided by Lockheed Martin~\cite{lockheedmartin}.  Simulink~\cite{chaturvedi2017modeling, DBLP:conf/ssbse/MatinnejadNBBP13} is a widely used language for  CPS specification and simulation. An example based on this system, simplified to have fewer inputs, is illustrated in Section~\ref{sec:example}.
The Simulink model of autopilot captures both the autopilot system, which includes the control logic and algorithms responsible for stabilizing and navigating the aircraft, as well as the simulator, which simulates the aircraft's physical dynamics and environmental factors such as wind and turbulence. The inputs to the autopilot system are signals related to the flight dynamics of the aircraft such as throttle, pitch angle, turning rate, heading and desired flight objectives such as a target altitude.  To ensure realism in our experiments, we cap the rate of change of input signals -- such as throttle and pitch angle -- between consecutive signal steps so that their differences remain within bounds imposed by the aircraft dynamics. The autopilot system is expected to satisfy the following system-level requirement: when the autopilot is enabled, the aircraft should reach the desired altitude within $500$ seconds. 

The Simulink model of the autopilot system is developed in compliance with the DO-178C standard~\cite{do-178c, autopilotbenchmark}, which prohibits non-determinism. Although this case study is deterministic -- and thus not susceptible to flakiness -- it nevertheless involves signal-based inputs, making it an interesting system for evaluating the accuracy and alignment of the test validators in RQ2 and RQ4, respectively.

\textbf{ADS systems.} We use two types of self-driving controllers as our ADS systems, both executed and tested using the BeamNG simulator -- a widely used open-source tool for ADS testing~\cite{beamng}. The first system is the autopilot controller of BeamNG, a classical self-driving controller~\cite{beamng, samak2021control}. The second is \textsc{Dave2}~\cite{dave2}, a deep neural network (DNN) model trained for end-to-end self-driving. To test the autopilot controller, we developed two simulation environments in BeamNG: (1)~a complex town map with multi-lane roads, other vehicles, and various static objects along the roads, and (2)~a simpler environment with a two-lane road without other vehicles and static objects. We refer to the setup that tests the autopilot controller on the town map as \textsc{AP--TWN}, and the one testing it on the simpler road as \textsc{AP--SNG}. We test \textsc{Dave2} using only the simpler road map, as \textsc{Dave2} is specifically trained for this environment. The inputs to \textsc{AP--SNG} and \textsc{Dave2} include road shape, weather conditions, time of day, the initial and target positions of the ego vehicle, as well as its speed and type. In addition to these, \textsc{AP--TWN} also takes as input the number of non-ego vehicles in the map, along with the initial and target positions, speeds, and types of both ego and non-ego vehicles. To ensure realism in our experiments, we enforce plausible vehicle positions, feasible target positions and avoid initial collisions caused by physically implausible starting positions or unsafe spacing between vehicles. The \textsc{Dave2} model and the single road setup used in our evaluation are provided by the CPS Testing Tool Competition track at the SBFT workshop~\cite{sbft}.

To determine whether a test passes or fails in our ADS setups, we consider the following system-level requirements adopted from prior studies~\cite{amini2024evaluating, sbft,BorgANJS21}: For \textsc{AP--SNG} and \textsc{Dave2}, tested in the single-road  environment, a test fails if the ego vehicle veers off the lane (\textbf{R1}). In the case of \textsc{AP--TWN}, operating in the complex town map, a test can fail not only for veering off the main road (\textbf{R1}) but also for three additional reasons: failing to maintain a safe distance from other vehicles (\textbf{R2}), failing to maintain a safe distance from static objects (\textbf{R3}), and not reaching the specified destination within the simulation duration (\textbf{R4}). Test execution time is approximately three minutes for \textsc{AP--TWN} and one minute for \textsc{AP--SNG} and \textsc{Dave2}. Flakiness in these ADS test setups  can arise due to  inconsistencies in timing between the simulator and ADS controller, which may lead to variations in the images or sensory data received by the ADS. Furthermore, the addition of white noise to the images passed to the ADS may contribute to  flakiness~\cite{amini2024evaluating}. In the complex town map, factors like the presence of non-ego vehicles and traffic lights introduce additional flakiness in the test outcomes for~\textsc{AP--TWN}.

Table~\ref{tab:studysubjects} outlines the key characteristics of our 
case-study systems. These systems include  the Router system, the aircraft autopilot (\textsc{AP--DHB}), the ADS autopilot  controller tested in a complete town (\textsc{AP--TWN}) and on a single-road map (\textsc{AP--SNG}), as well as the DNN-based controller tested on a single-road map (\textsc{Dave2}). For \textsc{AP--TWN}, tested against the above-mentioned requirements (R1--R4), we present the results for each requirement separately.

\begin{table}[t]
\caption{Key characteristics of our case-study systems. \emph{System} refers to the SUT aligned with the naming convention we adopt in the article. \emph{Simulator} indicates the environment or testbed used for test execution. \emph{Test Execution Time} indicates the approximate duration required to run a single test input on the corresponding simulator.}
\label{tab:studysubjects}
\scalebox{0.78}{
\begin{tabular}{|p{3.6cm}|p{3.7cm}|l|}
\hline
\textbf{System}    & \textbf{Simulator}      &  \textbf{Test Execution Time} \\ \hline
Router~\cite{enrich}             & Router Testbed                & $\sim 4.5$ min \\ \hline
Aircraft autopilot system of De Havilland Beaver aircraft (\textsc{AP--DHB})~\cite{autopilotbenchmark} & Simulink models of environmental factors (e.g., wind, turbulence, temperature, and atmospheric conditions) and aircraft's physical dynamics
& $\sim 0.5$ min \\ \hline
ADS autopilot tested in a complete town (\textsc{AP--TWN}) & \multirow{2}{*}{BeamNG} & $\sim 3$ min \\ \cline{1-1} \cline{3-3}
ADS autopilot tested on a single road map   (\textsc{AP--SNG})  & & $\sim 1$ min \\ \cline{1-1}
DNN self-driving controller tested on a single road map (\textsc{Dave2}) & &  \\ \hline
\end{tabular}}
\vspace*{-.3cm}
\end{table}

\subsection{RQ1 (Existence of Flakiness)}

\textbf{Experiment setting.} RQ1 measures the degree of flakiness in our case studies, namely the Router, \textsc{AP--TWN}, \textsc{AP--SNG}, \textsc{Dave2},  and \textsc{AP--DHB}. We randomly generate 100 test inputs for the router case study and 200 test inputs for each of our ADS-based and aerospace case studies.  Since the router is our most resource-intensive system to execute, we generate a smaller number of test inputs for it. Each test input is executed 10 times to detect any non-determinism in the test outcomes. We then prepare ten datasets for each case study where each dataset contains the verdicts from a distinct execution of test inputs. We refer to each dataset as $\mathit{TS}_{n}$ where $n$ denotes the $n$-th execution of the test inputs. Consequently, for each case study, we obtain datasets $\mathit{TS}_{1}$, $\mathit{TS}_{2}$, $\ldots$, $\mathit{TS}_{10}$.

\textbf{Results.}
Table~\ref{tab:rq1-1} shows the percentage of flaky tests observed for each case study across $\mathit{TS}_{1}$ to $\mathit{TS}_{10}$.    We consider a test case to be flaky unless all ten runs produce the same outcome. The percentage of flaky tests for \textsc{AP--DHB} is $0\%$, indicating the absence of flakiness in this system. In contrast, the percentage of flaky tests for Router is $11\%$.  The percentage of flakiness for the tests exercising \textsc{AP--TWN} for requirements R1 to R4 ranges between $21\%$ and $79\%$.

\begin{table}[t]
\centering
\caption{Percentage of flaky tests in the systems of Table~\ref{tab:studysubjects}.}
\label{tab:rq1-1}

\scalebox{0.89}{
\begin{tabular}{|c|c|cccc|c|c|}
\hline
\multirow{2}{*}{\textbf{Router}} & \multirow{2}{*}{\textbf{\textsc{AP--DHB}}}                                              & \multicolumn{4}{c|}{\textbf{\textsc{AP--TWN}}}                                                                                & \multirow{2}{*}{\textbf{\textsc{AP--SNG}}} & \multirow{2}{*}{\textbf{\textsc{Dave2}}} \\  \cline{3-6}
&                                  & \multicolumn{1}{c|}{\textbf{R1}} & \multicolumn{1}{c|}{\textbf{R2}} & \multicolumn{1}{c|}{\textbf{R3}} & \textbf{R4} &                                   &                                 \\ \hline
11\%         & 0\%                             & \multicolumn{1}{c|}{79\%}        & \multicolumn{1}{c|}{64\%}        & \multicolumn{1}{c|}{70\%}        & 21\%        & 1.5 \%                            & 33\%                            \\ \hline
\end{tabular}}
\vspace*{-.3cm}
\end{table}

\begin{tcolorbox}[breakable,colback=white,colframe=black!75!black]
\textbf{Finding.} Except for \textsc{AP--DHB}, all our case-study systems exhibit flaky tests, with rates ranging from 1.5\% to 79\%.
\end{tcolorbox}

\subsection{RQ2 (Accuracy)}
\label{sec:RQ2}
\textbf{Experiment setting.}
To answer RQ2, we generate test validators using the GP, DT, and DR alternatives by applying these methods to the training sets for each case study. Specifically, for each case study, we select one of the ten datasets from RQ1 to serve as the training set. This enables us to assess each method without regard to the variations caused by flakiness across the different datasets from RQ1. Analyzing the impact of the variations caused by flakiness is left for RQ3.  We tune the hyper-parameters of DT and DR using Bayesian Optimization~\cite{bayesian}. We configure GP using the parameters in Table~\ref{tab:parameters} and apply GP with each of the fitness functions from  Figure~\ref{fig:rankingmetrics}: Naish denoted by $\mathit{GP_N}$, Tarantula denoted by  $\mathit{GP_T}$, and Ochiai denoted by $\mathit{GP_O}$.   
To account for the randomness of GP, DT, and DR, we apply each technique 20 times to the training set for each case study.

In addition to considering the test validator generation methods individually, we also consider an \emph{ensemble} approach. Specifically, for each run of $\mathit{GP_N}$, $\mathit{GP_T}$, $\mathit{GP_O}$, DT, and DR, the ensemble method computes the union of the conditions generated by these techniques. Then, we derive a consistent set of assertions by applying the third step of \app\ (Section~\ref{subsec:consistency}) to this union. We then compare the assertion-based test validators generated by the ensemble with those generated by each method individually.

Recall from Definition~\ref{def:verdictthreshold} that each SUT is associated with a user-defined verdict threshold $\theta$, which specifies the minimum confidence level required for an assertion to be included in the test validator. 
We vary the user-defined verdict threshold $\theta$, from $0.5$ to $1$ in increments of $0.05$. We do not consider $\theta < 0.5$, since verdict predictions by assertions with less than $50\%$ confidence are unlikely to be trusted and  used in practice.

\begin{table}[t]
\caption{Parameters of GP: mutation rate (\emph{Mut\_rate}),  number of generations (\emph{Num\_gen}), population size (\emph{Pop\_size}), maximum value (\emph{Max\_const}), crossover rate (\emph{Cr\_rate}), tournament size (\emph{T\_size}),  max. tree depth (\emph{Max\_d}), minimum value (\emph{Min\_const}).}
\label{tab:parameters}
\begin{center}
\scalebox{0.7}{
\begin{tabular}{|c|c|c|c|c|c|c|c|}
\hline
\textbf{Parameter}          & \textbf{Value}                                & \textbf{Parameter}         & \textbf{Value}                               & \textbf{Parameter}          & \textbf{Value}                               & \textbf{Parameter}           & \textbf{Value}                      \\ \hline
                            & \cellcolor{cyan!20}                      &                            & \cellcolor{yellow!20}                     &                             & \cellcolor{cyan!20}                     &                              & \cellcolor{green!20}ADS: 100    \\ \cline{8-8} 
\multirow{-2}{*}{Mut\_rate} & \multirow{-2}{*}{\cellcolor{cyan!20}0.1} & \multirow{-2}{*}{Num\_gen} & \multirow{-2}{*}{\cellcolor{yellow!20}50} & \multirow{-2}{*}{Pop\_size} & \multirow{-2}{*}{\cellcolor{cyan!20}50} & \multirow{-2}{*}{Max\_const} & \cellcolor{green!20}Router: 400 \\ \cline{8-8} & \cellcolor{cyan!20} &  & \cellcolor{yellow!20} & &\cellcolor{cyan!20} & & \cellcolor{green!20}Autopilot: 45\\ \hline
Cr\_rate                    & \cellcolor{cyan!20}0.7                   & T\_size                    & \cellcolor{cyan!20}7                    & Max\_d                      & \cellcolor{cyan!20}5                    & Min\_const                   & \cellcolor{green!20}ADS: 0           \\ \cline{8-8} & \cellcolor{cyan!20} & & \cellcolor{cyan!20} & & \cellcolor{cyan!20} & & \cellcolor{green!20}Router: 0 \\ \cline{8-8} & \cellcolor{cyan!20} & & \cellcolor{cyan!20} & & \cellcolor{cyan!20} & & \cellcolor{green!20}Autopilot: -30 \\ \hline
\end{tabular}

}
\end{center}
\footnotesize{
\textbf{Note:} The values within the framed box \colorbox{cyan!20}{\phantom{00}} are from~\cite{li2024using,gaaloul2021combining,luke2006comparison}. The values within the framed box \colorbox{yellow!20}{\phantom{00}} are set by assessing the average number of generations required to reach a plateau. The values within the framed box \colorbox{green!20}{\phantom{00}} are based on the lowest and highest values that the input variables of our systems can assume.
}
\vspace*{-.5cm}
\end{table}

We generate each case study's test set by randomly creating inputs and executing them on the case-study system to obtain ground-truth verdicts. To mitigate the impact of flakiness on the ground-truth verdicts for test sets, we do as follows: (1)~For systems with a flaky test rate below 50\%, we include only tests that exhibit consistent behaviour in all ten runs. Any test that shows flakiness in those runs is excluded from the test set.   (2)~ For systems with a flaky test rate above $50\%$, since completely excluding flaky tests is cost-prohibitive, we include tests that produce consistent verdicts in at least eight out of ten runs. Their final verdicts are determined by majority voting.  We ensure that each test set for each case study ultimately contained $200$ elements.

\textbf{Results.}
To answer RQ2, we assess the accuracy of the generated test validators using the metrics from Section~\ref{sec:assertionbasedtestoracles}: accuracy, misprediction rates -- i.e.,  the rate of pass verdicts predicted as fail, and the rate of fail verdicts predicted as pass --  and relative accuracy. 
All statistical tests are performed using the Mann-Whitney U test~\cite{mann1947test} and the Vargha-Delaney's $\hat{A}_{12}$ effect size~\cite{vargha2000critique}. All statistical significance tests in RQ2 are reported with p-values adjusted using the Benjamini–Hochberg (BH) procedure~\cite{benjamini1995controlling}. We classify effect size values for accuracy and relative accuracy, where higher values indicate better performance, as follows: effect sizes are classified as small, medium, and large when their values are greater than or equal to 0.56, 0.64, and 0.71, respectively~\cite{vargha2000critique}. For misprediction rates, where lower values indicate better performance, effect sizes are classified as small, medium, and large when their values are lower than or equal to 0.44, 0.36, and 0.29, respectively~\cite{vargha2000critique}.

\begin{figure}[t]
    \centering
    
    \includegraphics[width=1.1\columnwidth]{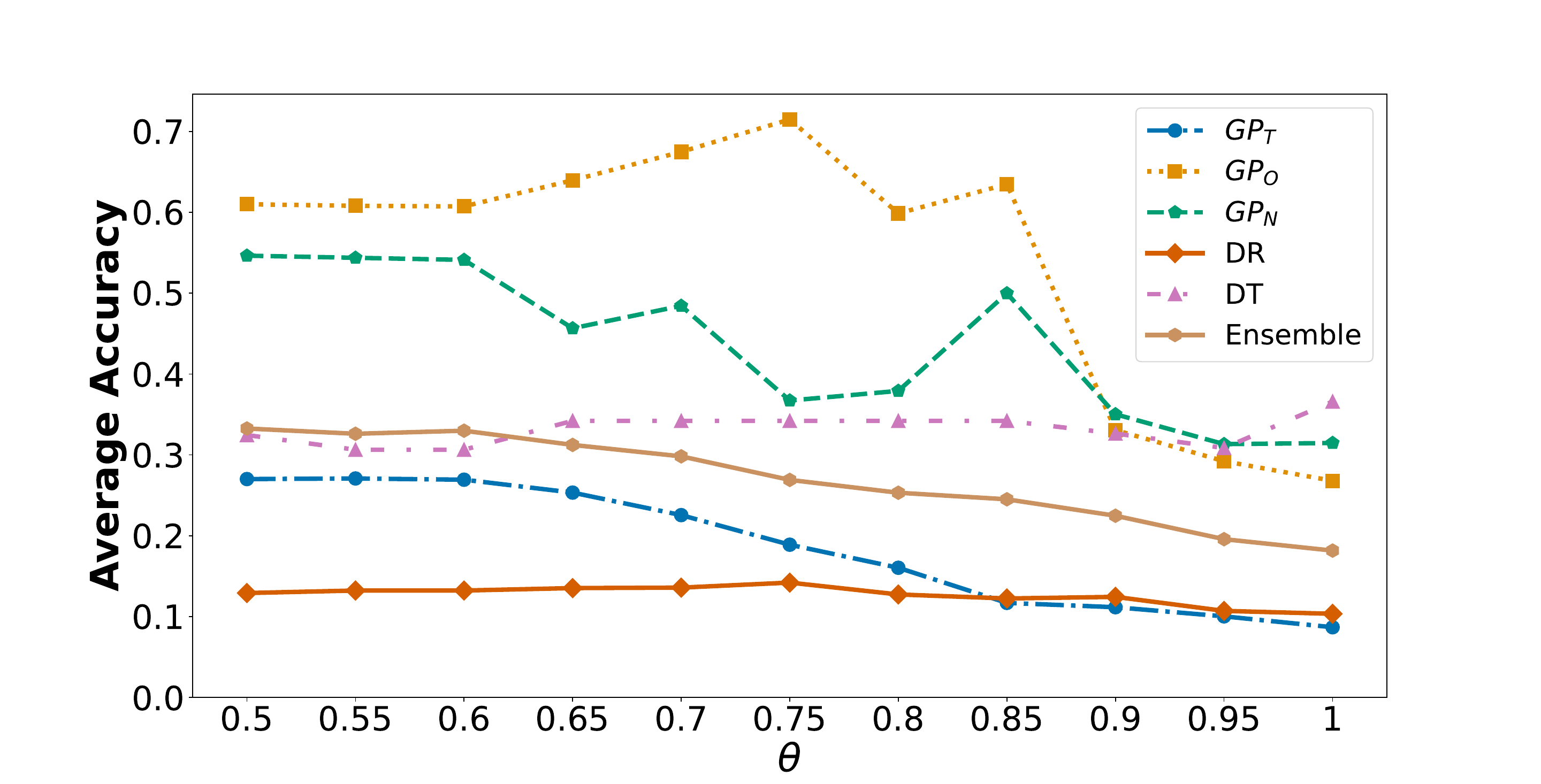}
    
    \caption{Average accuracies of the test validators generated  for all case studies  and for $0.5 \leq \theta \leq 1$.} 
    \label{fig:aar}
    \vspace*{-.4cm}
\end{figure}

\textbf{(1) Accuracy. } 
Figure~\ref{fig:aar} shows the average accuracy of the test validators generated by each technique for all case studies when $\theta$ varies from $0.5$ to $1$. The average accuracy of test validators generated by $\mathit{GP_T}$, $\mathit{GP_O}$, $\mathit{GP_N}$ and ensemble  decreases as $\theta$ increases because a higher value of $\theta$ results in fewer assertions in these test validators, as we only retain those with a confidence level of at least $\theta$. In contrast, since the  confidence levels of assertions produced by DT and DR are generally high (i.e., above $0.8$),  the accuracy of DT and DR remains relatively stable  as $\theta$ increases.

For $0.5 \leq \theta \leq 0.85$, $\mathit{GP_O}$ produces the most accurate test validators compared to other techniques. Statistical tests comparing the accuracy results in Figure~\ref{fig:aar} are provided in Table~\ref{tab:aarstat} in Appendix~\ref{apx:results}.  Based on this table, for \hbox{$0.5 \leq \theta \leq 0.85$}, test validators generated by $\mathit{GP_O}$ are significantly more accurate than those generated by other techniques. The effect-size values for the comparisons of $\mathit{GP_O}$ with $\mathit{GP_T}$, DT, DR, and the ensemble are all large, while the comparisons of $\mathit{GP_O}$ with $\mathit{GP_N}$ show both large and small effect sizes. For $\theta \geq 0.9$, there are no statistically significant differences in accuracy between $\mathit{GP_O}$ and $\mathit{GP_N}$, DT or the ensemble method.

Statistical tests comparing the accuracy results of $\mathit{GP_O}$ with $\mathit{GP_T}$, $\mathit{GP_N}$, DT, DR, and the ensemble for each case-study system across all values of $\theta$ are provided in Table~\ref{tab:aarstatperstudy} in Appendix~\ref{apx:results}. Based on these results, test validators from $\mathit{GP_O}$ are significantly more accurate than those from $\mathit{GP_T}$ and DR in all eight systems, outperform ensemble in seven, DT in six, and $\mathit{GP_N}$ in five of the eight case studies. The effect-size values for these comparisons are small, medium or large.


\begin{tcolorbox}[breakable,colback=white,colframe=black!75!black]
\textbf{Finding.} For $0.5 \leq \theta \leq 0.85$, test validators generated by GP with Ochiai achieve significantly better accuracy than other methods, outperforming the interpretable ML models (DT and DR) by at least $25\%$ in terms of accuracy on average. In contrast, for $\theta \geq 0.9$, there is no statistically significant difference in accuracy between GP with Ochiai and Naish, DT, or the ensemble method. 
\end{tcolorbox}

\begin{figure}[t]
    \centering
    
    \includegraphics[width=1.1\columnwidth]{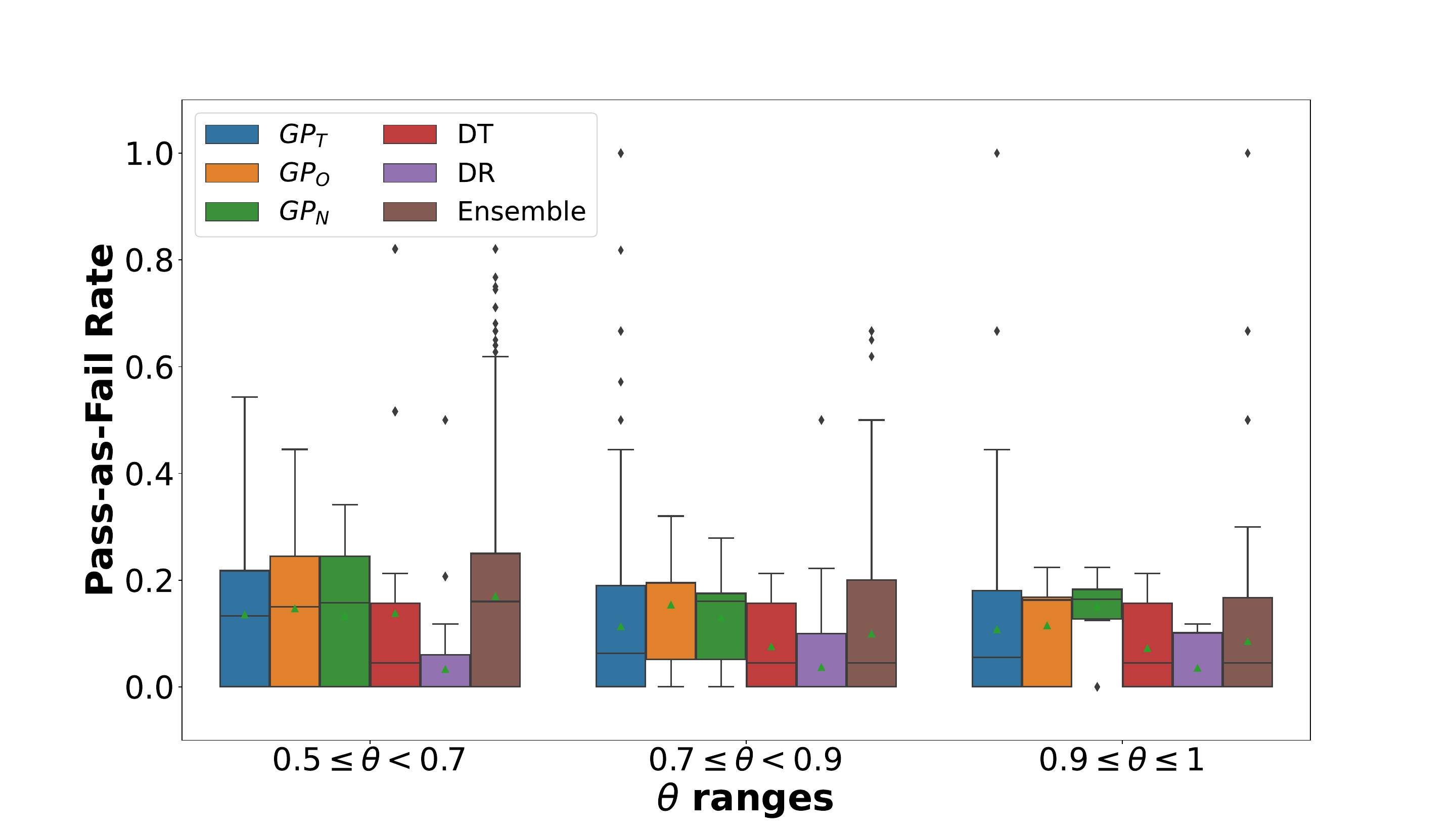}
    
    \caption{Pass-as-Fail rates of the test validators generated  for all case studies and for $ 0.5 \leq \theta \leq 1$.} 
    \label{fig:pasf}
    \vspace*{-.5cm}
\end{figure}

\begin{figure}[t]
    \centering
    
    \includegraphics[width=1.1\columnwidth]{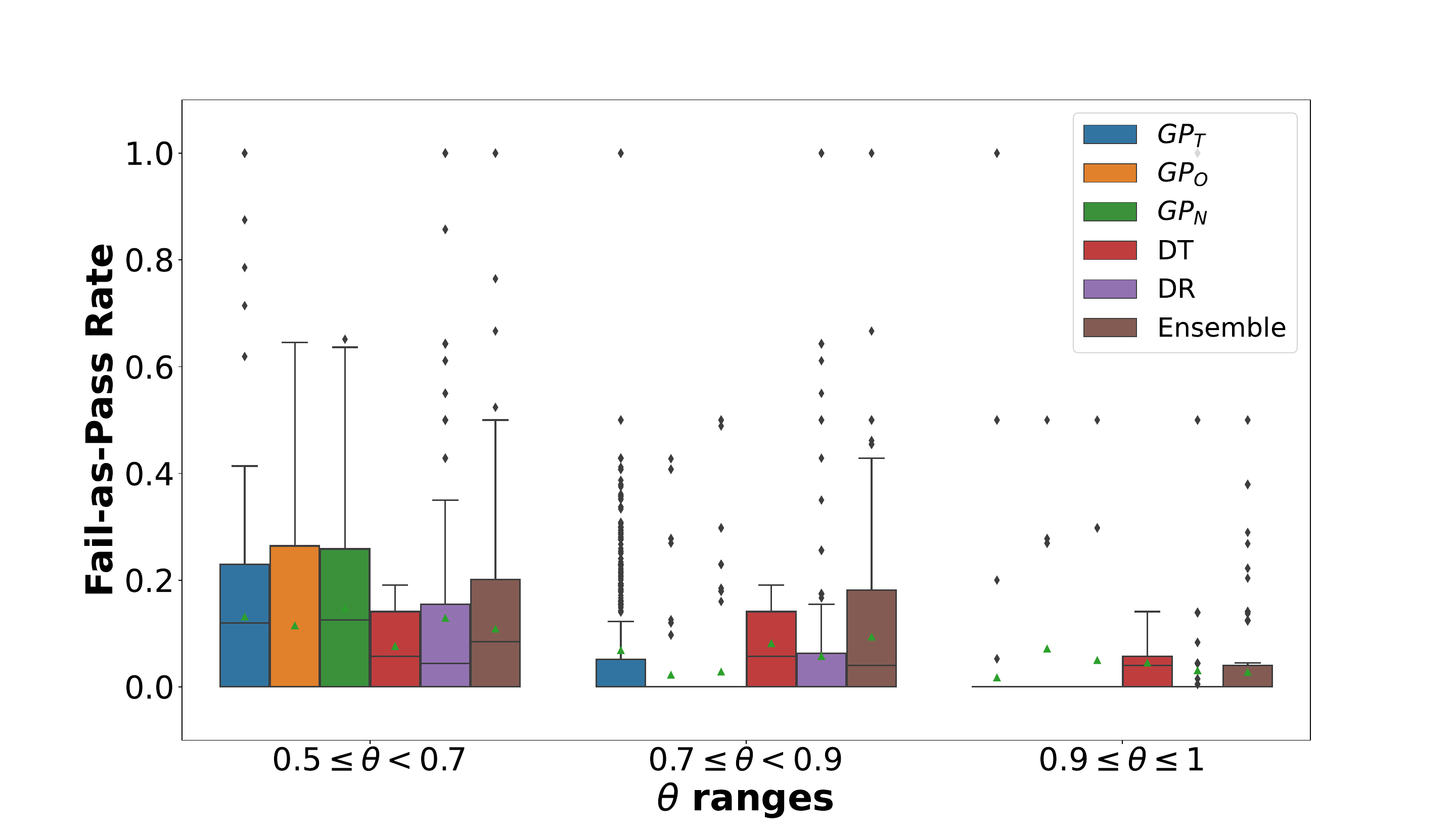}
    
    \caption{Fail-as-Pass rates of the test validators generated for all case studies  and for $ 0.5 \leq \theta \leq 1$.}
    \label{fig:fasp}
    \vspace*{-.5cm}
\end{figure}

\textbf{(2) Misprediction Rates. }
Figures~\ref{fig:pasf} and~\ref{fig:fasp} show the rate of actual pass verdicts predicted as fail, and the rate of actual fail verdicts predicted as pass, respectively, for all case studies when $\theta$ varies from $0.5$ to $1$. For clarity, we report these two rates across the following three aggregated \(\theta\)-ranges, since presenting them for each individual \(\theta\) does not provide a concise overview: low ($0.5 \leq \theta < 0.7$), medium ($0.7 \leq \theta < 0.9$), and high ($0.9 \leq \theta$).  We refer to the rate of pass verdicts predicted as fail as \emph{Pass-as-Fail}, and the rate of fail verdicts predicted as pass as \emph{Fail-as-Pass}.

As shown in Figure~\ref{fig:pasf}, across all \(\theta\) ranges, DR consistently achieves the lowest (best) average Pass-as-Fail rates compared to all GP techniques -- $\mathit{GP_T}$, $\mathit{GP_O}$, and $\mathit{GP_N}$ -- as well as the ensemble and DT. In contrast, for the Fail-as-Pass misprediction results,  the GP techniques overall, and in particular $\mathit{GP_O}$, produce better results compared to DT, DR, and the ensemble.  
Statistical tests comparing the Pass-as-Fail (Figure~\ref{fig:pasf}) and Fail-as-Pass (Figure~\ref{fig:fasp}) results are provided in Tables~\ref{tab:inaccstat}(a) and~(b), respectively, in Appendix~\ref{apx:results}.
Specifically, for the Pass-as-Fail rate, we compare the best-performing method for this metric, DR, with the other techniques, and for the Fail-as-Pass rate, we compare its best-performing method, $\mathit{GP_O}$, with the others. 
Based on Table~\ref{tab:inaccstat}(a), test validators generated by DR lead to a significantly lower rate of Pass-as-Fail compared to those obtained by other techniques with small, medium or large effect sizes.  Based on Table~\ref{tab:inaccstat}(b) for $0.5 \leq \theta < 0.9$, $\mathit{GP_O}$ either achieves a significantly lower Fail-as-Pass rate (with small, medium or large effect sizes) or shows no statistically significant difference in Fail-as-Pass rate compared to other techniques. For $0.9 \leq \theta \leq 1$, $\mathit{GP_T}$ results in a significantly lower Fail-as-Pass rate compared to other techniques.

Table~\ref{tab:inaccstatperstudy} in Appendix~\ref{apx:results} presents the statistical tests comparing the Pass-as-Fail rate of DR and the Fail-as-Pass rate of $\mathit{GP_O}$ with those of the other techniques, for each case-study system and across all values of $\theta$.  Based on Table~\ref{tab:inaccstatperstudy}(a), test validators generated by DR have a significantly lower Pass-as-Fail rate than those obtained by $\mathit{GP_N}$ in all eight case-study systems, by $\mathit{GP_T}$ and $\mathit{GP_O}$ in seven, by ensemble in six, and by DT in four out of eight case-study systems.  Based on Table~\ref{tab:inaccstatperstudy}(b), test validators generated by $\mathit{GP_O}$ have a significantly lower Fail-as-Pass rate than those obtained by DR in five case-study systems, and by $\mathit{GP_T}$ and ensemble in four case-study systems. The effect-size values in all the comparisons are negligible, small, medium or large.

\begin{figure}[t]
    \centering
    
    \includegraphics[width=1.1\columnwidth]{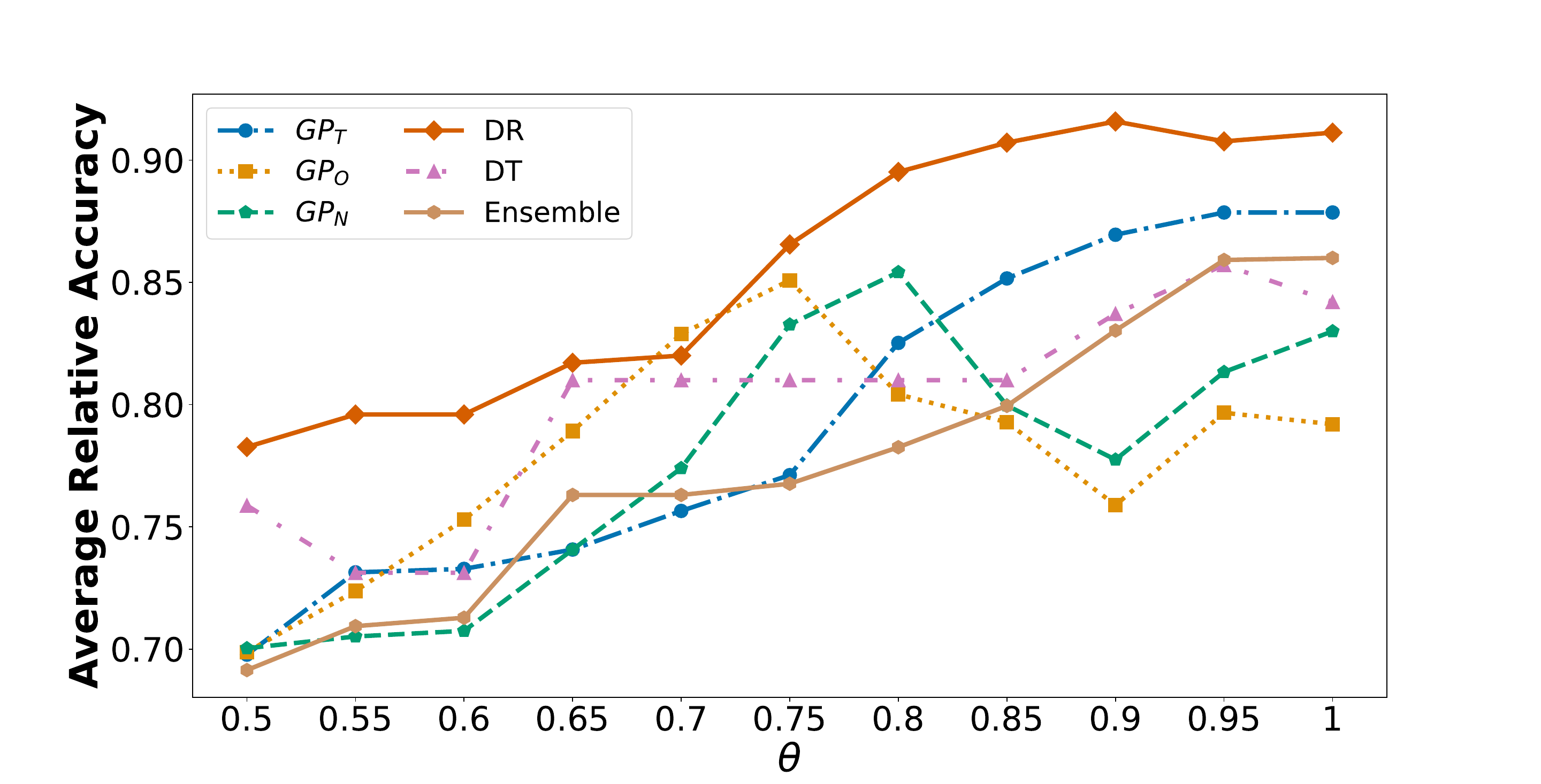}
    \caption{Average relative accuracies of the test validators generated for all case studies and for $ 0.5 \leq \theta \leq 1$.}
    \label{fig:relacc}
    \vspace*{-.3cm}
\end{figure}

\begin{tcolorbox}[breakable,colback =white,colframe=black!75!black]
\textbf{Finding.} Test validators generated by DR achieve a significantly lower Pass-as-Fail rate than those of other methods across all verdict thresholds $\theta$ between $0.5$ and $1$, outperforming GP-based techniques (GP with Tarantula, Ochiai, Naish) by at least $6\%$ on average. In contrast, for $0.5 \leq \theta \leq 1$, test validators generated by GP with Ochiai either result in a significantly lower Fail-as-Pass rate compared to interpretable ML models (DT and DR) and the ensemble method or exhibit no statistically significant difference in Fail-as-Pass rate compared to these techniques. 

\end{tcolorbox}

\textbf{(3) Relative Accuracy.} Figure~\ref{fig:relacc} presents the average \emph{relative} accuracy of the test validators generated by GP, DT, DR and ensemble for all our case studies and for $0.5 \leq \theta \leq 1$. Recall from Section~\ref{sec:example} that while accuracy is the percentage of correct predictions among all predictions, relative accuracy is the percentage of correct predictions excluding inconclusive predictions. Based on Figure~\ref{fig:relacc}, the average relative accuracies of GP, DT and DR, for all values of $\theta$, exceed $0.7$, indicating that all the compared methods have high levels of correctness when they make conclusive predictions.

The test validators generated by DR show consistently higher average relative accuracy compared to the other techniques across all values of $\theta$, with the exception of $\theta = 0.7$, where test validators generated by $\mathit{GP_O}$ achieve a higher average of relative accuracy than those of DR. This superior relative accuracy over GP techniques -- $\mathit{GP_T}$, $\mathit{GP_O}$ and $\mathit{GP_N}$ -- is because DR generates stronger assertions containing multiple logical terms. In contrast, GP techniques generate assertions with fewer logical terms. The weaker assertions generated by GP techniques can provide predictions for more test inputs compared to DR. However, the higher number of predictions made by GP techniques also increases their susceptibility to mispredictions compared to DR.

Statistical tests comparing the relative accuracy results in Figure~\ref{fig:relacc} are in Table~\ref{tab:relstat} in Appendix~\ref{apx:results}. These statistical tests are consistent with the average comparisons discussed above. Specifically,  test validators generated by DR lead to significantly higher relative accuracy compared to those obtained by the other techniques across all values of $\theta$, except for $\theta = 0.7$ and $\theta = 0.95$, where no statistically significant differences in relative accuracy are observed between DR and $\mathit{GP_O}$, and between DR and $\mathit{GP_T}$, respectively. 

Table~\ref{tab:relaccstatperstudy} in Appendix~\ref{apx:results} presents the statistical tests comparing the relative accuracy  of DR with $\mathit{GP_T}$, $\mathit{GP_O}$, $\mathit{GP_N}$, DT and the ensemble for each case-study system across all values of $\theta$. Based on this table, test validators generated by DR lead to a significantly higher relative accuracy than those generated by $\mathit{GP_O}$ and $\mathit{GP_N}$ in all eight case-study systems, by $\mathit{GP_T}$ and DT in six, and by ensemble in five case-study systems. For the router case study, test validators generated by $\mathit{GP_T}$ and the ensemble yield significantly higher relative accuracy compared to DR.

\begin{tcolorbox}[breakable,colback=white,colframe=black!75!black]
\textbf{Finding.} For most values of the verdict threshold $\theta$, test validators generated by DR achieve significantly higher relative accuracy compared to those obtained by other techniques. 
\end{tcolorbox}

\noindent
\tikz[baseline=(X.base)]{
  \node[
    rounded corners=1.5pt,
    inner sep=2pt,
    fill=blue!15
  ] (X) {\textcolor{red}{{\Large$\blacktriangleright$}}\ \textbf{Size of assertions}}} $\mathit{GP_O}$, and GP techniques in general, generate weaker assertions, characterized by a smaller number of relational terms. We measure the length of the generated assertions as the number of relational terms they contain.
Assertions generated by GP techniques contain, on average, $2.1$ relational terms involving one or two input variables. This pattern is consistent across both the autopilot case study, which has seven input variables, and the ADS case studies, which have eight input variables. In the router system, where assertions often involve arithmetic combinations of data flows, GP-generated assertions include summations over up to four of the eight input variables.
For the autopilot and ADS case studies, DT-generated assertions contain, on average, $3.5$ relational terms involving up to three input variables, while DR-generated assertions contain, on average, $2.3$ relational terms involving up to two input variables. For the router system, DT-generated assertions contain, on average, $3.2$ relational terms, involving summations over up to three of the eight input variables, while DR-generated assertions include, on average, $2.1$ relational terms, with summations over at most three input variables. Across all case-study systems, the generated assertions, include no more than five relational terms.

\noindent
\tikz[baseline=(X.base)]{
  \node[
    rounded corners=1.5pt,
    inner sep=2pt,
    fill=blue!15
  ] (X) {\textcolor{red}{{\Large$\blacktriangleright$}}\ \textbf{Recommendation on setting $\theta$}}} While  RQ2 compares different methods for assertion generation with $\theta$ varying between 0.5 and 1, when strong assurance of verdict prediction is required, we expect $\theta$ values higher than 0.9 to be considered. For $\theta \geq 0.9$, \app\ becomes inconclusive for many tests. However, due to the following two reasons, even with a moderate rate of conclusive verdicts, \app\ remains useful: 
  
  (1)~\emph{Conclusive predictions by our test validators at high verdict thresholds are most accurate, with a negligible rate of predicting actual failures as passes.} In particular, for our case-study systems, the rate of Fail-as-Pass is zero or negligible ($< 3\%$) when $\theta \geq 0.9$. This is particularly important in safety-critical systems, where minimizing the risk of misclassifying a failing test as passing is essential to avoid overlooking potential failures. 
  
  (2)~\emph{Even a moderate rate of conclusive verdicts yields substantial execution-time savings.} Given the high accuracy rate and the negligible incidence of Fail-as-Pass, even a moderate proportion of conclusive verdicts is valuable.  For instance, for $\theta \geq 0.9$, test validators generated by \app\ for checking car accidents with other vehicles (i.e., the \textsc{AP--TWN}(R2) case study) generate conclusive verdicts for $30\%$ of test cases, thus avoiding execution for these test cases. Given that ADS simulations have a high flakiness rate and are time-consuming (each simulation takes at least three minutes based on Table~\ref{tab:studysubjects}), avoiding execution for $30\%$ of tests represents significant time savings. For checking lane evasions (i.e., the \textsc{AP-SNG} case study), when $\theta \geq 0.9$, our test validators produce conclusive verdicts for $80\%$ of test cases,  meaning that only $20\%$ of  tests require execution. For evaluating users' quality of experience in the router system, when $\theta \geq 0.9$, our test validators produce conclusive verdicts for $15\%$ to $60\%$ of test cases, depending on the underlying technique used in \app\ for generating test validators. Given the router system's flakiness and the time-intensive nature of its simulations (each taking at least $4.5$ minutes based on Table~\ref{tab:studysubjects}), avoiding $15\%$ to $60\%$ of test executions results in significant time savings.

\subsection{RQ3 (Robustness to Flakiness)}
\label{sec:RQ3}
\textbf{Experiment setting.}
For RQ3, we use the ten datasets, $\mathit{TS_1}$ to $\mathit{TS}_{10}$,  from RQ1 for each flaky case-study system, i.e., Router, \textsc{AP--TWN} (R1) to (R4), \textsc{AP--SNG} and \textsc{Dave2}. As noted in RQ1, while these datasets contain identical test inputs for a given system, the verdicts assigned to these inputs may vary due to flakiness. In RQ3, we assess the robustness of each assertion-inference technique across these ten datasets for each case-study system. Specifically, we apply the $\mathit{GP_T}$, $\mathit{GP_O}$, $\mathit{GP_N}$, DT, DR, and ensemble methods to each dataset to generate assertion-based test validators. Each technique is configured using the same parameters as in RQ2, and similarly to RQ2,  we apply each technique 20 times to each dataset to account for randomness.

For each case study, we use the same test sets generated in RQ2 to measure the accuracy of the generated test validators. To assess variations in the accuracy of test validators, we report the average absolute deviation (AAD) of accuracy values for each case study. Given a distribution of accuracy values, AAD is computed as the average deviation of individual accuracy values from the mean accuracy.  A low AAD indicates that the test validators' accuracy is less impacted by variations in verdicts across the different training datasets.

Similar to RQ2, all statistical tests are performed using the Mann-Whitney U test and the Vargha-Delaney's effect size. All statistical significance tests in RQ3 are reported with p-values adjusted using the BH procedure~\cite{benjamini1995controlling}.

\textbf{Results.} Table~\ref{tab:RQ3avgacc} shows the average accuracy of the test validators for all verdict thresholds $\theta$ and computed using the ten datasets, $\mathit{TS_1}$ to $\mathit{TS}_{10}$, generated in RQ1. 
Based on this table, the average accuracy of the test validators generated by $\mathit{GP_O}$ surpasses that of other techniques across all case studies, except for $\textsc{Dave2}$, where the test validators generated by $\mathit{GP_N}$ achieve the highest average accuracy.

\begin{table}[t]
\caption{Average accuracy of $\mathit{GP_T}$, $\mathit{GP_O}$, $\mathit{GP_N}$, DT, DR and ensemble when datasets $TS_1$ to $TS_{10}$ are used for each case-study system in RQ3. The cells highlighted in blue represent the maximum average accuracy obtained for each case study.}
\label{tab:RQ3avgacc}
\scalebox{0.77}{
\begin{tabular}{c|c|cccc|c|c||c|}
\cline{2-9}
\multirow{2}{*}{}                       & \multirow{2}{*}{\textbf{Router}} & \multicolumn{4}{c|}{\textbf{\textsc{AP--TWN}}}                                                      & \multirow{2}{*}{\textbf{\textsc{AP--SNG}}} & \multirow{2}{*}{\textbf{\textsc{Dave2}}} & \multirow{2}{*}{\textit{\textbf{Average}}} \\ \cline{3-6}
                                        &                                  & \multicolumn{1}{c|}{\textbf{R1}} & \multicolumn{1}{c|}{\textbf{R2}} & \multicolumn{1}{c|}{\textbf{R3}} & \textbf{R4} &                                                             &                                                           &                                            \\ \hline
\multicolumn{1}{|c|}{\textbf{$\mathbf{GP_T}$}}   & 6\%                              & \multicolumn{1}{c|}{8\%}         & \multicolumn{1}{c|}{13\%}        & \multicolumn{1}{c|}{17\%}        & 9\%         & 8\%                                                         & 16\%                                                      & 11\%                                       \\ \hline
\multicolumn{1}{|c|}{\textbf{$\mathbf{GP_O}$}}   & \cellcolor[HTML]{96FFFB}57\%                             & \multicolumn{1}{c|}{\cellcolor[HTML]{96FFFB}60\%}        & \multicolumn{1}{c|}{\cellcolor[HTML]{96FFFB}41\%}        & \multicolumn{1}{c|}{\cellcolor[HTML]{96FFFB}97\%}        & \cellcolor[HTML]{96FFFB}54\%        & \cellcolor[HTML]{96FFFB}81\%                                                        & 37\%                                                      & \cellcolor[HTML]{96FFFB}60\%                                       \\ \hline
\multicolumn{1}{|c|}{\textbf{$\mathbf{GP_N}$}}   & 40\%                             & \multicolumn{1}{c|}{44\%}        & \multicolumn{1}{c|}{37\%}        & \multicolumn{1}{c|}{45\%}        & 36\%        & 42\%                                                        & \cellcolor[HTML]{96FFFB}48\%                                                      & 42\%                                       \\ \hline
\multicolumn{1}{|c|}{\textbf{DT}}       & 24\%                             & \multicolumn{1}{c|}{19\%}        & \multicolumn{1}{c|}{34\%}        & \multicolumn{1}{c|}{23\%}        & 15\%        & 18\%                                                        & 19\%                                                      & 22\%                                       \\ \hline
\multicolumn{1}{|c|}{\textbf{DR}}       & 14\%                             & \multicolumn{1}{c|}{8\%}         & \multicolumn{1}{c|}{10\%}        & \multicolumn{1}{c|}{8\%}         & 1\%         & 12\%                                                        & 22\%                                                      & 11\%                                       \\ \hline
\multicolumn{1}{|c|}{\textbf{Ensemble}} & 23\%                             & \multicolumn{1}{c|}{10\%}        & \multicolumn{1}{c|}{25\%}        & \multicolumn{1}{c|}{19\%}        & 6\%         & 17\%                                                        & 28\%                                                      & 18\%                                       \\ \hline
\end{tabular}}
\end{table}

Statistical tests comparing the accuracy results in Table~\ref{tab:RQ3avgacc} for each verdict threshold $\theta$ are provided in Table~\ref{tab:RQ3aarstat} in Appendix~\ref{apx:results}. 
As shown in this table, test validators generated by $\mathit{GP_O}$ lead to significantly higher accuracy compared to those obtained by other techniques across all values of $\theta$, except for $\theta = 1$ (i.e., 100\%), where no statistically significant differences in accuracy are observed between $\mathit{GP_O}$ and $\mathit{GP_N}$.

Table~\ref{tab:accdiff} presents the AAD in the accuracy of test validators generated based on the $\mathit{TS_1}$ to $\mathit{TS}_{10}$ datasets for our case studies. Based on this table, variations in the $\mathit{TS_1}$ to $\mathit{TS}_{10}$ datasets result in an average fluctuation of 4\% in the accuracy of test validators generated by $\mathit{GP_O}$. In contrast, validators produced by $\mathit{GP_T}$, $\mathit{GP_N}$, and ensemble exhibit larger fluctuations. Similarly, test validators generated by DT and DR have accuracy fluctuations of 5\% and 3\%, respectively, indicating they are less affected by flakiness in the datasets. However, as Tables \ref{tab:RQ3avgacc} and \ref{tab:RQ3aarstat} show, these test validators achieve lower accuracy than those generated by $\mathit{GP_O}$. Thus, $\mathit{GP_O}$ produces the most accurate and robust test validators.

\begin{table}[t]
\caption{Average Absolute Deviation (AAD) of the accuracy  for ten datasets generated by executing a set of test inputs ten times for the case studies in RQ3.}
\label{tab:accdiff}
\scalebox{0.8}{
\begin{tabular}{c|c|cccc|c|c|c|}
\cline{2-9}
\multirow{2}{*}{}                       & \multirow{2}{*}{\textbf{Router}} & \multicolumn{4}{c|}{\textbf{\textsc{AP--TWN}}}                                                      & \multirow{2}{*}{\textbf{\textsc{AP--SNG}}} & \multirow{2}{*}{\textbf{\textsc{Dave2}}} & \multirow{2}{*}{\textit{\textbf{Average}}} \\ \cline{3-6}
                                        &                                  & \multicolumn{1}{c|}{\textbf{R1}} & \multicolumn{1}{c|}{\textbf{R2}} & \multicolumn{1}{c|}{\textbf{R3}} & \textbf{R4} &                                                             &                                                           &                                            \\ \hline
\multicolumn{1}{|c|}{\textbf{$\mathbf{GP_T}$}}   & 7\%                              & \multicolumn{1}{c|}{4\%}         & \multicolumn{1}{c|}{9\%}         & \multicolumn{1}{c|}{10\%}        & 5\%         & 7\%                                                         & 9\%                                                       & 7\%                                        \\ \hline
\multicolumn{1}{|c|}{\textbf{$\mathbf{GP_O}$}}   & 7\%                             & \multicolumn{1}{c|}{ 5\%}         & \multicolumn{1}{c|}{10\%}        & \multicolumn{1}{c|}{0\%}         & 4\%         & 0\%                                                         & 4\%                                                       & \cellcolor[HTML]{96FFFB}4\%                                        \\ \hline
\multicolumn{1}{|c|}{\textbf{$\mathbf{GP_N}$}}   & 13\%                             & \multicolumn{1}{c|}{12\%}        & \multicolumn{1}{c|}{10\%}        & \multicolumn{1}{c|}{13\%}        & 9\%         & 13\%                                                        & 7\%                                                       & 11\%                                       \\ \hline
\multicolumn{1}{|c|}{\textbf{DT}}       & 15\%                             & \multicolumn{1}{c|}{4\%}         & \multicolumn{1}{c|}{6\%}         & \multicolumn{1}{c|}{3\%}         & 2\%         & 1\%                                                         & 4\%                                                       & \cellcolor[HTML]{96FFFB}5\%                                        \\ \hline
\multicolumn{1}{|c|}{\textbf{DR}}       & 9\%                              & \multicolumn{1}{c|}{3\%}         & \multicolumn{1}{c|}{1\%}         & \multicolumn{1}{c|}{1\%}         & 2\%         & 1\%                                                         & 6\%                                                       & \cellcolor[HTML]{96FFFB}3\%                                        \\ \hline
\multicolumn{1}{|c|}{\textbf{Ensemble}} & 17\%                             & \multicolumn{1}{c|}{7\%}         & \multicolumn{1}{c|}{10\%}        & \multicolumn{1}{c|}{12\%}        & 7\%         & 2\%                                                         & 7\%                                                       & 9\%                                        \\ \hline
\end{tabular}}
\vspace*{-.5cm}
\end{table}

\begin{tcolorbox}[breakable,colback=white,colframe=black!75!black]
\textbf{Finding.} Overall, our results indicate that test validators generated by GP with Ochiai are the most accurate and robust compared to those produced by other techniques.

\textbf{Take away~1.} SBFL ranking formulas integrated with GP generate accurate assertion-based test validators. In particular, Ochiai is well-suited for generating accurate test validators that are robust against flakiness.

\textbf{Take away~2.} When GP is used with Ochiai, flaky tests in the training set have only a negligible impact on the accuracy of the inferred test validators. Hence, removing flaky tests does not significantly alter the validators' accuracy. Therefore, practitioners can save effort when preparing training sets for test validator generation, as excluding flaky tests is not critical.
\end{tcolorbox}

\subsection{RQ4 (Alignment)}
\label{sec:rq4}
\textbf{Experiment setting.} We evaluate how closely the generated test validator assertions for our case studies align with the descriptions of precondition violations, ODD limits, and low-risk scenarios provided in the reference documentation for these case studies. Recall from Section~\ref{sec:studysubjects} that our case studies include a network router, different configurations of ADS, and an autopilot system. Table~\ref{tab:referencedoc} lists the reference documentation considered for each of these systems: For the router system, our sources are the technical standard on priority-based flow management~\cite{cakepaper,CAKE} and ITU-T Recommendation E.361~\cite{ITU-T}, which specifies network quality-of-service (QoS) parameters and performance requirements. For ADS, we rely on the SAE J3016 standard~\cite{J3016_202104} -- which defines the ADS taxonomy and the human driver's role at each automation level -- along with published ADS-related results based on simulations, field tests, and other expert-validated findings~\cite{yoneda2019automated, chen2024modeling, lou2024autonomous, wang2025openlka, wang2025empirical, bhandari2020driving, boyapati2023automated, abdessalem2018testing}. For \textsc{AP--DHB}, we draw on autopilot system handbooks~\cite{autopilothandbook, federal2009pilot}, which describe operational principles, performance limits, and safety constraints. 

\begin{table}[t]
\caption{Reference documentation for case-study systems.}
\label{tab:referencedoc}
\vspace*{-.1cm}
\begin{center}
\scalebox{0.85}{\begin{tabular}{|l|p{2.5cm}|p{4cm}|}
\hline
\textbf{System}      & \textbf{Type of Reference}             & \textbf{Documentation}                                                                                                                                                              \\ \hline
Router               & Industry standard                     & \begin{tabular}[c]{@{}l@{}} Traffic-shaping manual~\cite{cakepaper, CAKE}\\  ITU-T Recommendation E.361~\cite{ITU-T}\end{tabular} \\ \hline
\multirow{2}{*}{ADS} & Industry standard                     & SAE J3016~\cite{J3016_202104}                                                                                                                                           \\ \cline{2-3} 
                     & Published empirical and expert-validated results & ~\cite{yoneda2019automated, chen2024modeling, lou2024autonomous, wang2025openlka, wang2025empirical, bhandari2020driving, boyapati2023automated, abdessalem2018testing}                                                                                                                                                    \\ \hline
AP-DHB               & Industry standard                     & Autopilot system handbooks~\cite{autopilotbenchmark, federal2009pilot}                                                                                                                          \\ \hline
\end{tabular}}
\end{center}
\vspace*{-.3cm}
\end{table}

From the test validators generated by \app\ in RQ2, we select a subset of assertions for each case study. We consider high-confidence assertions (i.e., $\theta \geq 0.9$), following our conclusion from RQ2 that these yield the highest accuracy and the lowest rate of mispredicting failures as passes. We apply stratified sampling~\cite{singh1996stratified} to each of the eight case studies, selecting 20 high-confidence assertions per system (160 in total), stratified by their binary outcome labels (pass vs. fail), to ensure balanced representation of both categories. We develop textual assertions and evaluate their alignment with the reference documentation through a systematic human-subject study\footnote{Written informed consent was obtained from all participants prior to their participation in the study. Ethics approval for this human-subject study was granted by the University of Ottawa's Research Ethics Board (file number H-11-25-12283).}, as described in the following three steps:

\emph{Step~1: Translate logical assertions into natural language.}
We first mapped each variable and constant to its domain-specific meaning using a glossary derived from the reference documentation for each case study and validated by the authors. For variables, this involved translating variable names using domain terminology, such as translating $\mathit{pitchwheel}(t)$ into ``the angle of the nose of the aircraft.'' For constants, this involved translating numeric, Boolean, and categorical values into their domain meanings and units, for example, expressing $0.3$ as 30\% of maximum thrust.
We then translated arithmetic expressions, relational operators and logical connectives according to their semantics, while preserving precedence and parenthesization. 
We expressed pass/fail verdicts using the expected system outcome specified by the corresponding requirement. For example, fail verdicts for \textsc{AP-DHB} are written as ``the aircraft will not reach the required altitude in 500 seconds''. For temporal intervals, we expressed time bounds in natural language while preserving whether they are exclusive or inclusive. Using the above protocol, the first author drafted each translation and the remaining authors reviewed it for fidelity, clarity, and consistency. When alternative phrasings arose, they were discussed and reconciled through consensus.
All authors agreed on the final translations. Examples of these translations are shown in Table~\ref{tab:translation}, and the complete list of assertions, along with their translations, is in our supplementary material~\cite{github} and open for scrutiny.

\begin{table*}[t]
\caption{Assertion examples for \textsc{AP-DHB}, Router, \textsc{AP-TWN (R2)}, \textsc{Dave2} and \textsc{AP-SNG} case studies along with their natural-language translation, sentences retrieved automatically by GPT as being related to the textual assertions, and labels from human annotators. Segments highlighted in \colorbox{green!25}{green} show where the translated assertions align with the retrieved sentences, while segments highlighted in \colorbox{orange!25}{orange} in  the example assertion labelled as overlapping mark information that is missing from the assertion but present in the retrieved sentences.}
\label{tab:translation}
\vspace*{-.2cm}
\begin{center}
\scalebox{0.8}{\begin{tabular}{|p{1cm}|p{4.5cm}|p{6cm}|p{7cm}|p{1.5cm}|}
\hline
\textbf{Case Study} & \multicolumn{1}{l|}{\textbf{Assertion}} & \multicolumn{1}{l|}{\textbf{Assertions stated in Natural Language}} & \multicolumn{1}{l|}{\textbf{Retrieved Sentences}} & \multicolumn{1}{l|}{\textbf{Label}} \\ \hline
\multirow{2}{*}{\textsc{AP--DHB}} 
& $(\forall t \in [0, 500) : p(t) \leq 0) \land (\forall t \in [0, 250) : th(t) < 0.3) \Rightarrow \mathit{fail}$ 
& If the \colorbox{green!25}{nose of the aircraft is pointed downwards} for 500 seconds and 
\colorbox{green!25}{the thrust applied to the engine is low} (less than 30\% of maximum thrust) for 250 seconds, 
the aircraft will not reach the required altitude in 500 seconds. 
& ``Consequently, the tail is again pushed downward and the \colorbox{green!25}{nose rises into a climbing attitude}''~\cite{federal2009pilot}. \newline \newline ``If \colorbox{green!25}{thrust decreases} and airspeed decreases, lift will become less than weight and the aircraft will start to \colorbox{green!25}{descend}''~\cite{federal2009pilot}. \newline \newline
``Climb performance is directly dependent upon the ability to produce either \colorbox{green!25}{excess thrust} or excess power''~\cite{federal2009pilot}.
& Aligned \\ \hline
\multirow{2}{*}{Router} 
& $\mathit{flow}_{5} + \mathit{flow}_{6} + \mathit{flow}_{7} > 372 \Rightarrow  \mathit{fail}$ 
&  Attempting to use \colorbox{green!25}{high-priority DiffServ classes} (classes 5, 6, and 7) so that they jointly exceed 68\% of their \colorbox{green!25}{allocated bandwidth share} (more than 372 mb/s) degrades the quality of experience across the network.
& ``Secondly, it shows that the high-priority flows \colorbox{green!25}{are limited} so as to not use more than the \colorbox{green!25}{share of the bandwidth allocated} to the \colorbox{green!25}{high-priority DiffServ classes}''~\cite{cakepaper}. 
& Aligned
\\ \hline
\textsc{AP--TWN (R2)} 
& $\mathit{weather} = \mathit{foggy} \wedge  \mathit{time\_of\_day} = \mathit{night} \wedge \mathit{initial\_speed} > 90 \wedge \mathit{traffic\_density} = \mathit{high} \Rightarrow \mathit{fail}$ 
& While travelling at a high speed (more than 90 km/h) through a town on \colorbox{green!25}{a foggy night}, the ego-vehicle collides with nearby vehicles in dense traffic. 
& ``Visible light camera is vulnerable to bad conditions such as \colorbox{green!25}{fog} and are difficult to see without a light source \colorbox{green!25}{at night}''~\cite{yoneda2019automated}. \newline \newline
``According to reports that evaluated the measurement distance, it was confirmed that the LiDAR measurement distance decreased as \colorbox{green!25}{fog became darker}, and the visibility distance became shorter''~\cite{yoneda2019automated}. 
& Aligned \\ \hline
\textsc{Dave2} 
& $\mathit{weather} = \mathit{sunny} \wedge   \mathit{initial\_speed} < 10  \Rightarrow \mathit{pass}$ 
& While travelling at a low speed (less than 10 km/h) on a \colorbox{green!25}{sunny day}, the ego-vehicle keeps its lane. 
& 
``Both test vehicles were able to maintain lane keeping during all runs under Baseline conditions (\colorbox{orange!25}{ambient air temperatures between 20~$\degree$F and 100~$\degree$F, peak} \colorbox{orange!25}{wind speeds below 22.4 mph, \colorbox{green!25}{sun} position greater than} \colorbox{orange!25}{15~$\degree$ above the horizon,} \colorbox{green!25}{ambient daylight conditions with} \colorbox{green!25}{clear sky,} and \colorbox{orange!25}{dry and clear pavement})''~\cite{boyapati2023automated}.
& Overlapping \\ \hline
\textsc{AP--SNG} 
&  $\mathit{initial\_speed} < 6 \wedge \mathit{road\_angle} \leq 5 \wedge \mathit{initial\_position} = \mathit{centre\_of\_lane} \Rightarrow \mathit{pass}$ 
& When the ego vehicle is positioned at the center of the lane, is driving very slowly (less than 6 km/h), and the road is straight or slightly curved, the ego vehicle keeps its lane.
& ``Crash risk is minimal when a vehicle travels along the centerline of a lane, and the vehicle heading is parallel to the centerline''~\cite{chen2024modeling}. \newline \newline
``Lane-keeping-assist driving remains centered when the oncoming vehicle approaches the self-driving car, and that the human driver actively avoids the oncoming vehicle''~\cite{wang2025empirical}. 
&  Unrelated (but judged as plausible safe or low-risk scenario) \\ \hline
\end{tabular}}
\end{center}
\vspace*{-.3cm}
\end{table*}

\emph{Step~2: Identify sentences in the reference documentation that are related to each assertion.} For each textual assertion, we use OpenAI's API~\cite{openai} (GPT-5, medium reasoning, with a context window of 400k tokens) to find semantically related sentences in the system's reference documentation.  Using the template in Figure~\ref{fig:prompttemplate} (Appendix~\ref{apx:prompt}), we construct one-shot prompts -- one per natural-language assertion -- each containing (1) the assertion and (2) system reference documentation related to that assertion, provided in a vector database.  Each prompt instructs GPT to return two to five of the most related sentences for the assertion.  We ran the prompt twice for each assertion. For approximately 90\% of the assertions, the identified sentences were identical across the two runs. For the remaining assertions, we took the intersection of the identified sentences as the final set of related sentences. No assertion yielded an empty intersection.  Finally, we manually verified all selected sentences to ensure that they appeared verbatim in the documentation and were not hallucinations.

\emph{Step~3: Evaluate the extent to which each assertion aligns with the retrieved sentences  in Step~2.} To evaluate how closely assertions align with the retrieved sentences in Step~2, we collaborated with two third-party annotators (non-authors). Both are graduate students in computer science with over two years of experience in software testing and requirements analysis, and have conducted research on CPS, ADS, and Simulink models. Among the pool of potential participants, these two candidates were selected because their background most closely matched the required domain expertise.  Annotators were provided with  textual assertions along with the retrieved sentences for each assertion as well as the following four-point Likert scale: (1)~\textit{Aligned}: The situation described by the assertion either (a) matches the situation described in at least one retrieved sentence, or (b) is a more specific instance of it (i.e., the assertion stays entirely within the scope of the retrieved sentences and does not extend beyond what they entail). Further, the assertion is not  inconsistent with any of the retrieved sentences.  (2)~\textit{Overlapping}: The situation described by the assertion is strictly broader in some respect and only partially overlaps with the situations described in at least one retrieved sentence (i.e., the assertion generalizes the retrieved sentences or covers cases not supported by the retrieved sentences alone).  Further, the assertion is not  inconsistent with any of the retrieved sentences. (3)~\textit{Inconsistent}: The assertion is inconsistent with at least one of the retrieved sentences.  (4)~\textit{Unrelated}: The assertion and the retrieved sentences are not aligned, overlapping or inconsistent. For these assertions, we asked the annotators to judge whether the situation described by the assertion corresponds to (i) a safe or low-risk situation, (ii) a violation of the ODD or preconditions, or (iii) neither.

We divided the 160 assertions (8 case-study systems $\times$ 20 assertions per system) equally between the two annotators, with a 20\% overlap. As a result, each annotator evaluated 96 assertions in total. Assertions were randomly assigned and balanced across the case-study systems. We then conducted a three-hour training session to calibrate the annotators on the Likert scale, clarify labelling criteria, and ensure a consistent understanding of the annotation task. During the training, after introducing the Likert scale and discussing illustrative examples outside our experimental materials, we asked both annotators to independently label the 20\% overlapping assertions. We then calculated the disagreement rate, which was approximately 9\%, corresponding to a Cohen’s kappa ($\kappa$) value of 0.75, indicating substantial agreement~\cite{cohen1960coefficient}. A disagreement was counted whenever the two annotators assigned different Likert labels to the same assertion. All disagreements occurred when one annotator labelled an assertion as aligned and the other as overlapping. We subsequently discussed and resolved all disagreements with both annotators. At the end of the training meeting, after reaching consensus on the disagreements, both annotators felt confident in annotating their respective sets of non-overlapping assertions. The annotators then independently labelled the remaining assertions.

Table~\ref{tab:translation} shows examples of aligned, overlapping, and unrelated assertions based on labels from the annotators.  For \textsc{AP--DHB}, the assertion that the aircraft’s nose points downward while the engine operates at low thrust is \textit{aligned} with the retrieved sentences, which state that the nose must rise to a climbing attitude or that excess thrust is required to gain altitude. For \textsc{Dave2}, the assertion that the vehicle keeps its lane when driving at low speed on a sunny day is \textit{overlapping}, because the first retrieved sentence includes specific conditions regarding air temperature, wind speeds, sun position, and clear pavement in addition to daylight conditions and a clear sky, whereas the assertion omits these details and therefore goes beyond what that sentence specifies. For \textsc{AP--SNG}, the assertion is labeled as \textit{unrelated}, but it was nevertheless rated by our annotators as a plausible, low-risk lane-keeping situation. We note that no assertions in our study were labelled inconsistent by the annotators.

\textbf{Results.}
Table~\ref{tab:usefulness} shows, for each case-study system, the percentage of aligned, overlapping, and unrelated assertions among all the selected assertions for this research question.  At least 75\% of the assertions for each system are either aligned or overlapping with the reference documentation.  This high rate of alignment indicates that the generated assertions effectively capture situations that violate environmental assumptions or preconditions, describe low-risk operational states, and identify conditions that exceed the system's ODD limits.

\begin{table}[t]
\caption{The percentage of aligned, overlapping, and unrelated assertions. The inconsistent label is not shown because no assertions were rated inconsistent by the annotators.
}
\centering
\label{tab:usefulness}
\vspace*{-.1cm}
\scalebox{0.8}{
\begin{tabular}{|l|c|c|c|c|c|}
\hline
\multirow{2}{*}{\textbf{Case Study}} 
  & \multirow{2}{*}{\textbf{\makecell{Aligned}}} 
  & \multirow{2}{*}{\textbf{\makecell{Overlapping}}} 
  & \multicolumn{3}{c|}{\textbf{Unrelated}} \\ \cline{4-6}
  & & & \textbf{Low-risk} & \textbf{\makecell{ODD \\ Violation}} & \textbf{Neither} \\ \hline
Router              
  & 95\% & 0\%  & 5\% & 0\% & 0\% \\ \hline
\textsc{AP--DHB}      
  & 95\% & 0\%  & 0\% & 5\% & 0\% \\ \hline
\textsc{AP--TWN (R1)} 
  & 60\%  & 15\% & 25\% & 0\% & 0\% \\ \hline
\textsc{AP--TWN (R2)} 
  & 80\%  & 15\% & 0\% & 5\% & 0\% \\ \hline
\textsc{AP--TWN (R3)} 
  & 100\% & 0\%  & 0\% & 0\% & 0\% \\ \hline
\textsc{AP--TWN (R4)} 
  & 80\%  & 5\% & 5\% & 10\% & 0\% \\ \hline
\textsc{AP--SNG}      
  & 65\%  & 15\% & 20\% & 0\% & 0\% \\ \hline
\textsc{Dave2}       
  & 65\%  & 20\%  & 15\% & 0\% & 0\% \\ \hline
\emph{Average} & 80\% & 8.7\% & 8.7\% & 2.5\% & 0\%\\ \hline
\end{tabular}}
\vspace*{-.4cm}
\end{table}

\begin{tcolorbox}[breakable, enhanced jigsaw,
  colback=white, colframe=black!75!black]
\textbf{Finding.}
On average, 80\% of the high-confidence assertions ($\theta \geq 0.9$)  considered in our study are aligned with the reference documentation, and 8.7\% are overlapping. While the remaining 11.2\% of assertions are not explicitly covered in the documentation, i.e., labelled as unrelated, they still describe plausible low-risk situations or ODD-violation scenarios, according to our annotators.  No assertions in our study were found to contradict the information in the reference documentation. 
\end{tcolorbox}

\noindent
\tikz[baseline=(X.base)]{
  \node[
    rounded corners=1.5pt,
    inner sep=2pt,
    fill=blue!15
  ] (X) {\textcolor{red}{{\Large$\blacktriangleright$}}\ \textbf{Usefulness of the inferred assertions:}}} 
We identify four ways in which the assertions inferred by our approach are useful to practitioners: \emph{First}, assertions generated by \app\ provide SUT-specific quantified thresholds that are missing from the reference documentation, which typically only states qualitative ODD limits or high-level assumptions and leaves the choice of concrete thresholds to engineers. For example, for the router system, the assertion in Table~\ref{tab:translation} identifies an upper bound (threshold) on aggregate traffic that can flow through high-priority classes without compromising QoS, a value that is not specified in the reference documentation on QoS and priority-based flow management~\cite{cakepaper, CAKE, ITU-T}. Engineers of this system often have to select thresholds through limited ad-hoc observations rather than systematically collected data, which can yield inaccurate values that do not generalize beyond the scenarios engineers tried. Instead, \app\ infers thresholds from a broader set of system executions, making them more robust and more likely to generalize than those chosen through ad-hoc observations.
These thresholds can then guide system configuration and service-level agreements, for example, in the case of the router system, by assigning flows to priority classes so that the inferred limits are not exceeded.

\emph{Second}, assertions rated as unrelated capture plausible low-risk scenarios not covered explicitly in the reference documentation, where the system trivially behaves correctly.
For example, for the AP-SNG system, \app\ identifies a safe or low-risk situation in which the ego vehicle is placed in the centre of a lane, starts at a very low speed, and drives on a straight road. Under these conditions, the vehicle keeps its lane. This scenario is not described as a specific safe case in the reference documentation, but \app\ classifies it as low risk based on the observed executions used to infer the assertions. For these low-risk situations, invoking the simulator provides limited insights.

\emph{Third},  the inferred assertions can be integrated as runtime monitors -- similar to self-oracles~\cite{Stocco_2020_selforacle} -- to check whether the system exceeds its ODD and to alert the human operator upon ODD violation.  For example, for the \textsc{Dave2} system, \app\ infers that when the turning angle of the road exceeds 15 degrees, the ego vehicle fails to keep its lane. This assertion can be deployed as a runtime guard that detects entry into such road segments and issues a warning to the driver to take over or disengage the automated component.

\emph{Fourth}, our approach generates assertions in a formal, machine-analyzable notation, enabling their direct use in automation. This reduces the effort and errors in extracting these assertions from natural language.

\subsection{Threats to Validity}
\label{sec:threats}
\textbf{Internal Validity.}
The accuracy of the inferred assertions depends on the size and representativeness of the training set. To improve representativeness, we used adaptive random testing, which ensures diversity among generated tests~\cite{metaheuristicsbook}. Since training data generation is time- and resource-intensive, the size of the training set was determined by the available budget, and we allocated one month for training set generation.
To ensure a fair comparison among the different approaches, we use identical training sets and identical test sets across all experiments. 
We  subjected DT and DR to systematic  hyper-parameter tuning via Bayesian optimization~\cite{bayesian}, and configured GP based on the best-practice recommendations from prior studies~\cite{li2024using,gaaloul2021combining,luke2006comparison}. To enable DT and DR to generate assertions with the same expressive power as  those generated by GP, we performed feature engineering for DT and DR. This ensures that the conditions they learn for our case-study systems are comparable in structure and expressive power to those generated by our GP grammar. Consistent with prior research on identifying flaky tests~\cite{amini2024evaluating, khatiri2023simulation, khatiri2024simulation}, we re-executed each test ten times to distinguish flaky ones. The test sets used to assess test-validator accuracy in RQ2 and RQ3 contain no flaky tests for systems with flakiness rates below 50\%. Furthermore, for other systems, we minimized the presence of flaky tests in the test sets by including only those tests that achieved at least 80\% consistent verdicts across  re-executions. This approach to building test sets is not biased towards any specific test-validator generation technique. Further, since the accuracy of these techniques is assessed using the same test sets, we have not favoured any technique in the experiments for RQ3.

In RQ4, we use stratified sampling to select assertions for each case study. We chose a sample size of 20 because the experiments in RQ2 and RQ3 show that our test validators produce about 20 assertions on average in a single run. For RQ4, we require the logical assertions to be translated into natural language. Because accurate translation requires detailed knowledge of the domain concepts and system requirements -- for example, in \textsc{AP--DHB} the pitchwheel signal refers to the angle of the aircraft's nose, and in the router system, classes 5, 6, and 7 denote high-priority differentiated-services (DiffServ) classes -- the translations were performed by the authors rather than automated tools or third-party annotators, whose lack of domain knowledge about our case studies would limit their ability to produce accurate translations. To mitigate this threat, the lead author produced the translations, and the other authors independently reviewed them to identify any changes in meaning and to maintain consistent terminology within each domain. We provide the complete set of translations in the supplementary material~\cite{github}.

\textbf{Conclusion Validity.} Since running multiple statistical tests can increase the risk of Type I error inflation~\cite{arcuri2011practical}, we apply the Benjamini-Hochberg (BH) procedure~\cite{benjamini1995controlling} to control the false discovery rate.  In RQ2 and RQ3, all statistical significance tests were reported using BH-adjusted p-values.

\textbf{External Validity.} Our experiments are based on five different CPS and network systems. The aircraft autopilot system that we used in our evaluation is from the Lockheed Martin benchmark and has been previously used in the literature on testing CPS models~\cite{giannakopoulou2021automated, nejati2019evaluating}. 
Our router case study is one of the few examples of industrial network systems in the literature~\cite{enrich,tosem}.  The ADS systems we used in our evaluation  are based on  BeamNG, a leading open-source simulator widely referenced for virtual and hybrid testing in ADS research, and used by the software testing community for benchmarking and competitions~\cite{beamng, sbft}. In addition, one of our ADS relies on \textsc{Dave2} which is a DNN that has been successfully employed in real-world road testing conducted by NVIDIA~\cite{dave2roadtest}. Further experiments with a broader range of CPS would strengthen generalizability. 

Our signal encoding  assumes a piecewise constant interpolation function. Our evaluation of major CPS benchmarks and case studies from the literature indicates that this assumption is commonly made~\cite{khandait2024arch, lockheedmartin, cruisecontroller, clutchlockup, guidancecontrol, dcmotor}. While this choice is well suited to our case study domains, applying our encoding to systems in other domains may require adapting the interpolation strategy to better match domain-specific signal characteristics.

\textbf{Limitation.}
\app\ is data-driven  and the assertions it generates  are based on empirical data rather than formal correctness guarantees. As a result, although in RQ4 we observed that \app\ did not produce assertions inconsistent with the reference documentation, we cannot guarantee that \app\ never generates such inconsistent assertions. 
To our knowledge, no formal techniques currently exist for deriving human-readable assertions that characterize input validity for complex CPS, such as those in our case studies. As discussed in Section~\ref{sec:intro}, our work is related to prior data-driven approaches for determining the validity of test inputs for DL systems~\cite{WhenWhyTest2022, delaram, Dola_2021_DAIV} and for building self-oracles for ADS~\cite{Stocco_2020_selforacle}. While these approaches rely on the probability that an input is out of distribution to identify invalid inputs or ODD violations, \app\ uses high-confidence assertions for similar purposes. In contrast to the prior work, the assertions generated by \app\ are  human-readable. This enables domain experts to inspect them, compare them against reference documentation using the procedure described in RQ4, and identify situations where the inferred assertions may be inconsistent with the documented requirements.

\section{Related Work}
\label{sec:related}

\textit{\bfseries Test input validity.} Traditionally, assessing the validity of test inputs has focused on identifying the preconditions and environmental assumptions of a system -- i.e., the expectations a system makes about its operational context~\cite{giannakopoulou2002assumption}. Recently, test input validity for DL systems has been studied by defining invalid inputs as those underrepresented in the training set of a given DL model and by proposing out-of-distribution detection methods to identify such inputs~\cite{WhenWhyTest2022, Dola_2021_DAIV, Stocco_2020_selforacle}. Related work has likewise used the training-data distribution to guide test generation toward inputs that remain within that distribution, for example through manifold-based approaches that learn a latent representation of the training data and generate test inputs by sampling from the learned manifold~\cite{byun2020manifold}.
These approaches primarily focus on DL systems whose test inputs are images, and propose to construct  test validators using ML algorithms -- e.g., variational autoencoders~\cite{Dola_2021_DAIV, Stocco_2020_selforacle, WhenWhyTest2022, byun2020manifold}, vision systems metrics~\cite{hu2022if}, or a combination of both~\cite{delaram}.

Our test validators extend the existing test validators for DL systems in three ways. \emph{First}, our test validators are defined for systems with numeric inputs rather than those with image inputs. \emph{Second}, our test validators are specified using logical and interpretable assertions rather than DL models or non-interpretable classification algorithms. \emph{Third}, our test input validators are conditions on the inputs that determine whether the system passes or fails with high certainty,  
whereas test input validators for DL systems determine validity by checking whether a test input lies within the same distribution as the training data or remains within a human-tolerated range of visual changes.
The purpose of DL test validators is to recognize inputs that the model has not been exposed to during training, while our goal is to learn conditions that lead to violations of environmental assumptions, conditions characterizing safe or low-risk scenarios, or conditions that describe situations outside the SUT’s ODD limits.

\textit{\bfseries Inferring assertions or grammars to explain behaviours of SUT and their underlying causes.}  Several studies have explored inferring assertions or grammars that explain particular behaviours of a SUT~\cite{kapugama2022human, gopinath2020abstracting, soremekun2020inputs, kampmann2020does, tosem, gaaloul2021combining, enrich}. For instance, approaches that aim to explain the circumstances of failures typically start with one or more known examples. They then generate additional tests iteratively, and infer assertions or grammars that explain the underlying causes of these failures. 
In contrast to these approaches, our work is the first to define assertion-based test validators for CPS.  Our evaluation focuses on flaky CPS simulators where we examine the impact of flakiness on the accuracy of the test validators for these systems.

\textit{\bfseries Surrogate Modelling.} Surrogate models are used to skip system executions for test inputs whose outputs can be predicted with near certainty~\cite{tosem, matinnejad2014mil,NejatiSSFMM23}. To our knowledge,  DT and DR are the only interpretable surrogate  models  used in the literature~\cite{ben2016testing, tosem, yuan2022visual, zhu2022fuzzy}. In our evaluation, we have compared DT and DR with GP techniques in terms of both accuracy and robustness. Our results in RQ2 and RQ3 show that GP with Ochiai generates more accurate and robust test validators than DT or DR.

\textit{\bfseries Test Oracles.}  Our test validators can be regarded as test oracles, since they decide whether the system passes or fails for a given test input. To construct test oracles, similar to our work, several existing techniques rely on a data-driven approach and infer test oracles from labelled executions~\cite{barr2015oracle}. The key difference between these approaches and our work is that they infer test oracles based on a reference system rather than based on the SUT. Further, unlike most test oracles that require both inputs and outputs to determine verdicts, our test validators require only test inputs to issue verdicts.

\section{Conclusion}
\label{sec:conclusion}
In this article, we introduced assertion-based test validators and evaluated their accuracy, robustness, and alignment with descriptions of precondition violations, ODD limits, and low-risk scenarios provided in technical standards as well as in empirical and expert-validated studies~\cite{cakepaper, CAKE, ITU-T, J3016_202104, yoneda2019automated, chen2024modeling, lou2024autonomous, wang2025openlka, wang2025empirical, bhandari2020driving, boyapati2023automated, abdessalem2018testing, autopilotbenchmark, federal2009pilot}. Test validators ensure that testing resources are not wasted on inputs that do not meaningfully exercise the system under test. Future work will explore iteratively updating test validators with new system executions and versions, as well as enhancing the logical expressiveness of the assertions.

\section*{Acknowledgment}
We gratefully acknowledge funding from NSERC of Canada under the Discovery, Discovery Accelerator and I2I programs.

\bibliographystyle{IEEEtran}
\bibliography{bibliography}

\onecolumn
\appendices

\section{Pruning Algorithm to Obtain Consistent Assertion Set}
\label{apx:prune}
We present the formalization and full details of the pruning algorithm that was introduced in Section~\ref{subsec:consistency}. We first establish the necessary notation and definitions, then describe our pruning strategy.

\begin{definition}[\textbf{Bipartite Graph based on Assertion Conditions}]
\label{def:graph}
Let $A$ be a set of assertions. We define the bipartite graph $\mathcal{B} = (V, E)$ corresponding to $A$ as follows: each assertion in $A$ corresponds uniquely to a vertex in  $V$. The set $V$ is partitioned into two disjoint subsets $V_p$ and $V_f$, where $V_p$ represents conditions of  the  pass assertions  in $A$, and $V_f$ represents conditions of the fail assertion in $A$. An edge $(x, y) \in E$ connects a vertex $x \in V_p$ to a vertex $y \in V_f$ if and only if the conjunction of their associated assertion  conditions is satisfiable (SAT). More precisely, an edge between $x$ and $y$ indicates that the pair of assertions corresponding to $x$ and $y$ together represent an inconsistency within $A$  since they, respectively,  represent a passing assertion and a failing assertion that can simultaneously hold for some test input.
\end{definition}

Figure~\ref{fig:inconsistency} in Section~\ref{subsec:consistency} shows a bipartite graph where $a_1$, $a_2$, $a_3$, and $a_4$ represent conditions  of pass assertions (i.e., they belong to $V_p$), and $a'_1$, $a'_2$, and $a'_3$ represent conditions of fail assertions (i.e., they belong to $V_f$). The edges in this graph represent pairs of pass and fail assertions whose conjunctions are satisfiable. For example, there is an edge between $a_1$ and $a'_1$ because $x_2 < 43 \wedge x_1 + x_2 <10$ is SAT.

Let $x \in V$ be a vertex. We denote the \textit{length} of the condition associated with $x$ by $\mathit{len}(x)$, defined as the number of arithmetic and logical operators present in that condition. For example, in Figure~\ref{fig:inconsistency} in Section~\ref{subsec:consistency}, $\mathit{len}(a_1) = 1$ and $\mathit{len}(a_2) = 3$. We also denote the \textit{degree} of a vertex $x$ by $\mathit{deg}(x)$, defined as the number of edges connected to $x$. For instance, in Figure~\ref{fig:inconsistency} in Section~\ref{subsec:consistency}, $\mathit{deg}(a_1)$ is three, indicating that the passing assertion corresponding to $a_1$ is inconsistent with the failing assertions corresponding to $a'_1$, $a'_2$, and $a'_3$.

We now present the details for our pruning method, shown in Algorithm~\ref{alg:consistency},  which aims to eliminate inconsistencies (i.e., edges) between pass and fail conditions by removing vertices (i.e., conditions) from $V_p$ or $V_f$. Since there are multiple ways to eliminate inconsistencies, resulting in alternative consistent sets of assertions, we devise heuristics in Algorithm~\ref{alg:consistency} whose goal is to obtain a consistent set of assertions while minimizing the number of removed vertices.
Algorithm~\ref{alg:consistency} takes a potentially inconsistent set of assertions $A$ and generates a consistent subset $A'$ of assertions from $A$. Based on Definition~\ref{def:graph}, the algorithm represents $A$ as a bipartite graph $\mathcal{B} = (V, E)$, where each assertion condition becomes a vertex in $V$ and each satisfiable pass-fail pair is an edge in $E$ (line $1$ in Algorithm~\ref{alg:consistency}).  We use the Z3 SMT solver~\cite{z3} to check the satisfiability of the conjunctions of all pairs of pass and fail conditions to establish edges.

\begin{algorithm}[h]
\caption{Pruning strategy to obtain a consistent set of assertions}
\label{alg:consistency}
\flushleft
\footnotesize
\textbf{Input} $\mathit{A}$: A (potentially inconsistent) set of assertions \\
\textbf{Output} $A'$: A consistent subset of $\mathit{A}$ \\
\begin{algorithmic}[1]
\State \hspace{-0.1cm} Let $\mathcal{B} = (V, E)$ be a bipartite graph created based on Definition~\ref{def:graph}. 
\State \hspace{-0.1cm} \textbf{while} $E \neq \emptyset$ \textbf{do} 
\State \hspace{0.1cm} $S \leftarrow$ $\{ x \in V \mid \mathit{deg(x) \geq 1 \wedge \forall y \in V: \mathit{len(x)} \leq \mathit{len(y)}} \}$
\State \hspace{0.1cm} \textbf{if} $\mid S \mid \ == 1$ \
\State \hspace{0.1cm} \hspace{0.1cm}  $\mathit{vertexToRemove} \gets$ select the vertex in $S$ 
\State \hspace{0.1cm} \textbf{else}
\State \hspace{0.1cm} \hspace{0.1cm} $H \leftarrow$ $\{ x \in S \mid \forall y \in S: \mathit{deg(x)} \geq \mathit{deg(y)} \}$
\State \hspace{0.1cm} \hspace{0.1cm} \textbf{if} $\mid H \mid \ == 1$ 
\State \hspace{0.1cm} \hspace{0.1cm} \hspace{0.1cm}  $\mathit{vertexToRemove} \gets$ select the vertex in $H$ 
\State \hspace{0.1cm} \hspace{0.1cm} \textbf{elseif} $H \cap V_p \neq \emptyset$ 
\State \hspace{0.1cm} \hspace{0.1cm} \hspace{0.1cm} $\mathit{vertexToRemove} \gets$ randomly select a vertex in $H \cap V_p$ 
\State \hspace{0.1cm} \hspace{0.1cm}  \textbf{else}
\State \hspace{0.1cm} \hspace{0.1cm}  \hspace{0.1cm} $\mathit{vertexToRemove} \gets$ randomly select a vertex in $H$
\State \hspace{0.1cm} \hspace{0.1cm}  \textbf{end}
\State \hspace{0.1cm}  \textbf{end}
\State \hspace{0.1cm} $V \leftarrow V \setminus \{\mathit{vertexToRemove}\}$
\State \hspace{0.1cm}  $E \leftarrow E \setminus \{ (u, v) \in E \mid u = \mathit{vertexToRemove} \vee v = \mathit{vertexToRemove} \}$
\State \hspace{-0.1cm} \textbf{end}
\State \hspace{-0.1cm} $A' \leftarrow $ $\{ x \Rightarrow \mathit{fail} \mid x \in V_f\} \cup \{ x \Rightarrow \mathit{pass} \mid x \in V_p\}$
\State \hspace{-0.1cm} \textbf{return} $A'$
\end{algorithmic}
\end{algorithm}

In the while-loop from lines 2 to 18, Algorithm~\ref{alg:consistency} iteratively removes vertices until there are no remaining edges between the pass ($V_p$) and fail ($V_f$) partitions. The loop begins by identifying the set $S$ of vertices with a degree of at least one and the shortest length among such vertices (line 3 in Algorithm~\ref{alg:consistency}). If $S$ contains exactly one vertex, that vertex is selected and stored in the variable \textit{vertexToRemove} (lines 4--5 in Algorithm~\ref{alg:consistency}) for removal at the end of the while-loop iteration. This is because the  condition corresponding to this vertex is the least constrained, thus having a higher likelihood of conflicting with other conditions, making it a priority candidate for removal.

If multiple vertices exist in $S$, the algorithm computes the subset $H \subseteq S$ containing vertices with the highest degree (line 7 in Algorithm~\ref{alg:consistency}). If $H$ contains exactly one vertex, this vertex is selected for removal (lines 8--9 in Algorithm~\ref{alg:consistency}). Otherwise, when multiple vertices exist in $H$, the algorithm prioritizes vertices belonging to the pass set ($V_p$) by randomly selecting one vertex in $H \cap V_p$ (lines 10--11 in Algorithm~\ref{alg:consistency}). This prioritization is motivated by the observation that the fail partition ($V_f$) typically contains fewer vertices, making fail vertices more valuable to retain. If no vertex in $H$ belongs to the pass set, the algorithm randomly selects a vertex from $H$  (line 13 in Algorithm~\ref{alg:consistency}). Having stored a vertex in  \textit{vertexToRemove}, 
this vertex and all its incident edges are removed from the graph (lines 16--17 in Algorithm~\ref{alg:consistency}).  The assertions corresponding to the remaining vertices in $V_p$ and $V_f$ are collected into the set $A'$, which is then returned (lines 19--20 in Algorithm~\ref{alg:consistency}).

\section{Translation Rules and Equivalence of $\mathcal{G}$ and $\mathcal{L}$}
\label{apx:grammar}
In this section, first, we explain rewriting rules that translate assertion conditions over control points into constraints over signal variables (Section~\ref{subapx:translation}), then we prove that every formula in $\mathcal{L}$ has an equivalent condition in $\mathcal{G}$ and vice versa (Section~\ref{subapx:equivalence}).

\subsection{Translation of control-point conditions into signal constraints
}
\label{subapx:translation}
To translate assertion conditions defined over control points into constraints defined directly over signal variables, we use two sets of rewriting rules.
The first set of rewriting rules, based on Menghi et al.~\cite{menghi2019generating}, introduces a $\forall$ quantifier over each arithmetic expression, \synt{exp}, while replacing signal control points with signal variables in these expressions. The second set of rules consists of standard logic rewriting rules: one for merging universal quantifiers over disjoint domains, and another for conjunction of universal quantifiers over the same domain.

\textbf{Rules for replacing signal control points with signal variables~\cite{menghi2019generating}.} Let $u: \timedomain \rightarrow \mathbb{R}$ be an input signal.  Suppose we represent $u$ using the following $n_u$ control points:    $c_{u,0}$, $c_{u,1}$, \ldots,  $c_{u,n_u-2}$, $c_{u,n_u-1}$ such that each control point $c_{u,i}$ is positioned at position $i \cdot I$ where $I=\frac{b}{n_u-1}$.  Let \texttt{exp} be an arithmetic expression generated by the grammar $\mathcal{G}$ in Figure~\ref{fig:grammar} in Section~\ref{sec:approach}. Based on the definition of assertions over signals given above, \texttt{exp} contains only signal control points in the same position $j$. This expression can then  be rewritten into the following equivalent logical expression $\forall t \in [j\cdot I, (j+1)\cdot I): \texttt{exp}'$ where  $\texttt{exp}'$ is obtained by substituting each control point  $c_{u,j}$ with the expression  $u(t)$ representing the input signal $u$ at time~$t$. Note that this rewriting rule is valid because we assume that the interpolation function connecting the control points is piecewise constant.

For example, in Figure~\ref{fig:background} in Section~\ref{sec:example}, consider the control points $c_{e,0}$ to $c_{e,5}$ for the apeng signal $e$, and the control points $c_{p,0}$ to $c_{p,5}$ for the pitchwheel signal $p$. The first control point of both signals is at time 0s, the second at 100s, the third at 200s, and so on. Now, consider the following condition over these control points which can be generated by our grammar:

\begin{center}
    $(c_{e,0} - c_{p,0} \le 20) \wedge (c_{e,1} + c_{p,1} < 0)$
\end{center} 
The above condition is rewritten into the following logical formula over apeng ($e$) and pitchwheel ($p$) signals based on the rule discussed above: 
\begin{center}
    $
    (\forall t \in [0,100) : e(t) - p(t) \le 20)$\hspace{0.1cm} $\wedge$
$(\forall t \in [100,200) :
    e(t) + p(t) < 0)$
\end{center}

Note that for the control points at position zero, i.e., $c_{e,0}$  and $c_{p,0}$, we quantify the variable $t$ over the domain $[0,100)$, and for the control points at position one, i.e., $c_{e,1}$  and $c_{p,1}$, we quantify the variable $t$ over the next time slot $[100,200)$.

\textbf{Quantifier conjunction rules.} After introducing universal quantifiers and replacing control points with signal variables, we apply the following  standard logic rewriting rules: 
$$
\begin{array}{l}
\left( \forall t \in A : \texttt{exp} \right) \land \left( \forall t \in B : \texttt{exp} \right) \ \equiv \ \forall t \in A \cup B : \texttt{exp} \\
\left( \forall t \in A : \texttt{exp} \right) \land \left( \forall t \in A : \texttt{exp}' \right) \ \equiv \ \forall t \in A : \texttt{exp} \land \texttt{exp}' 
\end{array}
$$
\\
where $\texttt{exp}$ and $\texttt{exp}'$ are  arithmetic expressions containing signals over the time domain $\timedomain=[0,b]$, and $A$ and $B$ are two time domains that are subsets of $\timedomain$. 

\begin{figure*}[t]
    \centering
    \includegraphics[width=1\linewidth]{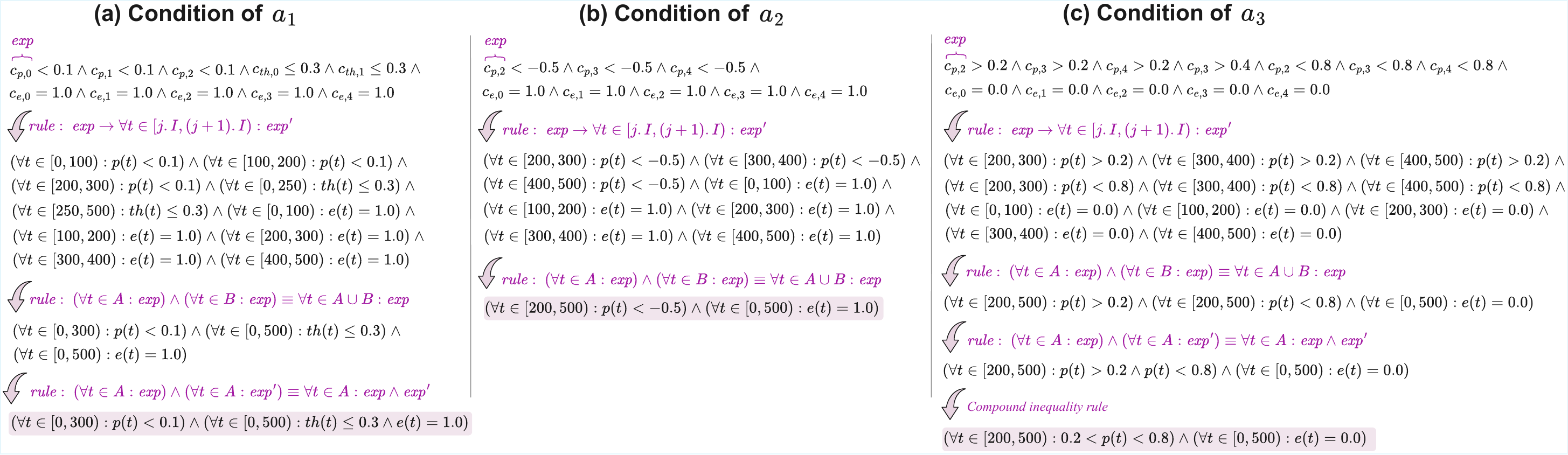}
    \caption{Step-by-step illustration of using rules to derive logical assertion conditions over the signals in Figure~\ref{fig:background} in Section~\ref{sec:example} from assertion conditions based on the control points of apeng ($e$), pitchwheel ($p$) and throttle ($th$).}
    \label{fig:conversion}
\end{figure*}

Figure~\ref{fig:conversion} illustrates the condition assertions generated using our grammar over the control points of the apeng ($e$), pitchwheel ($p$), and throttle ($\mathit{th}$) signals, along with the step-by-step translation of these conditions into logical formulas over their corresponding signal variables.

\subsection{Equivalence between logic fragment $\mathcal{L}$ and grammar $\mathcal{G}$}
\label{subapx:equivalence}
We show equivalence by constructing bidirectional transformations between any formula in $\mathcal{L}$ and conditions expressed by the grammar $\mathcal{G}$ defined in Figure~\ref{fig:grammar}.

\textbf{(1) Transforming a formula in $\mathcal{L}$ to a condition based on grammar $\mathcal{G}$.}
We show that any formula $\phi$ written in $\mathcal{L}$ can be expressed as a condition based on our grammar $\mathcal{G}$. 

\begin{proof}
    We proceed by structural induction on the structure of $\phi \in \mathcal{L}$. 
    
    \textbf{Base Cases.}
    \begin{itemize}
        \item \textbf{Base Case 1: $\mathbf{\phi = r \sim 0}$ }\\
        Let $\phi = r \sim 0$ where $\sim \in \{ <, \leq, >, \geq, =, \neq\}$. Based on the definition of $\mathcal{L}$ in Section~\ref{sec:example}, we have $r \in \mathbb{R}$. Therefore, $r$ maps to \synt{const} in grammar $\mathcal{G}$. Thus, $r \sim 0$ corresponds to \synt{const} $\sim$ 0, which is a \synt{rel-term} in grammar $\mathcal{G}$.

        \item \textbf{Base Case 2: $\mathbf{\phi = \forall t \in \langle n_1, n_2 \rangle: u(t) \sim 0}$ }\\
        Let $\phi = \forall t \in \langle n_1, n_2 \rangle: u(t) \sim 0$, where $u$ is a signal over the time domain $\mathbb{T} = [0, b]$, and $\langle n_1, n_2 \rangle$ is a subinterval of $\mathbb{T}$. Based on Section~\ref{sec:example}, $u$ is encoded using $n_u$ control points, i.e., $c_{u, 0}, \ldots, c_{u, n_u - 1}$ each placed at time instants $0, I, 2\cdot I, \ldots, (n_u - 1) \cdot I$, where $I = \frac{b}{n_u - 1}$. Under the piecewise constant interpolation assumption, the value of $u(t)$ is constant within each interval $[j \cdot I, (j+1) \cdot I)$ and equal to $c_{u,j}$.
        
        If the interval $\langle n_1, n_2 \rangle$ is contained within a single time interval $[j\cdot I, (j+1)\cdot I)$, then $\forall t \in \langle n_1, n_2 \rangle: u(t)$ maps to $c_{u,j}$ which is control point \synt{cp} in grammar $\mathcal{G}$. Thus, $\forall t \in \langle n_1, n_2 \rangle: u(t) \sim 0$ corresponds to \synt{cp} $\sim 0$ which is a \synt{rel-term} in grammar $\mathcal{G}$. 
        
        If $\langle n_1, n_2 \rangle$ spans multiple control-point intervals (e.g., $[j \cdot I, (j+k) \cdot I)$ for some $k \geq 2$), then $\forall t \in \langle n_1, n_2 \rangle: u(t) \sim 0$ can be decomposed into conjunctions over each unit interval:

        $\forall t \in [j\cdot I, (j+1)\cdot I) : u(t) \sim 0 \ \land \ldots \land \forall t \in [(j+k -1)\cdot I, (j+k)\cdot I) : u(t) \sim 0$. 

        Each subformula $\forall t \in [i \cdot I, (i+1) \cdot I): u(t) \sim 0$ is semantically equivalent to $c_{u,i} \sim 0$. Hence, the entire formula corresponds to:

        $c_{u, j} \sim 0 \ \land c_{u, j+1} \sim 0 \ \land \ldots \land c_{u, j+k-1} \sim 0$

        which is an \synt{and-term} over \synt{rel-term}s in grammar $\mathcal{G}$.

        \item \textbf{Base Case 3: $\mathbf{\phi = \forall t \in \langle n_1, n_2 \rangle: \rho \sim 0}$ }\\
        Based on the definition of $\mathcal{L}$ in Section~\ref{sec:example}, $\rho ::= u(t) \mid r \mid \rho_1 + \rho_2 \mid \rho_1 - \rho_2 \mid \rho_1 \times \rho_2 \mid \rho_1 / \rho_2$. By the inductive structure of $\rho$, $\rho_1$ and $\rho_2$ are each either $u(t)$ or a constant $r$, or further composed of arithmetic terms. We note that based on Section~\ref{sec:example}, $\mathcal{L}$ does not allow nested quantifiers, hence, no additional quantifiers (i.e., $\forall t$) appear in $\phi$.

        As in Base Cases 1 and 2, each occurrence of $u(t)$ within the interval $\langle n_1, n_2 \rangle$ can be replaced with the corresponding control point $c_{u,j}$.
        Further, each occurrence of $r$ remains as is. Therefore, $\rho_1$ and $\rho_2$ can be rewritten into arithmetic expressions over control points and constants, i.e., valid \synt{exp}s in $\mathcal{G}$. Hence, the  formula $\forall t \in \langle n_1, n_2 \rangle: \rho \sim 0$ corresponds to an expression like \synt{exp} $\sim 0$  over control points and constants, which is a  \synt{rel-term} in grammar $\mathcal{G}$.

        \end{itemize}

    \textbf{Inductive Cases.}    
    \begin{itemize}    

        \item \textbf{Inductive Case 1: $\mathbf{\phi = \phi_1 \wedge \phi_2}$}\\
        The inductive hypothesis is that $\phi_1$ and $\phi_2$ correspond to \synt{rel-term}s or \synt{and-terms} in  grammar $\mathcal{G}$. 
        Since we assume that no free time variables exist in $\phi_1$ and $\phi_2$ based on Section~\ref{sec:example}, $\phi_1$ and $\phi_2$ ultimately reduce to one of the base cases -- either a simple relational expression over constants (Base Case 1), over control points (Base Case 2) or over arithmetic expressions (Base Case 3).
        Hence, $\phi_1 \wedge \phi_2$ is a conjunction of expressions derived from base cases and hence corresponds to an \synt{and-term} in grammar $\mathcal{G}$.

        \item \textbf{Inductive Case 2: $\mathbf{\phi = \phi_1 \vee \phi_2}$}\\
        Similar to Inductive Case 1, by the inductive hypothesis, both $\phi_1$ and $\phi_2$ correspond to \synt{rel-term}s or \synt{and-terms} in  grammar $\mathcal{G}$. As there are no free time variables, all expressions in $\phi_1$ and $\phi_2$ are grounded in the base cases, making them structurally reducible to \synt{rel-term}s or \synt{and-term}s. Hence, $\phi_1 \vee \phi_2$ is a disjunction of expressions derived from base cases and hence 
        corresponds to an \synt{or-term} in grammar $\mathcal{G}$.
        
    \end{itemize}

    Therefore, every formula $\phi \in \mathcal{L}$ corresponds to a well-formed condition in grammar $\mathcal{G}$.
\end{proof}

\textbf{(2) Transforming a condition in $\mathcal{G}$ to a formula in $\mathcal{L}$. }
The transformation of any condition in $\mathcal{G}$ into a formula in $\mathcal{L}$ can be established analogously, using structural induction on the grammar $\mathcal{G}$. Briefly, 
\begin{itemize}
    \item Each control point \synt{cp} corresponds to a sub-interval of the time domain where the signal $u(t)$ is constant; thus, each control point maps to a formula $\forall  t \in \langle n_1, n_2 \rangle: u(t)$ in $\mathcal{L}$ by the piecewise constant assumption. Further, each \synt{const} corresponds to $r$ in logic $\mathcal{L}$.
    \item Each \synt{rel-term} corresponds directly to a formula of the form $\rho \sim 0$ in $\mathcal{L}$, where $\rho$ is an arithmetic expression over signals and constants. 
    \item Each \synt{and-term} corresponds to conjunction $\phi_1 \wedge \phi_2$, and each \synt{or-term} corresponds to disjunction $\psi_1 \vee \psi_2$ in $\mathcal{L}$.
    
\end{itemize}
Hence, the transformation from grammar $\mathcal{G}$ to logic $\mathcal{L}$ follows the same structural induction as in (1) and is omitted for brevity. 
\\
\\
\textbf{Conclusion.} Based on (1) and (2), we have equivalence between logic fragment $\mathcal{L}$ and grammar $\mathcal{G}$, meaning that for every formula in $\mathcal{L}$ there exists a  condition in $\mathcal{G}$, and for every condition in $\mathcal{G}$ there exists a  formula in $\mathcal{L}$, with both transformations preserving the original semantics.

\newpage

\section{Supplementary results for RQ2 and RQ3}
\label{apx:results}
This section presents the statistical-test results for RQ2 and RQ3. Detailed discussions of these tables are provided in Sections~\ref{sec:RQ2} and ~\ref{sec:RQ3}.

\begin{table*}[ht]
\caption{Statistical tests comparing the accuracy results of $\mathit{GP_O}$ against those of $\mathit{GP_T}$, $\mathit{GP_N}$, DT, DR and ensemble. The p-values highlighted in blue represent cases where $\mathit{GP_O}$ significantly outperforms the compared alternative. The significance level is $0.05$.}
\label{tab:aarstat}

\centering
\scalebox{0.9}{
\begin{tabular}{|c|cc|cc|cc|cc|cc|}
\hline
                                    & \multicolumn{2}{c|}{\textbf{$\mathit{GP_O}$ vs $\mathit{GP_T}$}}                       & \multicolumn{2}{c|}{\textbf{$\mathit{GP_O}$ vs $\mathit{GP_N}$}}                       & \multicolumn{2}{c|}{\textbf{$\mathit{GP_O}$ vs DT}}                           & \multicolumn{2}{c|}{\textbf{$\mathit{GP_O}$ vs DR}}                           & \multicolumn{2}{c|}{\textbf{$\mathit{GP_O}$ vs Ensemble}}                     \\ \cline{2-11} 
\multirow{-2}{*}{\textbf{$\theta$}} & \multicolumn{1}{c|}{\textbf{P-value}}                 & $\mathbf{\hat{A}_{12}}$ & \multicolumn{1}{c|}{\textbf{P-value}}                 & $\mathbf{\hat{A}_{12}}$ & \multicolumn{1}{c|}{\textbf{P-value}}                 & $\mathbf{\hat{A}_{12}}$ & \multicolumn{1}{c|}{\textbf{P-value}}                 & $\mathbf{\hat{A}_{12}}$ & \multicolumn{1}{c|}{\textbf{P-value}}                 & $\mathbf{\hat{A}_{12}}$ \\ \hline
\textbf{0.5}                        & \multicolumn{1}{c|}{\cellcolor[HTML]{96FFFB}\makecell{1.05E-25 \\ \tiny{$GP_O > GP_T$}} } & 0.86 (L)        & \multicolumn{1}{c|}{\cellcolor[HTML]{96FFFB} \makecell{0.03 \\ \tiny{$GP_O > GP_N$}}} & 0.57 (S)        & \multicolumn{1}{c|}{\cellcolor[HTML]{96FFFB}\makecell{1.59E-25 \\ \tiny{$GP_O > DT$}}} & 0.85 (L)       & \multicolumn{1}{c|}{\cellcolor[HTML]{96FFFB}\makecell{4.3E-45 \\ \tiny{$GP_O > DR$}}} & 0.98 (L)        & \multicolumn{1}{c|}{\cellcolor[HTML]{96FFFB}\makecell{7.83E-21 \\ \tiny{$GP_O > Ensemble$}}} & 0.82 (L)        \\ \hline
\textbf{0.55}                       & \multicolumn{1}{c|}{\cellcolor[HTML]{96FFFB}\makecell{3.59E-26 \\ \tiny{$GP_O > GP_T$}}} & 0.87 (L)        & \multicolumn{1}{c|}{\cellcolor[HTML]{96FFFB} \makecell{0.01 \\ \tiny{$GP_O > GP_N$}}} & 0.59 (S)       & \multicolumn{1}{c|}{\cellcolor[HTML]{96FFFB}\makecell{2.14E-25 \\ \tiny{$GP_O > DT$}}} & 0.85 (L)       & \multicolumn{1}{c|}{\cellcolor[HTML]{96FFFB}\makecell{2.56E-44 \\ \tiny{$GP_O > DR$}}} & 0.97 (L)        & \multicolumn{1}{c|}{\cellcolor[HTML]{96FFFB}\makecell{1.98E-19 \\ \tiny{$GP_O > Ensemble$}}} & 0.81 (L)        \\ \hline
\textbf{0.6}                        & \multicolumn{1}{c|}{\cellcolor[HTML]{96FFFB}\makecell{1.5E-21 \\ \tiny{$GP_O > GP_T$}}} & 0.84 (L)        & \multicolumn{1}{c|}{\cellcolor[HTML]{96FFFB} \makecell{0.04 \\ \tiny{$GP_O > GP_N$}}} & 0.57 (S)        & \multicolumn{1}{c|}{\cellcolor[HTML]{96FFFB}\makecell{3.33E-22 \\ \tiny{$GP_O > DT$}}} & 0.84 (L)        & \multicolumn{1}{c|}{\cellcolor[HTML]{96FFFB}\makecell{5.07E-41 \\ \tiny{$GP_N > DR$}}} & 0.98 (L)        & \multicolumn{1}{c|}{\cellcolor[HTML]{96FFFB}\makecell{7E-16 \\ \tiny{$GP_O > Ensemble$}}} & 0.79 (L)        \\ \hline
\textbf{0.65}                       & \multicolumn{1}{c|}{\cellcolor[HTML]{96FFFB}\makecell{9.03E-23 \\ \tiny{$GP_O > GP_T$}}} & 0.88 (L)        & \multicolumn{1}{c|}{\cellcolor[HTML]{96FFFB} \makecell{5.55E-08 \\ \tiny{$GP_O > GP_N$}}} & 0.71  (L)& \multicolumn{1}{c|}{\cellcolor[HTML]{96FFFB}\makecell{7.74E-18 \\ \tiny{$GP_O > DT$}}} & 0.83  (L)       & \multicolumn{1}{c|}{\cellcolor[HTML]{96FFFB}\makecell{3.68E-36 \\ \tiny{$GP_O > DR$}}} & 0.98  (L)       & \multicolumn{1}{c|}{\cellcolor[HTML]{96FFFB}\makecell{9.48E-18 \\ \tiny{$GP_O > Ensemble$}}} & 0.83  (L)       \\ \hline
\textbf{0.7}                        & \multicolumn{1}{c|}{\cellcolor[HTML]{96FFFB}\makecell{1.19E-24 \\ \tiny{$GP_O > GP_T$}}} & 0.92 (L)        & \multicolumn{1}{c|}{\cellcolor[HTML]{96FFFB} \makecell{1.51E-07 \\ \tiny{$GP_O > GP_N$}}} & 0.73 (L)        & \multicolumn{1}{c|}{\cellcolor[HTML]{96FFFB}\makecell{3.37E-16 \\ \tiny{$GP_O > DT$}}} & 0.84 (L)        & \multicolumn{1}{c|}{\cellcolor[HTML]{96FFFB}\makecell{8.42E-32\\ \tiny{$GP_O > DR$}}} & 0.98 (L)        & \multicolumn{1}{c|}{\cellcolor[HTML]{96FFFB}\makecell{1.6E-20 \\ \tiny{$GP_O > Ensemble$}}} & 0.88 (L)        \\ \hline
\textbf{0.75}                       & \multicolumn{1}{c|}{\cellcolor[HTML]{96FFFB}\makecell{1.24E-19 \\ \tiny{$GP_O > GP_T$}}} & 0.94 (L)        & \multicolumn{1}{c|}{\cellcolor[HTML]{96FFFB} \makecell{4.42E-11 \\ \tiny{$GP_O > GP_N$}}} & 0.85 (L)& \multicolumn{1}{c|}{\cellcolor[HTML]{96FFFB}\makecell{1.09E-14 \\ \tiny{$GP_O > DT$}}} & 0.87 (L)        & \multicolumn{1}{c|}{\cellcolor[HTML]{96FFFB}\makecell{1.14E-22 \\ \tiny{$GP_O > DR$}}} & 0.97 (L)        & \multicolumn{1}{c|}{\cellcolor[HTML]{96FFFB}\makecell{1.51E-17 \\ \tiny{$GP_O > Ensemble$}}} & 0.91 (L)        \\ \hline
\textbf{0.8}                        & \multicolumn{1}{c|}{\cellcolor[HTML]{96FFFB}\makecell{2.9E-14 \\ \tiny{$GP_O > GP_T$}}} & 0.94 (L)        & \multicolumn{1}{c|}{\cellcolor[HTML]{96FFFB} \makecell{6.45E-06 \\ \tiny{$GP_O > GP_N$}}} & 0.78  (L)       & \multicolumn{1}{c|}{\cellcolor[HTML]{96FFFB}\makecell{3.33E-08 \\ \tiny{$GP_O > DT$}}} & 0.81 (L)        & \multicolumn{1}{c|}{\cellcolor[HTML]{96FFFB}\makecell{6.42E-17 \\ \tiny{$GP_O > DR$}}} & 0.97  (L)       & \multicolumn{1}{c|}{\cellcolor[HTML]{96FFFB}\makecell{2.33E-11 \\ \tiny{$GP_O > Ensemble$}}} & 0.88 (L)        \\ \hline
\textbf{0.85}                       & \multicolumn{1}{c|}{\cellcolor[HTML]{96FFFB}\makecell{3.61E-13 \\ \tiny{$GP_O > GP_T$}}} & 0.96 (L)        & \multicolumn{1}{c|}{\cellcolor[HTML]{96FFFB} \makecell{0.004 \\ \tiny{$GP_O > GP_N$}}} & 0.72 (L)        & \multicolumn{1}{c|}{\cellcolor[HTML]{96FFFB}\makecell{3.15E-08 \\ \tiny{$GP_O > DT$}}} & 0.84  (L)       & \multicolumn{1}{c|}{\cellcolor[HTML]{96FFFB}\makecell{2.47E-14 \\ \tiny{$GP_O > DR$}}} & 0.97  (L)       & \multicolumn{1}{c|}{\cellcolor[HTML]{96FFFB}\makecell{3.3E-10 \\ \tiny{$GP_O > Ensemble$}}} & 0.89 (L)        \\ \hline
\textbf{0.9}                        & \multicolumn{1}{c|}{\cellcolor[HTML]{96FFFB}\makecell{0.0004 \\ \tiny{$GP_O > GP_T$}}} & 0.85 (L)         & \multicolumn{1}{c|}{\makecell{0.97 \\ \tiny{$GP_O \approx GP_N$}}}                           & 0.50 (N)& \multicolumn{1}{c|}{\makecell{0.81 \\ \tiny{$GP_O \approx DT$}}}                             & 0.52 (N)& \multicolumn{1}{c|}{\cellcolor[HTML]{96FFFB}\makecell{0.0003 \\ \tiny{$GP_O > DR$}}} & 0.85 (L)        & \multicolumn{1}{c|}{\makecell{0.12 \\ \tiny{$GP_O \approx Ensemble$}}}                             & 0.65 (M)        \\ \hline
\textbf{0.95}                       & \multicolumn{1}{c|}{\cellcolor[HTML]{96FFFB}\makecell{0.001 \\ \tiny{$GP_O > GP_T$}}} & 0.89 (L)        & \multicolumn{1}{c|}{\makecell{0.67 \\ \tiny{$GP_O \approx GP_N$}}}                         & 0.57 (S)& \multicolumn{1}{c|}{\makecell{0.84 \\ \tiny{$GP_O \approx DT$}}}                             & 0.47 (N)& \multicolumn{1}{c|}{\cellcolor[HTML]{96FFFB}\makecell{0.0002 \\ \tiny{$GP_O > DR$}}} & 0.93 (L)        & \multicolumn{1}{c|}{\makecell{0.16 \\ \tiny{$GP_O \approx Ensemble$}}}                             & 0.66 (M)\\ \hline
\textbf{1}                          & \multicolumn{1}{c|}{\cellcolor[HTML]{96FFFB}\makecell{0.001 \\ \tiny{$GP_O > GP_T$}}} & 0.91 (L)        & \multicolumn{1}{c|}{\makecell{0.74 \\ \tiny{$GP_O \approx GP_N$}}}                             & 0.57 (S)& \multicolumn{1}{c|}{\makecell{0.28 \\ \tiny{$GP_O \approx DT$}}}                             & 0.36 (M)& \multicolumn{1}{c|}{\cellcolor[HTML]{96FFFB}\makecell{0.001 \\ \tiny{$GP_O > DR$}}} & 0.91 (L)        & \multicolumn{1}{c|}{\makecell{0.18 \\ \tiny{$GP_O \approx Ensemble$}}}                             & 0.67 (M)\\ \hline
\end{tabular}}

\end{table*}

\begin{table*}[h]
\caption{Statistical tests comparing the accuracy results of $\mathit{GP_O}$ against those of $\mathit{GP_T}$, $\mathit{GP_N}$, DT, DR and ensemble for each case-study system. The p-values highlighted in blue represent cases where $\mathit{GP_O}$ significantly outperforms the compared alternative. The p-values highlighted in orange indicate cases where the compared alternative outperforms $\mathit{GP_O}$. The cells highlighted in yellow represent cases where DT is not applicable. The significance level is $0.05$.}
\label{tab:aarstatperstudy}

\centering
\scalebox{0.9}{
\begin{tabular}{|c|cc|cc|cc|cc|cc|}
\hline
                                    & \multicolumn{2}{c|}{\textbf{$\mathit{GP_O}$ vs $\mathit{GP_T}$}}                       & \multicolumn{2}{c|}{\textbf{$\mathit{GP_O}$ vs $\mathit{GP_N}$}}                       & \multicolumn{2}{c|}{\textbf{$\mathit{GP_O}$ vs DT}}                           & \multicolumn{2}{c|}{\textbf{$\mathit{GP_O}$ vs DR}}                           & \multicolumn{2}{c|}{\textbf{$\mathit{GP_O}$ vs Ensemble}}                     \\ \cline{2-11} 
\multirow{-2}{*}{\textbf{Case-Study System}} & \multicolumn{1}{c|}{\textbf{P-value}}                 & $\mathbf{\hat{A}_{12}}$ & \multicolumn{1}{c|}{\textbf{P-value}}                 & $\mathbf{\hat{A}_{12}}$ & \multicolumn{1}{c|}{\textbf{P-value}}                 & $\mathbf{\hat{A}_{12}}$ & \multicolumn{1}{c|}{\textbf{P-value}}                 & $\mathbf{\hat{A}_{12}}$ & \multicolumn{1}{c|}{\textbf{P-value}}                 & $\mathbf{\hat{A}_{12}}$ \\ \hline
\textbf{Router}                        & \multicolumn{1}{c|}{\cellcolor[HTML]{96FFFB}\makecell{1.9E-33 \\ \tiny{$GP_O > GP_T$}}} & 0.87 (L)        & \multicolumn{1}{c|}{\cellcolor[HTML]{96FFFB} \makecell{8.4E-05 \\ \tiny{$GP_O > GP_N$}}} & 0.62 (S)       & \multicolumn{1}{c|}{\cellcolor[HTML]{F8A102}\makecell{4.55E-15 \\ \tiny{$DT > GP_O$}}} & 0.29 (L)        & \multicolumn{1}{c|}{\cellcolor[HTML]{96FFFB}\makecell{4.72E-51 \\ \tiny{$GP_O > DR$}}} & 0.91  (L)       & \multicolumn{1}{c|}{\cellcolor[HTML]{96FFFB} \makecell{6.58E-07 \\ \tiny{$GP_O > Ensemble$}}} & 0.65 (M)        \\ \hline
\textbf{\textsc{AP--DHB}}                       & \multicolumn{1}{c|}{\cellcolor[HTML]{96FFFB}\makecell{0.0001 \\ \tiny{$GP_O > GP_T$}}} & 0.68 (L)        & \multicolumn{1}{c|}{\makecell{0.16 \\ \tiny{$GP_N \approx GP_O$}}} & 0.43 (S)        & \multicolumn{1}{c|}{\cellcolor[HTML]{96FFFB}\makecell{1.64E-59 \\ \tiny{$GP_O > DT$}}} & 1  (L)       & \multicolumn{1}{c|}{\cellcolor[HTML]{96FFFB}\makecell{2.75E-26 \\ \tiny{$GP_O > DR$}}} & 1  (L)       & \multicolumn{1}{c|}{\cellcolor[HTML]{96FFFB} \makecell{1.93E-09 \\ \tiny{$GP_O > Ensemble$}}} & 0.77 (L)        \\ \hline

\textbf{\textsc{AP--TWN (R1)}}                        & \multicolumn{1}{c|}{\cellcolor[HTML]{96FFFB}\makecell{1.37E-11 \\ \tiny{$GP_O > GP_T$}}} & 1 (L)        & \multicolumn{1}{c|}{\cellcolor[HTML]{96FFFB} \makecell{6.69E-11 \\ \tiny{$GP_O > GP_N$}}} & 1 (L)        & \multicolumn{1}{c|}{\cellcolor[HTML]{96FFFB}\makecell{9.01E-12 \\ \tiny{$GP_O > DT$}}} & 1 (L)        & \multicolumn{1}{c|}{\cellcolor[HTML]{96FFFB}\makecell{4.42E-13 \\ \tiny{$GP_O > DR$}}} & 1 (L)        & \multicolumn{1}{c|}{\cellcolor[HTML]{96FFFB}\makecell{2.06E-11 \\ \tiny{$GP_O > Ensemble$}}} & 1 (L)        \\ \hline
\textbf{\textsc{AP--TWN (R2)}}                       & \multicolumn{1}{c|}{\cellcolor[HTML]{96FFFB}\makecell{0.002 \\ \tiny{$GP_O > GP_T$}}} & 0.60 (S)        & \multicolumn{1}{c|}{\makecell{0.17 \\ \tiny{$GP_N \approx GP_O$}}} & 0.54  (N)       & \multicolumn{1}{c|}{\cellcolor[HTML]{96FFFB}\makecell{1.83E-12 \\ \tiny{$GP_O > DT$}}} & 0.72  (L)       & \multicolumn{1}{c|}{\cellcolor[HTML]{96FFFB}\makecell{7.52E-66 \\ \tiny{$GP_O > DR$}}} & 1  (L)       & \multicolumn{1}{c|}{\cellcolor[HTML]{F8A102} \makecell{0.01 \\ \tiny{$Ensemble > GP_O$}}} & 0.41  (S)       \\ \hline
\textbf{\textsc{AP--TWN (R3)}}                        & \multicolumn{1}{c|}{\cellcolor[HTML]{96FFFB}\makecell{9.1E-33 \\ \tiny{$GP_O > GP_T$}}} & 1 (L)        & \multicolumn{1}{c|}{\cellcolor[HTML]{96FFFB} \makecell{1.01E-22 \\ \tiny{$GP_O > GP_N$}}} & 0.91 (L)        & \multicolumn{1}{c|}{\cellcolor[HTML]{96FFFB}\makecell{3.66E-69 \\ \tiny{$GP_O > DT$}}} & 1 (L)        & \multicolumn{1}{c|}{\cellcolor[HTML]{96FFFB}\makecell{4.72E-51 \\ \tiny{$GP_O > DR$}}} & 1 (L)        & \multicolumn{1}{c|}{\cellcolor[HTML]{96FFFB} \makecell{3.48E-43 \\ \tiny{$GP_O > Ensemble$}}} & 1 (L)        \\ \hline
\textbf{\textsc{AP--TWN (R4)}}                       & \multicolumn{1}{c|}{\cellcolor[HTML]{96FFFB}\makecell{1.41E-33 \\ \tiny{$GP_O > GP_T$}}} & 1 (L)        & \multicolumn{1}{c|}{\cellcolor[HTML]{96FFFB} \makecell{2.03E-15 \\ \tiny{$GP_O > GP_N$}}} & 0.82 (L)        & \multicolumn{2}{c|}{\cellcolor[HTML]{FFFC9E} DT is not applicable}          & \multicolumn{1}{c|}{\cellcolor[HTML]{96FFFB}\makecell{6.13E-21 \\ \tiny{$GP_O > DR$}}} & 1 (L)        & \multicolumn{1}{c|}{\cellcolor[HTML]{96FFFB} \makecell{8.76E-33 \\ \tiny{$GP_O > Ensemble$}}} & 1 (L)        \\ \hline
\textbf{\textsc{AP--SNG}}                        & \multicolumn{1}{c|}{\cellcolor[HTML]{96FFFB}\makecell{8.5E-46 \\ \tiny{$GP_O > GP_T$}}} & 1 (L)        & \multicolumn{1}{c|}{\cellcolor[HTML]{96FFFB} \makecell{2.48E-42 \\ \tiny{$GP_O > GP_N$}}} & 1 (L)        & \multicolumn{1}{c|}{\cellcolor[HTML]{96FFFB}\makecell{7.42E-68 \\ \tiny{$GP_O > DT$}}} & 1 (L)       & \multicolumn{1}{c|}{\cellcolor[HTML]{96FFFB}\makecell{1.17E-50 \\ \tiny{$GP_O > DR$}}} & 1 (L)        & \multicolumn{1}{c|}{\cellcolor[HTML]{96FFFB}\makecell{1.83E-58 \\ \tiny{$GP_O > Ensemble$}}} & 1 (L)        \\ \hline
\textbf{\textsc{Dave2}}                       & \multicolumn{1}{c|}{\cellcolor[HTML]{96FFFB}\makecell{7.11E-13 \\ \tiny{$GP_O > GP_T$}}} & 0.78 (L)        & \multicolumn{1}{c|}{\cellcolor[HTML]{F8A102} \makecell{0.02 \\ \tiny{$GP_N > GP_O$}}} & 0.40 (S)       & \multicolumn{1}{c|}{\cellcolor[HTML]{96FFFB}\makecell{1.24E-46 \\ \tiny{$GP_O > DT$}}} & 1 (L)       & \multicolumn{1}{c|}{\cellcolor[HTML]{96FFFB}\makecell{1.23E-22 \\ \tiny{$GP_O > DR$}}} & 0.86 (L)        & \multicolumn{1}{c|}{\cellcolor[HTML]{96FFFB}\makecell{6.78E-09 \\ \tiny{$GP_O > Ensemble$}}} & 0.72 (L)        \\ \hline
\end{tabular}}

\end{table*}

\begin{table*}[ht]
\caption{Statistical test results comparing (a) the rate of pass verdicts predicted as fail achieved with DR against those of $\mathit{GP_T}$, $\mathit{GP_O}$, $\mathit{GP_N}$, DT and ensemble (b) the rate of fail verdicts predicted as pass achieved with $\mathit{GP_O}$ against those of $\mathit{GP_T}$, $\mathit{GP_N}$, DT, DR and ensemble. The p-values highlighted in blue represent cases where the method on the left significantly outperforms the compared alternative. The p-values highlighted in orange indicate cases where the compared alternative significantly outperforms the method on the left. The significance level is $0.05$. }
\label{tab:inaccstat}
\centering
\begin{subtable}{\linewidth}
\centering
\subcaption{Rate of pass verdicts predicted as fail}
\scalebox{0.9}{
\begin{tabular}{|c|cc|cc|cc|cc|cc|}
\hline
                                    & \multicolumn{2}{c|}{\textbf{DR vs $\mathit{GP_T}$}}                       & \multicolumn{2}{c|}{\textbf{DR vs $\mathit{GP_O}$}}                       & \multicolumn{2}{c|}{\textbf{DR vs $\mathit{GP_N}$}}                           & \multicolumn{2}{c|}{\textbf{DR vs DT}}                           & \multicolumn{2}{c|}{\textbf{DR vs Ensemble}}                     \\ \cline{2-11} 
\multirow{-2}{*}{\textbf{$\theta$}} & \multicolumn{1}{c|}{\textbf{P-value}}                 & $\mathbf{\hat{A}_{12}}$ & \multicolumn{1}{c|}{\textbf{P-value}}                 & $\mathbf{\hat{A}_{12}}$ & \multicolumn{1}{c|}{\textbf{P-value}}                 & $\mathbf{\hat{A}_{12}}$ & \multicolumn{1}{c|}{\textbf{P-value}}                 & $\mathbf{\hat{A}_{12}}$ & \multicolumn{1}{c|}{\textbf{P-value}}                 & $\mathbf{\hat{A}_{12}}$ \\ \hline
\textbf{0.5}                        & \multicolumn{1}{c|}{\cellcolor[HTML]{96FFFB}\makecell{2.85E-11 \\ \tiny{$DR > GP_T$}}} & 0.29 (L)        & \multicolumn{1}{c|}{\cellcolor[HTML]{96FFFB}\makecell{3.56E-15 \\ \tiny{$DR > GP_O$}}} & 0.24 (L)        & \multicolumn{1}{c|}{\cellcolor[HTML]{96FFFB}\makecell{6.28E-18 \\ \tiny{$DR > GP_N$}}} & 0.22 (L)       & \multicolumn{1}{c|}{\cellcolor[HTML]{96FFFB}\makecell{1.11E-10 \\ \tiny{$DR > DT$}}} & 0.30 (M)        & \multicolumn{1}{c|}{\cellcolor[HTML]{96FFFB}\makecell{1.32E-12 \\ \tiny{$DR > Ensemble$}}} & 0.28 (L)        \\ \hline
\textbf{0.55}                       & \multicolumn{1}{c|}{\cellcolor[HTML]{96FFFB}\makecell{3.72E-11 \\ \tiny{$DR > GP_T$}}} & 0.29 (L)        & \multicolumn{1}{c|}{\cellcolor[HTML]{96FFFB}\makecell{5.52E-15 \\ \tiny{$DR > GP_O$}}} & 0.24 (L)       & \multicolumn{1}{c|}{\cellcolor[HTML]{96FFFB}\makecell{3.83E-17 \\ \tiny{$DR > GP_N$}}} & 0.22 (L)       & \multicolumn{1}{c|}{\cellcolor[HTML]{96FFFB}\makecell{3E-10 \\ \tiny{$DR > DT$}}} & 0.30 (M)        & \multicolumn{1}{c|}{\cellcolor[HTML]{96FFFB}\makecell{1.1E-12 \\ \tiny{$DR > Ensemble$}}} & 0.28 (L)        \\ \hline
\textbf{0.6}                        & \multicolumn{1}{c|}{\cellcolor[HTML]{96FFFB}\makecell{9.75E-11 \\ \tiny{$DR > GP_T$}}} & 0.30 (M)        & \multicolumn{1}{c|}{\cellcolor[HTML]{96FFFB}\makecell{3.78E-14 \\ \tiny{$DR > GP_O$}}} & 0.24 (L)        & \multicolumn{1}{c|}{\cellcolor[HTML]{96FFFB}\makecell{3.83E-17 \\ \tiny{$DR > GP_N$}}} & 0.22 (L)        & \multicolumn{1}{c|}{\cellcolor[HTML]{96FFFB}\makecell{3E-10 \\ \tiny{$DR > DT$}}} & 0.30 (M)        & \multicolumn{1}{c|}{\cellcolor[HTML]{96FFFB}\makecell{1.09E-12 \\ \tiny{$DR > Ensemble$}}} & 0.28 (L)        \\ \hline
\textbf{0.65}                       & \multicolumn{1}{c|}{\cellcolor[HTML]{96FFFB}\makecell{3.24E-10 \\ \tiny{$DR > GP_T$}}} & 0.30 (M)        & \multicolumn{1}{c|}{\cellcolor[HTML]{96FFFB}\makecell{2.2E-13 \\ \tiny{$DR > GP_O$}}} & 0.23  (L)& \multicolumn{1}{c|}{\cellcolor[HTML]{96FFFB}\makecell{2.31E-08 \\ \tiny{$DR > GP_N$}}} & 0.31  (M)       & \multicolumn{1}{c|}{\cellcolor[HTML]{96FFFB}\makecell{5.28E-06 \\ \tiny{$DR > DT$}}} & 0.35  (M)       & \multicolumn{1}{c|}{\cellcolor[HTML]{96FFFB}\makecell{2.09E-07 \\ \tiny{$DR > Ensemble$}}} & 0.34  (M)       \\ \hline
\textbf{0.7}                        & \multicolumn{1}{c|}{\cellcolor[HTML]{96FFFB}\makecell{3.94E-08 \\ \tiny{$DR > GP_T$}}} & 0.32 (M)        & \multicolumn{1}{c|}{\cellcolor[HTML]{96FFFB}\makecell{5.93E-22 \\ \tiny{$DR > GP_O$}}} & 0.11 (L)        & \multicolumn{1}{c|}{\cellcolor[HTML]{96FFFB}\makecell{3.68E-20 \\ \tiny{$DR > GP_N$}}} & 0.16 (L)        & \multicolumn{1}{c|}{\cellcolor[HTML]{96FFFB}\makecell{5.01E-06 \\ \tiny{$DR > DT$}}} & 0.35 (M)        & \multicolumn{1}{c|}{\cellcolor[HTML]{96FFFB}\makecell{1.83E-07 \\ \tiny{$DR > Ensemble$}}} & 0.34 (M)        \\ \hline
\textbf{0.75}                       & \multicolumn{1}{c|}{\cellcolor[HTML]{96FFFB}\makecell{1.88E-05 \\ \tiny{$DR > GP_T$}}} & 0.36 (M)        & \multicolumn{1}{c|}{\cellcolor[HTML]{96FFFB}\makecell{2.76E-12 \\ \tiny{$DR > GP_O$}}} & 0.17 (L)& \multicolumn{1}{c|}{\cellcolor[HTML]{96FFFB}\makecell{2.52E-15 \\ \tiny{$DR > GP_N$}}} & 0.19 (L)        & \multicolumn{1}{c|}{\cellcolor[HTML]{96FFFB}\makecell{1.99E-05 \\ \tiny{$DR > DT$}}} & 0.36 (M)        & \multicolumn{1}{c|}{\cellcolor[HTML]{96FFFB}\makecell{0.0007 \\ \tiny{$DR > Ensemble$}}} & 0.39 (S)        \\ \hline
\textbf{0.8}                        & \multicolumn{1}{c|}{\cellcolor[HTML]{96FFFB}\makecell{0.0001 \\ \tiny{$DR > GP_T$}}} & 0.37 (S)        & \multicolumn{1}{c|}{\cellcolor[HTML]{96FFFB}\makecell{1.64E-13 \\ \tiny{$DR > GP_O$}}} & 0.10  (L)       & \multicolumn{1}{c|}{\cellcolor[HTML]{96FFFB}\makecell{3.82E-13 \\ \tiny{$DR > GP_N$}}} & 0.20 (L)        & \multicolumn{1}{c|}{\cellcolor[HTML]{96FFFB}\makecell{1.9E-05 \\ \tiny{$DR > DT$}}} & 0.36  (M)       & \multicolumn{1}{c|}{\cellcolor[HTML]{96FFFB}\makecell{6.5E-05 \\ \tiny{$DR > Ensemble$}}} & 0.37 (S)        \\ \hline
\textbf{0.85}                       & \multicolumn{1}{c|}{\cellcolor[HTML]{96FFFB}\makecell{0.0003 \\ \tiny{$DR > GP_T$}}} & 0.37 (S)        & \multicolumn{1}{c|}{\cellcolor[HTML]{96FFFB}\makecell{9.42E-14 \\ \tiny{$DR > GP_O$}}} & 0.06 (L)        & \multicolumn{1}{c|}{\cellcolor[HTML]{96FFFB}\makecell{3.19E-15 \\ \tiny{$DR > GP_N$}}} & 0.05  (L)       & \multicolumn{1}{c|}{\cellcolor[HTML]{96FFFB}\makecell{1.23E-05 \\ \tiny{$DR > DT$}}} & 0.35  (M)       & \multicolumn{1}{c|}{\cellcolor[HTML]{96FFFB}\makecell{7.44E-05 \\ \tiny{$DR > Ensemble$}}} & 0.37 (S)        \\ \hline
\textbf{0.9}                        & \multicolumn{1}{c|}{\cellcolor[HTML]{96FFFB}\makecell{4.3E-05 \\ \tiny{$DR > GP_T$}}} & 0.35 (M)         & \multicolumn{1}{c|}{\cellcolor[HTML]{96FFFB}\makecell{0.002 \\ \tiny{$DR > GP_O$}}}                             & 0.23 (L)& \multicolumn{1}{c|}{\cellcolor[HTML]{96FFFB}\makecell{8.94E-07 \\ \tiny{$DR > GP_N$}}}                             & 0.11 (L)& \multicolumn{1}{c|}{\cellcolor[HTML]{96FFFB}\makecell{3.26E-06 \\ \tiny{$DR > DT$}}} & 0.34 (M)        & \multicolumn{1}{c|}{\cellcolor[HTML]{96FFFB}\makecell{5.12E-05 \\ \tiny{$DR > Ensemble$}}}                             & 0.36 (M)        \\ \hline
\textbf{0.95}                       & \multicolumn{1}{c|}{\cellcolor[HTML]{96FFFB}\makecell{0.0006 \\ \tiny{$DR > GP_T$}}} & 0.37 (S)        & \multicolumn{1}{c|}{\cellcolor[HTML]{96FFFB} \makecell{0.01 \\ \tiny{$DR > GP_O$}}}                             & 0.23 (L)& \multicolumn{1}{c|}{\cellcolor[HTML]{96FFFB}\makecell{6.06E-06 \\ \tiny{$DR > GP_N$}}}                             & 0.07 (L)& \multicolumn{1}{c|}{\cellcolor[HTML]{96FFFB}\makecell{0.0003 \\ \tiny{$DR > DT$}}} & 0.37 (S)        & \multicolumn{1}{c|}{\cellcolor[HTML]{96FFFB}\makecell{0.003 \\ \tiny{$DR > Ensemble$}}}                             & 0.39 (S)\\ \hline
\textbf{1}                          & \multicolumn{1}{c|}{\cellcolor[HTML]{96FFFB}\makecell{2.82E-05 \\ \tiny{$DR > GP_T$}}} & 0.33 (M)        & \multicolumn{1}{c|}{\cellcolor[HTML]{96FFFB} \makecell{0.03 \\ \tiny{$DR > GP_O$}}}                             & 0.26 (L)& \multicolumn{1}{c|}{\cellcolor[HTML]{96FFFB}\makecell{3.69E-05 \\ \tiny{$DR > GP_N$}}}                             & 0.09 (L)& \multicolumn{1}{c|}{\cellcolor[HTML]{96FFFB}\makecell{2.13E-05 \\ \tiny{$DR > DT$}}} & 0.33 (M)        & \multicolumn{1}{c|}{\cellcolor[HTML]{96FFFB}\makecell{0.001 \\ \tiny{$DR > Ensemble$}}}                             & 0.38 (S)\\ \hline
\end{tabular}}
\end{subtable}

\vspace{0.5cm}

\begin{subtable}{\linewidth}
\centering
\subcaption{Rate of fail verdicts predicted as pass}
\scalebox{0.9}{
\begin{tabular}{|c|cc|cc|cc|cc|cc|}
\hline
                                    & \multicolumn{2}{c|}{\textbf{$\mathit{GP_O}$ vs $\mathit{GP_T}$}}                       & \multicolumn{2}{c|}{\textbf{$\mathit{GP_O}$ vs $\mathit{GP_N}$}}                       & \multicolumn{2}{c|}{\textbf{$\mathit{GP_O}$ vs DT}}                           & \multicolumn{2}{c|}{\textbf{$\mathit{GP_O}$ vs DR}}                           & \multicolumn{2}{c|}{\textbf{$\mathit{GP_O}$ vs Ensemble}}                     \\ \cline{2-11} 
\multirow{-2}{*}{\textbf{$\theta$}} & \multicolumn{1}{c|}{\textbf{P-value}}                 & $\mathbf{\hat{A}_{12}}$ & \multicolumn{1}{c|}{\textbf{P-value}}                 & $\mathbf{\hat{A}_{12}}$ & \multicolumn{1}{c|}{\textbf{P-value}}                 & $\mathbf{\hat{A}_{12}}$ & \multicolumn{1}{c|}{\textbf{P-value}}                 & $\mathbf{\hat{A}_{12}}$ & \multicolumn{1}{c|}{\textbf{P-value}}                 & $\mathbf{\hat{A}_{12}}$ \\ \hline
\textbf{0.5}                        & \multicolumn{1}{c|}{\cellcolor[HTML]{96FFFB} \makecell{0.03 \\ \tiny{$GP_O > GP_T$}}} & 0.43 (S)        & \multicolumn{1}{c|}{\makecell{0.07 \\ \tiny{$GP_O \approx GP_N$}}} & 0.43 (S)        & \multicolumn{1}{c|}{\cellcolor[HTML]{96FFFB} \makecell{0.02 \\ \tiny{$GP_O > DT$}}} & 0.42 (S)       & \multicolumn{1}{c|}{\cellcolor[HTML]{96FFFB}\makecell{0.0006 \\ \tiny{$GP_O > DR$}}} & 0.38 (S)        & \multicolumn{1}{c|}{\makecell{0.08 \\ \tiny{$GP_O \approx Ensemble$}}} & 0.44 (N)        \\ \hline
\textbf{0.55}                       & \multicolumn{1}{c|}{\makecell{0.61 \\ \tiny{$GP_O \approx GP_T$}}} & 0.48 (N)        & \multicolumn{1}{c|}{\makecell{0.06 \\ \tiny{$GP_O \approx GP_N$}}} & 0.43 (S)       & \multicolumn{1}{c|}{\makecell{0.72 \\ \tiny{$GP_O \approx DT$}}} & 0.51 (N)       & \multicolumn{1}{c|}{\makecell{0.20 \\ \tiny{$GP_O \approx DR$}}} & 0.45 (N)        & \multicolumn{1}{c|}{\makecell{0.88 \\ \tiny{$GP_O \approx Ensemble$}}} & 0.50 (N)        \\ \hline
\textbf{0.6}                        & \multicolumn{1}{c|}{\cellcolor[HTML]{96FFFB} \makecell{0.03 \\ \tiny{$GP_O > GP_T$}}} & 0.43 (S)        & \multicolumn{1}{c|}{\cellcolor[HTML]{96FFFB}\makecell{0.001 \\ \tiny{$GP_O > GP_N$}}} & 0.39 (S)        & \multicolumn{1}{c|}{\cellcolor[HTML]{96FFFB} \makecell{0.03 \\ \tiny{$GP_O > DT$}}} & 0.43 (S)        & \multicolumn{1}{c|}{\cellcolor[HTML]{96FFFB}\makecell{0.0001 \\ \tiny{$GP_O > DR$}}} & 0.36 (M)        & \multicolumn{1}{c|}{\makecell{0.06 \\ \tiny{$GP_O \approx Ensemble$}}} & 0.43 (S)        \\ \hline
\textbf{0.65}                       & \multicolumn{1}{c|}{\cellcolor[HTML]{96FFFB} \makecell{0.02 \\ \tiny{$GP_O > GP_T$}}} & 0.42 (S)        & \multicolumn{1}{c|}{ \makecell{0.03 \\ \tiny{$GP_O \approx GP_N$}}} & 0.42  (S)& \multicolumn{1}{c|}{\cellcolor[HTML]{96FFFB}\makecell{0.0005 \\ \tiny{$GP_O > DT$}}} & 0.37  (S)       & \multicolumn{1}{c|}{\cellcolor[HTML]{96FFFB}\makecell{0.0002 \\ \tiny{$GP_O > DR$}}} & 0.36  (M)       & \multicolumn{1}{c|}{\cellcolor[HTML]{96FFFB} \makecell{0.01 \\ \tiny{$GP_O > Ensemble$}}} & 0.41  (S)\\ \hline
\textbf{0.7}                        & \multicolumn{1}{c|}{\cellcolor[HTML]{96FFFB}\makecell{3.47E-06 \\ \tiny{$GP_O > GP_T$}}} & 0.34 (M)        & \multicolumn{1}{c|}{\cellcolor[HTML]{96FFFB}\makecell{0.008 \\ \tiny{$GP_O > GP_N$}}} & 0.42 (S)& \multicolumn{1}{c|}{\cellcolor[HTML]{96FFFB}\makecell{9.12E-15 \\ \tiny{$GP_O > DT$}}} & 0.19 (L)        & \multicolumn{1}{c|}{\cellcolor[HTML]{96FFFB}\makecell{1.91E-13 \\ \tiny{$GP_O > DR$}}} & 0.21 (L)        & \multicolumn{1}{c|}{\cellcolor[HTML]{96FFFB}\makecell{8.52E-10 \\ \tiny{$GP_O > Ensemble$}}} & 0.27 (L)        \\ \hline
\textbf{0.75}                       & \multicolumn{1}{c|}{\cellcolor[HTML]{96FFFB}\makecell{0.001 \\ \tiny{$GP_O > GP_T$}}} & 0.38 (S)        & \multicolumn{1}{c|}{\makecell{0.27 \\ \tiny{$GP_O \approx GP_N$}}} & 0.52 (N)& \multicolumn{1}{c|}{\cellcolor[HTML]{96FFFB}\makecell{7.6E-11 \\ \tiny{$GP_O > DT$}}} & 0.19 (L)& \multicolumn{1}{c|}{\cellcolor[HTML]{96FFFB}\makecell{7.67E-08 \\ \tiny{$GP_O > DR$}}} & 0.25 (L)        & \multicolumn{1}{c|}{\cellcolor[HTML]{96FFFB}\makecell{6.19E-07 \\ \tiny{$GP_O > Ensemble$}}} & 0.28 (L)\\ \hline
\textbf{0.8}                        & \multicolumn{1}{c|}{\makecell{0.32 \\ \tiny{$GP_O \approx GP_T$}}} & 0.46 (N)        & \multicolumn{1}{c|}{\makecell{0.05 \\ \tiny{$GP_O \approx GP_N$}}} & 0.54  (N)& \multicolumn{1}{c|}{\cellcolor[HTML]{96FFFB}\makecell{7.79E-07 \\ \tiny{$GP_O > DT$}}} & 0.22 (L)& \multicolumn{1}{c|}{\makecell{0.19 \\ \tiny{$GP_O \approx DR$}}} & 0.44  (N)       & \multicolumn{1}{c|}{\cellcolor[HTML]{96FFFB}\makecell{0.0005 \\ \tiny{$GP_O > Ensemble$}}} & 0.32 (M)        \\ \hline
\textbf{0.85}                       & \multicolumn{1}{c|}{\makecell{0.89 \\ \tiny{$GP_O \approx GP_T$}}} & 0.50 (N)        & \multicolumn{1}{c|}{\makecell{0.63 \\ \tiny{$GP_O \approx GP_N$}}} & 0.52 (N)& \multicolumn{1}{c|}{\cellcolor[HTML]{96FFFB}\makecell{8.38E-06 \\ \tiny{$GP_O > DT$}}} & 0.22  (L)       & \multicolumn{1}{c|}{\makecell{0.45 \\ \tiny{$GP_O \approx DR$}}} & 0.46  (N)       & \multicolumn{1}{c|}{\cellcolor[HTML]{96FFFB}\makecell{0.002 \\ \tiny{$GP_O > Ensemble$}}} & 0.32 (M)        \\ \hline
\textbf{0.9}                        & \multicolumn{1}{c|}{\cellcolor[HTML]{F8A102}\makecell{0.002 \\ \tiny{$GP_T > GP_O$}}} & 0.64 (M)         & \multicolumn{1}{c|}{\makecell{0.50 \\ \tiny{$GP_O \approx GP_N$}}}                             & 0.56 (S)& \multicolumn{1}{c|}{\makecell{0.80 \\ \tiny{$GP_O \approx DT$}}}                             & 0.47 (N)& \multicolumn{1}{c|}{\makecell{0.18 \\ \tiny{$GP_O \approx DR$}}} & 0.59 (S)        & \multicolumn{1}{c|}{\makecell{0.56 \\ \tiny{$GP_O \approx Ensemble$}}}                             & 0.54 (N)\\ \hline
\textbf{0.95}                       & \multicolumn{1}{c|}{\cellcolor[HTML]{F8A102}\makecell{0.0003 \\ \tiny{$GP_T > GP_O$}}} & 0.65 (M)        & \multicolumn{1}{c|}{\makecell{0.44 \\ \tiny{$GP_O \approx GP_N$}}}                             & 0.59 (S)& \multicolumn{1}{c|}{\makecell{0.83 \\ \tiny{$GP_O \approx DT$}}}                             & 0.52 (N)& \multicolumn{1}{c|}{\makecell{0.39 \\ \tiny{$GP_O \approx DR$}}} & 0.57 (S)        & \multicolumn{1}{c|}{\makecell{0.26 \\ \tiny{$GP_O \approx Ensemble$}}}                             & 0.59 (S)        \\ \hline
\textbf{1}                          & \multicolumn{1}{c|}{\cellcolor[HTML]{F8A102}\makecell{6.99E-05 \\ \tiny{$GP_T > GP_O$}}} & 0.68 (M)        & \multicolumn{1}{c|}{\makecell{0.52 \\ \tiny{$GP_O \approx GP_N$}}}                             & 0.60 (S)& \multicolumn{1}{c|}{\makecell{0.88 \\ \tiny{$GP_O \approx DT$}}}                             & 0.52 (N)& \multicolumn{1}{c|}{\makecell{0.34 \\ \tiny{$GP_O \approx DR$}}} & 0.60 (S)        & \multicolumn{1}{c|}{\makecell{0.24 \\ \tiny{$GP_O \approx Ensemble$}}}                             & 0.61 (S)        \\ \hline
\end{tabular}}

\end{subtable}

\end{table*}
\begin{table*}[ht]
\caption{Statistical test results comparing (a) the rate of pass verdicts predicted as fail achieved with DR against those of $\mathit{GP_T}$, $\mathit{GP_O}$, $\mathit{GP_N}$, DT and ensemble (b) the rate of fail verdicts predicted as pass achieved with $\mathit{GP_O}$ against those of $\mathit{GP_T}$, $\mathit{GP_N}$, DT, DR and ensemble for each case-study system. The p-values highlighted in blue represent cases where the method on the left significantly outperforms the compared alternative. The p-values highlighted in orange indicate cases where the compared alternative significantly outperforms the method on the left. The cells highlighted in yellow represent cases where DT is not applicable. The significance level is $0.05$.}
\label{tab:inaccstatperstudy}
\centering
\begin{subtable}{\linewidth}
\centering
\subcaption{Rate of pass verdicts predicted as fail}
\scalebox{0.9}{
\begin{tabular}{|c|cc|cc|cc|cc|cc|}
\hline
                                    & \multicolumn{2}{c|}{\textbf{DR vs $\mathit{GP_T}$}}                       & \multicolumn{2}{c|}{\textbf{DR vs $\mathit{GP_O}$}}                       & \multicolumn{2}{c|}{\textbf{DR vs $\mathit{GP_N}$}}                           & \multicolumn{2}{c|}{\textbf{DR vs DT}}                           & \multicolumn{2}{c|}{\textbf{DR vs Ensemble}}                     \\ \cline{2-11} 
\multirow{-2}{*}{\textbf{Case-Study System}} & \multicolumn{1}{c|}{\textbf{P-value}}                 & $\mathbf{\hat{A}_{12}}$ & \multicolumn{1}{c|}{\textbf{P-value}}                 & $\mathbf{\hat{A}_{12}}$ & \multicolumn{1}{c|}{\textbf{P-value}}                 & $\mathbf{\hat{A}_{12}}$ & \multicolumn{1}{c|}{\textbf{P-value}}                 & $\mathbf{\hat{A}_{12}}$ & \multicolumn{1}{c|}{\textbf{P-value}}                 & $\mathbf{\hat{A}_{12}}$ \\ \hline
\textbf{Router}                        & \multicolumn{1}{c|}{\cellcolor[HTML]{96FFFB}\makecell{1.37E-49 \\ \tiny{$DR > GP_T$}}} & 0.13 (L)        & \multicolumn{1}{c|}{\cellcolor[HTML]{96FFFB} \makecell{4.95E-14 \\ \tiny{$DR > GP_O$}}} & 0.29 (L)       & \multicolumn{1}{c|}{\cellcolor[HTML]{96FFFB}\makecell{7.06E-17 \\ \tiny{$DR > GP_N$}}} & 0.27 (L)        & \multicolumn{1}{c|}{\cellcolor[HTML]{96FFFB}\makecell{1.18E-62 \\ \tiny{$DR > DT$}}} & 0.16  (L)       & \multicolumn{1}{c|}{\cellcolor[HTML]{96FFFB} \makecell{1.75E-58 \\ \tiny{$DR > Ensemble$}}} & 0.1 (L)        \\ \hline
\textbf{\textsc{AP--DHB}}                       & \multicolumn{1}{c|}{\cellcolor[HTML]{96FFFB}\makecell{2.34E-46 \\ \tiny{$DR > GP_T$}}} & 0.001 (L)        & \multicolumn{1}{c|}{\cellcolor[HTML]{96FFFB}{\makecell{1.14E-24 \\ \tiny{$DR > GP_O$}}}} & 0.01 (L)        & \multicolumn{1}{c|}{\cellcolor[HTML]{96FFFB}\makecell{5.19E-44 \\ \tiny{$DR > GP_N$}}} & 0  (L)       & \multicolumn{1}{c|}{\cellcolor[HTML]{96FFFB}\makecell{3.07E-67 \\ \tiny{$DR > DT$}}} & 0.03  (L)       & \multicolumn{1}{c|}{\cellcolor[HTML]{96FFFB} \makecell{8.41E-59 \\ \tiny{$DR > Ensemble$}}} & 0.01 (L)        \\ \hline

\textbf{\textsc{AP--TWN (R1)}}                        & \multicolumn{1}{c|}{\cellcolor[HTML]{96FFFB}\makecell{5.88E-38 \\ \tiny{$DR > GP_T$}}} & 0.19 (L)        & \multicolumn{1}{c|}{ \makecell{1 \\ \tiny{$DR \approx GP_O$}}} & 0.5 (N)        & \multicolumn{1}{c|}{\cellcolor[HTML]{96FFFB}\makecell{6.48E-36 \\ \tiny{$DR > GP_N$}}} & 0.18 (L)        & \multicolumn{1}{c|}{\cellcolor[HTML]{96FFFB}\makecell{2.09E-59 \\ \tiny{$DR > DT$}}} & 0 (L)        & \multicolumn{1}{c|}{\cellcolor[HTML]{96FFFB}\makecell{9.9E-38 \\ \tiny{$DR > Ensemble$}}} & 0.19 (L)        \\ \hline
\textbf{\textsc{AP--TWN (R2)}}                       & \multicolumn{1}{c|}{\makecell{1 \\ \tiny{$DR \approx GP_T$}}} & 0.5 (N)        & \multicolumn{1}{c|}{\cellcolor[HTML]{96FFFB}\makecell{1.61E-21 \\ \tiny{$DR > GP_O$}}} & 0.31  (M)       & \multicolumn{1}{c|}{\cellcolor[HTML]{96FFFB}\makecell{2.35E-25 \\ \tiny{$DR > GP_N$}}} & 0.28  (L)       & \multicolumn{1}{c|}{\makecell{1 \\ \tiny{$DR \approx DT$}}} & 0.5 (N)       & \multicolumn{1}{c|}{\makecell{1 \\ \tiny{$DR \approx Ensemble$}}} & 0.5  (N)       \\ \hline
\textbf{\textsc{AP--TWN (R3)}}                        & \multicolumn{1}{c|}{\cellcolor[HTML]{96FFFB}\makecell{0.005 \\ \tiny{$DR > GP_T$}}} & 0.48 (N)        & \multicolumn{1}{c|}{\cellcolor[HTML]{96FFFB} \makecell{7.63E-63 \\ \tiny{$DR > GP_O$}}} & 0 (L)        & \multicolumn{1}{c|}{\cellcolor[HTML]{96FFFB}\makecell{1.2E-50 \\ \tiny{$DR > GP_N$}}} & 0 (L)        & \multicolumn{1}{c|}{\makecell{1 \\ \tiny{$DR \approx DT$}}} & 0.5 (N)        & \multicolumn{1}{c|}{\cellcolor[HTML]{96FFFB} \makecell{0.004 \\ \tiny{$DR > Ensemble$}}} & 0.47 (N)        \\ \hline
\textbf{\textsc{AP--TWN (R4)}}                       & \multicolumn{1}{c|}{\cellcolor[HTML]{96FFFB}\makecell{0.0009 \\ \tiny{$DR > GP_T$}}} & 0.30 (M)        & \multicolumn{1}{c|}{\cellcolor[HTML]{96FFFB} \makecell{0.0001 \\ \tiny{$DR > GP_O$}}} & 0.29 (L)        & \multicolumn{1}{c|}{\cellcolor[HTML]{96FFFB}\makecell{0.001 \\ \tiny{$DR > GP_N$}}} & 0.29 (L)          & \multicolumn{2}{c|}{\cellcolor[HTML]{FFFC9E} DT is not applicable } &          \multicolumn{1}{c|}{\cellcolor[HTML]{96FFFB} \makecell{0.001 \\ \tiny{$DR > Ensemble$}}} & 0.30 (M)        \\ \hline
\textbf{\textsc{AP--SNG}}                        & \multicolumn{1}{c|}{\cellcolor[HTML]{96FFFB}\makecell{2.15E-13 \\ \tiny{$DR > GP_T$}}} & 0.27 (L)        & \multicolumn{1}{c|}{\cellcolor[HTML]{96FFFB} \makecell{6.34E-44 \\ \tiny{$DR > GP_O$}}} & 0.03 (L)        & \multicolumn{1}{c|}{\cellcolor[HTML]{96FFFB}\makecell{1.16E-45 \\ \tiny{$DR > GP_N$}}} & 0.03 (L)       & \multicolumn{1}{c|}{\cellcolor[HTML]{96FFFB}\makecell{2.42E-76 \\ \tiny{$DR > DT$}}} & 0.005 (L)        & \multicolumn{1}{c|}{\cellcolor[HTML]{96FFFB}\makecell{1.06E-71 \\ \tiny{$DR > Ensemble$}}} & 0.008 (L)        \\ \hline
\textbf{\textsc{Dave2}}                       & \multicolumn{1}{c|}{\cellcolor[HTML]{96FFFB}\makecell{1.23E-10 \\ \tiny{$DR > GP_T$}}} & 0.39 (S)        & \multicolumn{1}{c|}{\cellcolor[HTML]{96FFFB} \makecell{3.99E-42 \\ \tiny{$DR > GP_O$}}} & 0.09 (L)       & \multicolumn{1}{c|}{\cellcolor[HTML]{96FFFB}\makecell{3.11E-32 \\ \tiny{$DR > GP_N$}}} & 0.20 (L)       & \multicolumn{1}{c|}{\makecell{1 \\ \tiny{$DR \approx DT$}}} & 0.5 (N)        & \multicolumn{1}{c|}{\makecell{0.19 \\ \tiny{$DR \approx Ensemble$}}} & 0.49 (N)        \\ \hline
\end{tabular}}
\end{subtable}

\vspace{0.5cm}

\begin{subtable}{\linewidth}
\centering
\subcaption{Rate of fail verdicts predicted as pass}
\scalebox{0.9}{
\begin{tabular}{|c|cc|cc|cc|cc|cc|}
\hline
                                    & \multicolumn{2}{c|}{\textbf{$\mathit{GP_O}$ vs $\mathit{GP_T}$}}                       & \multicolumn{2}{c|}{\textbf{$\mathit{GP_O}$ vs $\mathit{GP_N}$}}                       & \multicolumn{2}{c|}{\textbf{$\mathit{GP_O}$ vs DT}}                           & \multicolumn{2}{c|}{\textbf{$\mathit{GP_O}$ vs DR}}                           & \multicolumn{2}{c|}{\textbf{$\mathit{GP_O}$ vs Ensemble}}                     \\ \cline{2-11} 
\multirow{-2}{*}{\textbf{Case-Study System}} & \multicolumn{1}{c|}{\textbf{P-value}}                 & $\mathbf{\hat{A}_{12}}$ & \multicolumn{1}{c|}{\textbf{P-value}}                 & $\mathbf{\hat{A}_{12}}$ & \multicolumn{1}{c|}{\textbf{P-value}}                 & $\mathbf{\hat{A}_{12}}$ & \multicolumn{1}{c|}{\textbf{P-value}}                 & $\mathbf{\hat{A}_{12}}$ & \multicolumn{1}{c|}{\textbf{P-value}}                 & $\mathbf{\hat{A}_{12}}$ \\ \hline
\textbf{Router}                        & \multicolumn{1}{c|}{\cellcolor[HTML]{F8A102}\makecell{1.88E-19 \\ \tiny{$GP_T > GP_O$}}} & 0.70 (M)        & \multicolumn{1}{c|}{ \makecell{0.18 \\ \tiny{$GP_O \approx GP_N$}}} & 0.53 (N)       & \multicolumn{1}{c|}{\cellcolor[HTML]{96FFFB}\makecell{0.006 \\ \tiny{$GP_O > DT$}}} & 0.42 (S)        & \multicolumn{1}{c|}{\cellcolor[HTML]{96FFFB}\makecell{0.0007 \\ \tiny{$GP_O > DR$}}} & 0.40  (S)       & \multicolumn{1}{c|}{\cellcolor[HTML]{F8A102} \makecell{9.37E-13 \\ \tiny{$Ensemble > GP_O$}}} & 0.67 (M)        \\ \hline
\textbf{\textsc{AP--DHB}}                       & \multicolumn{1}{c|}{\cellcolor[HTML]{96FFFB}\makecell{0.0007 \\ \tiny{$GP_O > GP_T$}}} & 0.33 (M)        & \multicolumn{1}{c|}{\makecell{0.04 \\ \tiny{$GP_O \approx GP_N$}}} & 0.40 (S)        & \multicolumn{1}{c|}{\cellcolor[HTML]{96FFFB}\makecell{0.02 \\ \tiny{$GP_O > DT$}}} & 0.42  (S)       & \multicolumn{1}{c|}{\cellcolor[HTML]{F8A102}\makecell{3.2E-10 \\ \tiny{$DR > GP_O$}}} & 0.78  (L)       & \multicolumn{1}{c|}{\cellcolor[HTML]{96FFFB} \makecell{0.002 \\ \tiny{$GP_O > Ensemble$}}} & 0.36 (M)        \\ \hline

\textbf{\textsc{AP--TWN (R1)}}                        & \multicolumn{1}{c|}{\makecell{0.19 \\ \tiny{$GP_O \approx GP_T$}}} & 0.59 (S)        & \multicolumn{1}{c|}{\cellcolor[HTML]{F8A102} \makecell{0.01 \\ \tiny{$GP_N > GP_O$}}} & 0.69 (M)        & \multicolumn{1}{c|}{\cellcolor[HTML]{F8A102}\makecell{9.02E-12 \\ \tiny{$DT > GP_O$}}} & 1 (L)        & \multicolumn{1}{c|}{\cellcolor[HTML]{F8A102}\makecell{5.06E-11 \\ \tiny{$DR > GP_O$}}} & 0.95 (L)        & \multicolumn{1}{c|}{\cellcolor[HTML]{F8A102}\makecell{0.04 \\ \tiny{$Ensemble > GP_O$}}} & 0.64 (M)        \\ \hline
\textbf{\textsc{AP--TWN (R2)}}                       & \multicolumn{1}{c|}{\cellcolor[HTML]{F8A102}\makecell{1.41E-06 \\ \tiny{$GP_T > GP_O$}}} & 0.66 (M)        & \multicolumn{1}{c|}{\makecell{0.24 \\ \tiny{$GP_O \approx GP_N$}}} & 0.54  (N)       & \multicolumn{1}{c|}{\cellcolor[HTML]{F8A102}\makecell{1.82E-07 \\ \tiny{$DT > GP_O$}}} & 0.66  (M)       & \multicolumn{1}{c|}{\cellcolor[HTML]{F8A102}\makecell{9.25E-36 \\ \tiny{$DR > GP_O$}}} & 0.81  (L)       & \multicolumn{1}{c|}{\cellcolor[HTML]{F8A102} \makecell{3.4E-06 \\ \tiny{$Ensemble > GP_O$}}} & 0.66  (M)       \\ \hline
\textbf{\textsc{AP--TWN (R3)}}                        & \multicolumn{1}{c|}{\makecell{0.1 \\ \tiny{$GP_O \approx GP_T$}}} & 0.48 (N)        & \multicolumn{1}{c|}{\makecell{1 \\ \tiny{$GP_O \approx GP_N$}}} & 0.5 (N)        & \multicolumn{1}{c|}{\makecell{1 \\ \tiny{$GP_O \approx DT$}}} & 0.5 (N)        & \multicolumn{1}{c|}{\cellcolor[HTML]{96FFFB}\makecell{0.0008 \\ \tiny{$GP_O > DR$}}} & 0.44 (S)        & \multicolumn{1}{c|}{\cellcolor[HTML]{96FFFB} \makecell{0.0003 \\ \tiny{$GP_O > Ensemble$}}} & 0.43 (S)        \\ \hline
\textbf{\textsc{AP--TWN (R4)}}                       & \multicolumn{1}{c|}{\cellcolor[HTML]{96FFFB}\makecell{0.04 \\ \tiny{$GP_O > GP_T$}}} & 0.46 (N)        & \multicolumn{1}{c|}{ \makecell{1 \\ \tiny{$GP_O \approx GP_N$}}} & 0.5 (N)        & \multicolumn{2}{c|}{\cellcolor[HTML]{FFFC9E} DT is not applicable}          & \multicolumn{1}{c|}{\cellcolor[HTML]{96FFFB}\makecell{7.01E-14 \\ \tiny{$GP_O > DR$}}} & 0.14 (L)        & \multicolumn{1}{c|}{\cellcolor[HTML]{96FFFB} \makecell{0.005 \\ \tiny{$GP_O > Ensemble$}}} & 0.44 (S)        \\ \hline
\textbf{\textsc{AP--SNG}}                        & \multicolumn{1}{c|}{\cellcolor[HTML]{96FFFB}\makecell{0.002 \\ \tiny{$GP_O > GP_T$}}} & 0.45 (S)        & \multicolumn{1}{c|}{ \makecell{1 \\ \tiny{$GP_O \approx GP_N$}}} & 0.5 (N)        & \multicolumn{1}{c|}{\makecell{1 \\ \tiny{$GP_O \approx DT$}}} & 0.5 (N)       & \multicolumn{1}{c|}{\cellcolor[HTML]{96FFFB}\makecell{2.61E-20 \\ \tiny{$GP_O > DR$}}} & 0.22 (L)        & \multicolumn{1}{c|}{\makecell{1 \\ \tiny{$GP_O \approx Ensemble$}}} & 0.5 (N)        \\ \hline
\textbf{\textsc{Dave2}}                       & \multicolumn{1}{c|}{\cellcolor[HTML]{96FFFB}\makecell{0.002 \\ \tiny{$GP_O > GP_T$}}} & 0.39 (S)        & \multicolumn{1}{c|}{\cellcolor[HTML]{96FFFB} \makecell{3.14E-05 \\ \tiny{$GP_O > GP_N$}}} & 0.34 (M)       & \multicolumn{1}{c|}{\cellcolor[HTML]{96FFFB}\makecell{5.33E-11 \\ \tiny{$GP_O > DT$}}} & 0.27 (L)       & \multicolumn{1}{c|}{\cellcolor[HTML]{96FFFB}\makecell{6.38E-07 \\ \tiny{$GP_O > DR$}}} & 0.31 (M)        & \multicolumn{1}{c|}{\cellcolor[HTML]{96FFFB}\makecell{4.78E-08 \\ \tiny{$GP_O > Ensemble$}}} & 0.29 (L)        \\ \hline
\end{tabular}}

\end{subtable}

\end{table*}

\begin{table*}[ht]
\caption{Statistical tests comparing the relative accuracy results of DR against those of $\mathit{GP_T}$, $\mathit{GP_O}$, $\mathit{GP_N}$, DT and ensemble. The p-values highlighted in blue represent cases where DR significantly outperforms the compared alternative. The significance level is $0.05$. }
\label{tab:relstat}
\centering
\scalebox{0.9}{
\begin{tabular}{|c|cc|cc|cc|cc|cc|}
\hline
                                    & \multicolumn{2}{c|}{\textbf{DR vs $\mathit{GP_T}$}}                       & \multicolumn{2}{c|}{\textbf{DR vs $\mathit{GP_O}$}}                       & \multicolumn{2}{c|}{\textbf{DR vs $\mathit{GP_N}$}}                           & \multicolumn{2}{c|}{\textbf{DR vs DT}}                           & \multicolumn{2}{c|}{\textbf{DR vs Ensemble}}                     \\ \cline{2-11} 
\multirow{-2}{*}{\textbf{$\theta$}} & \multicolumn{1}{c|}{\textbf{P-value}}                 & $\mathbf{\hat{A}_{12}}$ & \multicolumn{1}{c|}{\textbf{P-value}}                 & $\mathbf{\hat{A}_{12}}$ & \multicolumn{1}{c|}{\textbf{P-value}}                 & $\mathbf{\hat{A}_{12}}$ & \multicolumn{1}{c|}{\textbf{P-value}}                 & $\mathbf{\hat{A}_{12}}$ & \multicolumn{1}{c|}{\textbf{P-value}}                 & $\mathbf{\hat{A}_{12}}$ \\ \hline
\textbf{0.5}                        & \multicolumn{1}{c|}{\cellcolor[HTML]{96FFFB}\makecell{3.31E-07 \\ \tiny{$DR > GP_T$}}} & 0.66 (M)        & \multicolumn{1}{c|}{\cellcolor[HTML]{96FFFB}\makecell{5.89E-08 \\ \tiny{$DR > GP_O$}}} & 0.68 (M)& \multicolumn{1}{c|}{\cellcolor[HTML]{96FFFB}\makecell{2.38E-09 \\ \tiny{$DR > GP_N$}}} & 0.70 (M)& \multicolumn{1}{c|}{\cellcolor[HTML]{96FFFB}\makecell{0.001 \\ \tiny{$DR > DT$}}} & 0.60 (S)& \multicolumn{1}{c|}{\cellcolor[HTML]{96FFFB}\makecell{1.33E-08 \\ \tiny{$DR > Ensemble$}}} & 0.68 (M)\\ \hline
\textbf{0.55}                       & \multicolumn{1}{c|}{\cellcolor[HTML]{96FFFB}\makecell{1.39E-05 \\ \tiny{$DR > GP_T$}}} & 0.64 (M)        & \multicolumn{1}{c|}{\cellcolor[HTML]{96FFFB}\makecell{1.83E-07 \\ \tiny{$DR > GP_O$}}} & 0.67 (M)& \multicolumn{1}{c|}{\cellcolor[HTML]{96FFFB}\makecell{3.4E-10 \\ \tiny{$DR > GP_N$}}} & 0.71 (L)& \multicolumn{1}{c|}{\cellcolor[HTML]{96FFFB}\makecell{8.32E-05 \\ \tiny{$DR > DT$}}} & 0.62 (S)& \multicolumn{1}{c|}{\cellcolor[HTML]{96FFFB}\makecell{2.37E-07 \\ \tiny{$DR > Ensemble$}}} & 0.66 (M)\\ \hline
\textbf{0.6}                        & \multicolumn{1}{c|}{\cellcolor[HTML]{96FFFB}\makecell{2.15E-05 \\ \tiny{$DR > GP_T$}}} & 0.63 (S)        & \multicolumn{1}{c|}{\cellcolor[HTML]{96FFFB}\makecell{3.4E-05 \\ \tiny{$DR > GP_O$}}} & 0.64 (M)& \multicolumn{1}{c|}{\cellcolor[HTML]{96FFFB}\makecell{6.58E-10 \\ \tiny{$DR > GP_N$}}} & 0.70 (M)& \multicolumn{1}{c|}{\cellcolor[HTML]{96FFFB}\makecell{8.32E-05 \\ \tiny{$DR > DT$}}} & 0.62 (S)& \multicolumn{1}{c|}{\cellcolor[HTML]{96FFFB}\makecell{2.36E-07 \\ \tiny{$DR > Ensemble$}}} & 0.66 (M)\\ \hline
\textbf{0.65}                       & \multicolumn{1}{c|}{\cellcolor[HTML]{96FFFB}\makecell{2.09E-06 \\ \tiny{$DR > GP_T$}}} & 0.65 (M)        & \multicolumn{1}{c|}{\cellcolor[HTML]{96FFFB}\makecell{0.001 \\ \tiny{$DR > GP_O$}}} & 0.62  (S)& \multicolumn{1}{c|}{\cellcolor[HTML]{96FFFB}\makecell{1.13E-10 \\ \tiny{$DR > GP_N$}}} & 0.72 (L)& \multicolumn{1}{c|}{\cellcolor[HTML]{96FFFB}\makecell{0.0009 \\ \tiny{$DR > DT$}}} & 0.61  (S)& \multicolumn{1}{c|}{\cellcolor[HTML]{96FFFB}\makecell{2.98E-07 \\ \tiny{$DR > Ensemble$}}} & 0.66  (M)\\ \hline
\textbf{0.7}                        & \multicolumn{1}{c|}{\cellcolor[HTML]{96FFFB}\makecell{9.62E-06 \\ \tiny{$DR > GP_T$}}} & 0.64 (M)        & \multicolumn{1}{c|}{\makecell{0.07 \\ \tiny{$DR \approx GP_O$}}} & 0.57 (S)& \multicolumn{1}{c|}{\cellcolor[HTML]{96FFFB}\makecell{2.79E-06 \\ \tiny{$DR > GP_N$}}} & 0.67 (M)& \multicolumn{1}{c|}{\cellcolor[HTML]{96FFFB}\makecell{0.0003 \\ \tiny{$DR > DT$}}} & 0.61 (S)& \multicolumn{1}{c|}{\cellcolor[HTML]{96FFFB}\makecell{8.32E-08 \\ \tiny{$DR > Ensemble$}}} & 0.67 (M)\\ \hline
\textbf{0.75}                       & \multicolumn{1}{c|}{\cellcolor[HTML]{96FFFB}\makecell{2.89E-06 \\ \tiny{$DR > GP_T$}}} & 0.66 (M)        & \multicolumn{1}{c|}{\cellcolor[HTML]{96FFFB} \makecell{0.04 \\ \tiny{$DR > GP_O$}}} & 0.59 (S)& \multicolumn{1}{c|}{\cellcolor[HTML]{96FFFB}\makecell{0.005 \\ \tiny{$DR > GP_N$}}} & 0.61 (S)& \multicolumn{1}{c|}{\cellcolor[HTML]{96FFFB}\makecell{5.57E-09 \\ \tiny{$DR > DT$}}} & 0.69 (M)& \multicolumn{1}{c|}{\cellcolor[HTML]{96FFFB}\makecell{1.54E-13 \\ \tiny{$DR > Ensemble$}}} & 0.74 (L)\\ \hline
\textbf{0.8}                        & \multicolumn{1}{c|}{\cellcolor[HTML]{96FFFB}\makecell{0.0002 \\ \tiny{$DR > GP_T$}}} & 0.62 (S)        & \multicolumn{1}{c|}{\cellcolor[HTML]{96FFFB}\makecell{5.82E-08 \\ \tiny{$DR > GP_O$}}} & 0.80  (L)& \multicolumn{1}{c|}{\cellcolor[HTML]{96FFFB}\makecell{8.56E-05 \\ \tiny{$DR > GP_N$}}} & 0.66 (M)        & \multicolumn{1}{c|}{\cellcolor[HTML]{96FFFB}\makecell{1.38E-12 \\ \tiny{$DR > DT$}}} & 0.74  (L)& \multicolumn{1}{c|}{\cellcolor[HTML]{96FFFB}\makecell{1.24E-14 \\ \tiny{$DR > Ensemble$}}} & 0.76 (L)\\ \hline
\textbf{0.85}                       & \multicolumn{1}{c|}{\cellcolor[HTML]{96FFFB}\makecell{0.002 \\ \tiny{$DR > GP_T$}}} & 0.60 (S)        & \multicolumn{1}{c|}{\cellcolor[HTML]{96FFFB}\makecell{4.08E-09 \\ \tiny{$DR > GP_O$}}} & 0.85 (L)& \multicolumn{1}{c|}{\cellcolor[HTML]{96FFFB}\makecell{2.92E-09 \\ \tiny{$DR > GP_N$}}} & 0.83  (L)       & \multicolumn{1}{c|}{\cellcolor[HTML]{96FFFB}\makecell{3.01E-15 \\ \tiny{$DR > DT$}}} & 0.76  (L)& \multicolumn{1}{c|}{\cellcolor[HTML]{96FFFB}\makecell{1.57E-14 \\ \tiny{$DR > Ensemble$}}} & 0.76 (L)        \\ \hline
\textbf{0.9}                        & \multicolumn{1}{c|}{\cellcolor[HTML]{96FFFB} \makecell{0.01 \\ \tiny{$DR > GP_T$}}} & 0.59 (S)         & \multicolumn{1}{c|}{\cellcolor[HTML]{96FFFB}\makecell{2.88E-05 \\ \tiny{$DR > GP_O$}}}                             & 0.88 (L)        & \multicolumn{1}{c|}{\cellcolor[HTML]{96FFFB}\makecell{6.35E-06 \\ \tiny{$DR > GP_N$}}}                             & 0.86 (L)        & \multicolumn{1}{c|}{\cellcolor[HTML]{96FFFB}\makecell{1.19E-09 \\ \tiny{$DR > DT$}}} & 0.70 (M)& \multicolumn{1}{c|}{\cellcolor[HTML]{96FFFB}\makecell{1.68E-08 \\ \tiny{$DR > Ensemble$}}}                             & 0.69 (M)\\ \hline
\textbf{0.95}                       & \multicolumn{1}{c|}{\makecell{0.09 \\ \tiny{$DR \approx GP_T$}}} & 0.56 (N)        & \multicolumn{1}{c|}{\cellcolor[HTML]{96FFFB}\makecell{0.0007 \\ \tiny{$DR > GP_O$}}}                             & 0.88 (L)        & \multicolumn{1}{c|}{\cellcolor[HTML]{96FFFB}\makecell{0.0002 \\ \tiny{$DR > GP_N$}}}                             & 0.84 (L)        & \multicolumn{1}{c|}{\cellcolor[HTML]{96FFFB}\makecell{0.0004 \\ \tiny{$DR > DT$}}} & 0.62 (S)& \multicolumn{1}{c|}{\cellcolor[HTML]{96FFFB}\makecell{0.006 \\ \tiny{$DR > Ensemble$}}}                             & 0.59 (S)        \\ \hline
\textbf{1}                          & \multicolumn{1}{c|}{\cellcolor[HTML]{96FFFB} \makecell{0.04 \\ \tiny{$DR > GP_T$}}} & 0.57 (S)        & \multicolumn{1}{c|}{\cellcolor[HTML]{96FFFB}\makecell{0.001 \\ \tiny{$DR > GP_O$}}}                             & 0.87 (L)        & \multicolumn{1}{c|}{\cellcolor[HTML]{96FFFB}\makecell{0.004 \\ \tiny{$DR > GP_N$}}}                             & 0.79 (L)         & \multicolumn{1}{c|}{\cellcolor[HTML]{96FFFB}\makecell{4.59E-05 \\ \tiny{$DR > DT$}}} & 0.66 (M)& \multicolumn{1}{c|}{\cellcolor[HTML]{96FFFB}\makecell{0.01 \\ \tiny{$DR > Ensemble$}}}                             & 0.59 (S)        \\ \hline
\end{tabular}}

\end{table*}
\begin{table*}[ht]
\caption{Statistical tests comparing the relative accuracy results of DR against those of $\mathit{GP_T}$, $\mathit{GP_O}$, $\mathit{GP_N}$, DT and ensemble for each case-study system. The p-values highlighted in blue represent cases where $DR$ significantly outperforms the compared alternative. The p-values highlighted in orange indicate cases where the compared alternative outperforms $DR$. The cells highlighted in yellow represent cases where DT is not applicable. The significance level is $0.05$.}
\label{tab:relaccstatperstudy}

\centering
\scalebox{0.9}{
\begin{tabular}{|c|cc|cc|cc|cc|cc|}
\hline
                                    & \multicolumn{2}{c|}{\textbf{DR vs $\mathit{GP_T}$}}                       & \multicolumn{2}{c|}{\textbf{DR vs $\mathit{GP_O}$}}                       & \multicolumn{2}{c|}{\textbf{DR vs $G_N$}}                           & \multicolumn{2}{c|}{\textbf{DR vs DT}}                           & \multicolumn{2}{c|}{\textbf{DR vs Ensemble}}                     \\ \cline{2-11} 
\multirow{-2}{*}{\textbf{Case-Study System}} & \multicolumn{1}{c|}{\textbf{P-value}}                 & $\mathbf{\hat{A}_{12}}$ & \multicolumn{1}{c|}{\textbf{P-value}}                 & $\mathbf{\hat{A}_{12}}$ & \multicolumn{1}{c|}{\textbf{P-value}}                 & $\mathbf{\hat{A}_{12}}$ & \multicolumn{1}{c|}{\textbf{P-value}}                 & $\mathbf{\hat{A}_{12}}$ & \multicolumn{1}{c|}{\textbf{P-value}}                 & $\mathbf{\hat{A}_{12}}$ \\ \hline
\textbf{Router}                        & \multicolumn{1}{c|}{\cellcolor[HTML]{F8A102}\makecell{0.0001 \\ \tiny{$GP_T > DR$}}} & 0.40 (S)        & \multicolumn{1}{c|}{\cellcolor[HTML]{96FFFB} \makecell{6.72E-18 \\ \tiny{$DR > GP_O$}}} & 0.74 (L)       & \multicolumn{1}{c|}{\cellcolor[HTML]{96FFFB}\makecell{4.3E-09 \\ \tiny{$DR > GP_N$}}} & 0.65 (M)        & \multicolumn{1}{c|}{\cellcolor[HTML]{96FFFB}\makecell{3.99E-14 \\ \tiny{$DR > DT$}}} & 0.65 (M)       & \multicolumn{1}{c|}{\cellcolor[HTML]{F8A102} \makecell{2.6E-05 \\ \tiny{$Ensemble > DR$}}} & 0.39 (S)        \\ \hline
\textbf{\textsc{AP--DHB}}                       & \multicolumn{1}{c|}{\cellcolor[HTML]{96FFFB}\makecell{0.04 \\ \tiny{$DR > GP_T$}}} & 0.56 (S)        & \multicolumn{1}{c|}{\cellcolor[HTML]{96FFFB}\makecell{0.01 \\ \tiny{$DR > GP_O$}}} & 0.61 (S)        & \multicolumn{1}{c|}{\cellcolor[HTML]{96FFFB}\makecell{0.01 \\ \tiny{$DR > GP_N$}}} & 0.59  (S)       & \multicolumn{1}{c|}{\makecell{0.23 \\ \tiny{$DR \approx DT$}}} & 0.53  (N)       & \multicolumn{1}{c|}{\makecell{0.06 \\ \tiny{$DR \approx Ensemble$}}} & 0.55 (N)        \\ \hline

\textbf{\textsc{AP--TWN (R1)}}                        & \multicolumn{1}{c|}{\cellcolor[HTML]{96FFFB}\makecell{3.43E-39 \\ \tiny{$DR > GP_T$}}} & 0.89 (L)        & \multicolumn{1}{c|}{\cellcolor[HTML]{96FFFB} \makecell{5.06E-11 \\ \tiny{$DR > GP_O$}}} & 0.95 (L)        & \multicolumn{1}{c|}{\cellcolor[HTML]{96FFFB}\makecell{1.85E-37 \\ \tiny{$DR > GP_N$}}} & 0.98 (L)        & \multicolumn{1}{c|}{\cellcolor[HTML]{96FFFB}\makecell{2.64E-34 \\ \tiny{$DR > DT$}}} & 1 (L)        & \multicolumn{1}{c|}{\cellcolor[HTML]{96FFFB}\makecell{4.35E-51 \\ \tiny{$DR > Ensemble$}}} & 0.97 (L)        \\ \hline
\textbf{\textsc{AP--TWN (R2)}}                       & \multicolumn{1}{c|}{\cellcolor[HTML]{96FFFB}\makecell{1.17E-49 \\ \tiny{$DR > GP_T$}}} & 0.85 (L)        & \multicolumn{1}{c|}{\cellcolor[HTML]{96FFFB}\makecell{7.52E-66 \\ \tiny{$DR > GP_O$}}} & 1  (L)       & \multicolumn{1}{c|}{\cellcolor[HTML]{96FFFB}\makecell{1.68E-66 \\ \tiny{$DR > GP_N$}}} & 1  (L)       & \multicolumn{1}{c|}{\cellcolor[HTML]{96FFFB}\makecell{2.36E-66 \\ \tiny{$DR > DT$}}} & 0.90  (L)       & \multicolumn{1}{c|}{\cellcolor[HTML]{96FFFB} \makecell{6.82E-62 \\ \tiny{$DR > Ensemble$}}} & 0.91  (L)       \\ \hline
\textbf{\textsc{AP--TWN (R3)}}                        & \multicolumn{1}{c|}{\makecell{0.08 \\ \tiny{$DR \approx GP_T$}}} & 0.47 (N)        & \multicolumn{1}{c|}{\cellcolor[HTML]{96FFFB} \makecell{2.74E-33 \\ \tiny{$DR > GP_O$}}} & 0.88 (L)        & \multicolumn{1}{c|}{\cellcolor[HTML]{96FFFB}\makecell{2.06E-21 \\ \tiny{$DR > GP_N$}}} & 0.88 (L)        & \multicolumn{1}{c|}{\cellcolor[HTML]{96FFFB}\makecell{2.4E-07 \\ \tiny{$DR > DT$}}} & 0.44 (S)        & \multicolumn{1}{c|}{ \makecell{0.23 \\ \tiny{$DR \approx Ensemble$}}} & 0.52 (N)        \\ \hline
\textbf{\textsc{AP--TWN (R4)}}                       & \multicolumn{1}{c|}{\cellcolor[HTML]{96FFFB}\makecell{2.02E-15 \\ \tiny{$DR > GP_T$}}} & 0.02 (L)        & \multicolumn{1}{c|}{\cellcolor[HTML]{96FFFB} \makecell{5.77E-21 \\ \tiny{$DR > GP_O$}}} & 0 (L)        & \multicolumn{1}{c|}{\cellcolor[HTML]{96FFFB} \makecell{1.31E-15 \\ \tiny{$DR > GP_N$}}} & 0 (L)         & \multicolumn{2}{c|}{\cellcolor[HTML]{FFFC9E} DT is not applicable}         & \multicolumn{1}{c|}{\cellcolor[HTML]{96FFFB} \makecell{5.77E-10 \\ \tiny{$DR > Ensemble$}}} & 0.12 (L)        \\ \hline
\textbf{\textsc{AP--SNG}}                        & \multicolumn{1}{c|}{\cellcolor[HTML]{96FFFB}\makecell{8.78E-08 \\ \tiny{$DR > GP_T$}}} & 0.66 (M)        & \multicolumn{1}{c|}{\cellcolor[HTML]{96FFFB} \makecell{2.69E-40\\ \tiny{$DR > GP_O$}}} & 0.94 (L)        & \multicolumn{1}{c|}{\cellcolor[HTML]{96FFFB}\makecell{6.89E-42 \\ \tiny{$DR > GP_N$}}} & 0.94 (L)       & \multicolumn{1}{c|}{\cellcolor[HTML]{96FFFB}\makecell{8.92E-62 \\ \tiny{$DR > DT$}}} & 0.94 (L)        & \multicolumn{1}{c|}{\cellcolor[HTML]{96FFFB}\makecell{2.97E-58 \\ \tiny{$DR > Ensemble$}}} & 0.94 (L)        \\ \hline
\textbf{\textsc{Dave2}}                       & \multicolumn{1}{c|}{\cellcolor[HTML]{96FFFB}\makecell{0.002 \\ \tiny{$DR > GP_T$}}} & 0.59 (S)        & \multicolumn{1}{c|}{\cellcolor[HTML]{96FFFB} \makecell{1.11E-10 \\ \tiny{$DR > GP_O$}}} & 0.74 (L)       & \multicolumn{1}{c|}{\cellcolor[HTML]{96FFFB}\makecell{9.29E-13 \\ \tiny{$DR > GP_N$}}} & 0.72 (L)       & \multicolumn{1}{c|}{\cellcolor[HTML]{96FFFB}\makecell{1.38E-31 \\ \tiny{$DR > DT$}}} & 0.82 (L)        & \multicolumn{1}{c|}{\cellcolor[HTML]{96FFFB}\makecell{3.73E-12 \\ \tiny{$DR > Ensemble$}}} & 0.70 (L)        \\ \hline
\end{tabular}}

\end{table*}

\begin{table*}[ht]
\centering
\caption{Statistical test results comparing the accuracy of $\mathit{GP_O}$ against $\mathit{GP_T}$, $\mathit{GP_N}$, DT, DR, and the ensemble when datasets $TS_1$ to $TS_{10}$ are used. The p-values highlighted in blue indicate cases where $\mathit{GP_O}$ significantly outperforms the alternative it is being compared to. The significance level is $0.05$.}
\label{tab:RQ3aarstat}
\scalebox{0.89}{
\begin{tabular}{|c|cc|cc|cc|cc|cc|}
\hline
                                    & \multicolumn{2}{c|}{\textbf{$\mathit{GP_O}$ vs $\mathit{GP_T}$}}                       & \multicolumn{2}{c|}{\textbf{$\mathit{GP_O}$ vs $\mathit{GP_N}$}}                       & \multicolumn{2}{c|}{\textbf{$\mathit{GP_O}$ vs DT}}                           & \multicolumn{2}{c|}{\textbf{$\mathit{GP_O}$ vs DR}}                           & \multicolumn{2}{c|}{\textbf{$\mathit{GP_O}$ vs Ensemble}}                     \\ \cline{2-11} 
\multirow{-2}{*}{\textbf{$\theta$}} & \multicolumn{1}{c|}{\textbf{P-value}}                 & $\mathbf{\hat{A}_{12}}$ & \multicolumn{1}{c|}{\textbf{P-value}}                 & $\mathbf{\hat{A}_{12}}$ & \multicolumn{1}{c|}{\textbf{P-value}}                 & $\mathbf{\hat{A}_{12}}$ & \multicolumn{1}{c|}{\textbf{P-value}}                 & $\mathbf{\hat{A}_{12}}$ & \multicolumn{1}{c|}{\textbf{P-value}}                 & $\mathbf{\hat{A}_{12}}$ \\ \hline
\textbf{0.5}                        & \multicolumn{1}{c|}{\cellcolor[HTML]{96FFFB}\makecell{1E-23 \\ \tiny{$GP_O > GP_T$}}} & 0.94 (L)& \multicolumn{1}{c|}{\cellcolor[HTML]{96FFFB}\makecell{2.15E-05 \\ \tiny{$GP_O > GP_N$}}} & 0.60 (S)& \multicolumn{1}{c|}{\cellcolor[HTML]{96FFFB}\makecell{3.29E-20 \\ \tiny{$GP_O > DT$}}} & 0.92 (L)& \multicolumn{1}{c|}{\cellcolor[HTML]{96FFFB}\makecell{1.47E-25 \\ \tiny{$GP_O > DR$}}} & 0.98 (L)& \multicolumn{1}{c|}{\cellcolor[HTML]{96FFFB}\makecell{2.74E-21 \\ \tiny{$GP_O > Ensemble$}}} & 0.89 (L)\\ \hline
\textbf{0.55}                       & \multicolumn{1}{c|}{\cellcolor[HTML]{96FFFB}\makecell{3.73E-23 \\ \tiny{$GP_O > GP_T$}}} & 0.94 (L)& \multicolumn{1}{c|}{\cellcolor[HTML]{96FFFB}\makecell{3.29E-06 \\ \tiny{$GP_O > GP_N$}}} & 0.61 (S)& \multicolumn{1}{c|}{\cellcolor[HTML]{96FFFB}\makecell{2.22E-19 \\ \tiny{$GP_O > DT$}}} & 0.91 (L)& \multicolumn{1}{c|}{\cellcolor[HTML]{96FFFB}\makecell{3.71E-25 \\ \tiny{$GP_O > DR$}}} & 0.97 (L)& \multicolumn{1}{c|}{\cellcolor[HTML]{96FFFB}\makecell{4.74E-21 \\ \tiny{$GP_O > Ensemble$}}} & 0.88 (L)\\ \hline
\textbf{0.6}                        & \multicolumn{1}{c|}{\cellcolor[HTML]{96FFFB}\makecell{1.23E-21 \\ \tiny{$GP_O > GP_T$}}} & 0.94 (L)& \multicolumn{1}{c|}{\cellcolor[HTML]{96FFFB}\makecell{3.83E-06 \\ \tiny{$GP_O > GP_N$}}} & 0.66 (M)& \multicolumn{1}{c|}{\cellcolor[HTML]{96FFFB}\makecell{6.28E-18 \\ \tiny{$GP_O > DT$}}} & 0.91 (L)& \multicolumn{1}{c|}{\cellcolor[HTML]{96FFFB}\makecell{2.63E-23 \\ \tiny{$GP_O > DR$}}} & 0.97 (L)& \multicolumn{1}{c|}{\cellcolor[HTML]{96FFFB}\makecell{1.29E-19 \\ \tiny{$GP_O > Ensemble$}}} & 0.89 (L)\\ \hline
\textbf{0.65}                       & \multicolumn{1}{c|}{\cellcolor[HTML]{96FFFB}\makecell{7.65E-20 \\ \tiny{$GP_O > GP_T$}}} & 0.95 (L)& \multicolumn{1}{c|}{\cellcolor[HTML]{96FFFB}\makecell{8.94E-06 \\ \tiny{$GP_O > GP_N$}}} & 0.70 (M)& \multicolumn{1}{c|}{\cellcolor[HTML]{96FFFB}\makecell{5.28E-16 \\ \tiny{$GP_O > DT$}}} & 0.90 (L)& \multicolumn{1}{c|}{\cellcolor[HTML]{96FFFB}\makecell{1.47E-20 \\ \tiny{$GP_O > DR$}}} & 0.96 (L)& \multicolumn{1}{c|}{\cellcolor[HTML]{96FFFB}\makecell{5.07E-17 \\ \tiny{$GP_O > Ensemble$}}} & 0.90 (L)\\ \hline
\textbf{0.7}                        & \multicolumn{1}{c|}{\cellcolor[HTML]{96FFFB}\makecell{5.14E-18 \\ \tiny{$GP_O > GP_T$}}} & 0.97 (L)& \multicolumn{1}{c|}{\cellcolor[HTML]{96FFFB}\makecell{1.97E-06 \\ \tiny{$GP_O > GP_N$}}} & 0.78 (L)& \multicolumn{1}{c|}{\cellcolor[HTML]{96FFFB}\makecell{2.79E-14 \\ \tiny{$GP_O > DT$}}} & 0.93 (L)& \multicolumn{1}{c|}{\cellcolor[HTML]{96FFFB}\makecell{1.13E-17 \\ \tiny{$GP_O > DR$}}} & 0.96 (L)& \multicolumn{1}{c|}{\cellcolor[HTML]{96FFFB}\makecell{8.92E-15 \\ \tiny{$GP_O > Ensemble$}}} & 0.92 (L)\\ \hline
\textbf{0.75}                       & \multicolumn{1}{c|}{\cellcolor[HTML]{96FFFB}\makecell{8.1E-14 \\ \tiny{$GP_O > GP_T$}}} & 0.98 (L)& \multicolumn{1}{c|}{\cellcolor[HTML]{96FFFB}\makecell{5.34E-08 \\ \tiny{$GP_O > GP_N$}}} & 0.86 (L)& \multicolumn{1}{c|}{\cellcolor[HTML]{96FFFB}\makecell{3.37E-12 \\ \tiny{$GP_O > DT$}}} & 0.94 (L)& \multicolumn{1}{c|}{\cellcolor[HTML]{96FFFB}\makecell{8.78E-14 \\ \tiny{$GP_O > DR$}}} & 0.98 (L)& \multicolumn{1}{c|}{\cellcolor[HTML]{96FFFB}\makecell{7.87E-13 \\ \tiny{$GP_O > Ensemble$}}} & 0.95 (L)\\ \hline
\textbf{0.8}                        & \multicolumn{1}{c|}{\cellcolor[HTML]{96FFFB}\makecell{1.1E-11 \\ \tiny{$GP_O > GP_T$}}} & 0.98 (L)& \multicolumn{1}{c|}{\cellcolor[HTML]{96FFFB}\makecell{1.71E-07 \\ \tiny{$GP_O > GP_N$}}} & 0.86 (L)& \multicolumn{1}{c|}{\cellcolor[HTML]{96FFFB}\makecell{1.61E-10 \\ \tiny{$GP_O > DT$}}} & 0.94 (L)& \multicolumn{1}{c|}{\cellcolor[HTML]{96FFFB}\makecell{7.45E-12 \\ \tiny{$GP_O > DR$}}} & 0.98 (L)& \multicolumn{1}{c|}{\cellcolor[HTML]{96FFFB}\makecell{2.81E-11 \\ \tiny{$GP_O > Ensemble$}}} & 0.96 (L)\\ \hline
\textbf{0.85}                       & \multicolumn{1}{c|}{\cellcolor[HTML]{96FFFB}\makecell{5.21E-10 \\ \tiny{$GP_O > GP_T$}}} & 0.99 (L)& \multicolumn{1}{c|}{\cellcolor[HTML]{96FFFB}\makecell{5.49E-06 \\ \tiny{$GP_O > GP_N$}}} & 0.84 (L)& \multicolumn{1}{c|}{\cellcolor[HTML]{96FFFB}\makecell{2.06E-09 \\ \tiny{$GP_O > DT$}}} & 0.97 (L)& \multicolumn{1}{c|}{\cellcolor[HTML]{96FFFB}\makecell{2.9E-10 \\ \tiny{$GP_O > DR$}}} & 0.99 (L)& \multicolumn{1}{c|}{\cellcolor[HTML]{96FFFB}\makecell{5.21E-10 \\ \tiny{$GP_O > Ensemble$}}} & 0.99 (L)\\ \hline
\textbf{0.9}                        & \multicolumn{1}{c|}{\cellcolor[HTML]{96FFFB}\makecell{2.1E-04 \\ \tiny{$GP_O > GP_T$}}} & 0.99 (L)& \multicolumn{1}{c|}{\cellcolor[HTML]{96FFFB}\makecell{0.01 \\ \tiny{$GP_O > GP_N$}}}                             & 0.89 (L)& \multicolumn{1}{c|}{\cellcolor[HTML]{96FFFB}\makecell{0.001 \\ \tiny{$GP_O > DT$}}}                             & 0.95 (L)& \multicolumn{1}{c|}{\cellcolor[HTML]{96FFFB}\makecell{0.0001 \\ \tiny{$GP_O > DR$}}} & 0.99 (L)& \multicolumn{1}{c|}{\cellcolor[HTML]{96FFFB}\makecell{1.39E-06 \\ \tiny{$GP_O > Ensemble$}}}                             & 0.97 (L)\\ \hline
\textbf{0.95}                       & \multicolumn{1}{c|}{\cellcolor[HTML]{96FFFB}\makecell{0.00021 \\ \tiny{$GP_O > GP_T$}}} & 0.99 (L)& \multicolumn{1}{c|}{\cellcolor[HTML]{96FFFB}\makecell{0.01 \\ \tiny{$GP_O > GP_N$}}}                             & 0.82 (L)& \multicolumn{1}{c|}{\cellcolor[HTML]{96FFFB}\makecell{0.001 \\ \tiny{$GP_O > DT$}}}                             & 0.92 (L)& \multicolumn{1}{c|}{\cellcolor[HTML]{96FFFB}\makecell{0.0001 \\ \tiny{$GP_O > DR$}}} & 0.99 (L)& \multicolumn{1}{c|}{\cellcolor[HTML]{96FFFB}\makecell{0.0002 \\ \tiny{$GP_O > Ensemble$}}}                             & 0.95 (L)\\ \hline
\textbf{1}                          & \multicolumn{1}{c|}{\cellcolor[HTML]{96FFFB}\makecell{0.009 \\ \tiny{$GP_O > GP_T$}}} & 0.92 (L)& \multicolumn{1}{c|}{\makecell{$0.08$ \\ \tiny{$GP_O \approx GP_N$}}}                             & 0.83 (L)& \multicolumn{1}{c|}{\cellcolor[HTML]{96FFFB}\makecell{0.02 \\ \tiny{$GP_O > DT$}}}                             & 0.91 (L)& \multicolumn{1}{c|}{\cellcolor[HTML]{96FFFB}\makecell{0.003 \\ \tiny{$GP_O > DR$}}} & 0.99 (L)& \multicolumn{1}{c|}{\cellcolor[HTML]{96FFFB}\makecell{0.005 \\ \tiny{$GP_O > Ensemble$}}}                             & 0.95 (L)\\ \hline
\end{tabular}}
\end{table*}

\clearpage
\section{Prompt template for RQ4}
\label{apx:prompt}
Figure~\ref{fig:prompttemplate} presents the outline of the prompt used to identify related sentences from the reference documentation for an assertion.

\begin{figure}[h]
  \centering
  \begin{tcolorbox}[
    width=\linewidth,
    colback=white,
    colframe=black,
    boxrule=0.6pt,
    arc=0pt,
    left=4mm,right=4mm,top=2mm,bottom=2mm,
  ]
  Context: {\color{blue}\texttt{\{System description\}}}

  \vspace{4pt}

  Task: Given a natural-language assertion and a collection of documents, your task is to identify and return sentences that address the same condition, behaviour, or outcome as the assertion.

  \vspace{4pt}
    
  Inputs:
  \begin{itemize}
      \item Assertion: {\color{blue}\texttt{\{Assertion text\}}}
      \item Documentation: {\color{blue}\texttt{\{List of documents}\}}
  \end{itemize}

    \vspace{4pt}
    Instructions:
    \begin{enumerate}
        \item Retrieve between 2 and 5 sentences from the documentation that best support the assertion. Return 2--3 sentences if only a few are highly related. Return 4--5 sentences if several are strongly related.
        \item Exclude sentences that are loosely related or tangential.
        \item Preserve the original sentence wording exactly as it appears in the document.
        \item Output only the selected sentences, each on a new line; no explanations or metadata.
    \end{enumerate}

    Example: {\color{blue}\texttt{\{One-shot example}\}}

  \end{tcolorbox}
  \caption{Our prompt template}
  \label{fig:prompttemplate}
\end{figure}

\end{document}